\documentclass[aps,prc,preprint,showpacs]{revtex4}

\usepackage{graphicx}
\usepackage{dcolumn}
\usepackage{bm}
\usepackage{color}
\usepackage{subfigure}
\usepackage{epsfig}
\usepackage{amssymb}
\usepackage{amsmath}

\begin{document}

\title{Relativistic eikonal description of $A(p,pN)$ reactions}

\author{B. Van Overmeire}
\email[]{Bart.VanOvermeire@UGent.be}
\author{W. Cosyn}
\author{P. Lava}
\author{J. Ryckebusch}
\affiliation{Department of Subatomic and Radiation Physics, \\
Ghent University, Proeftuinstraat 86, B-9000 Gent, Belgium}

\date{\today}

\begin{abstract}
The authors present a relativistic and cross-section factorized
framework for computing quasielastic $A(p,pN)$ observables at
intermediate and high energies.  The model is based on the eikonal
approximation and can accomodate both optical potentials and the
Glauber method for dealing with the initial- and final-state interactions
(IFSI).  At lower nucleon energies, the optical-potential philosophy
is preferred, whereas at higher energies the Glauber method is more
natural.  This versatility in dealing with the IFSI allows one to describe
$A(p,pN)$ reactions in a wide energy range.  Most results presented here
use optical potentials as this approach is argued to be the optimum choice
for the kinematics of the experiments considered in the present paper.  The
properties of the IFSI factor, a function wherein the entire effect of the
IFSI is contained, are studied in detail.  The predictions of the presented
framework are compared with two kinematically different experiments.  First,
differential cross sections for quasielastic proton scattering at $1$~GeV
off $^{12}$C, $^{16}$O, and $^{40}$Ca target nuclei are computed and
compared to data from PNPI.  Second, the formalism is applied to the analysis
of a $^{4}$He$(p,2p)$ experiment at $250$~MeV.  The optical-potential
calculations are found to be in good agreement with the data from both
experiments, showing the reliability of the adopted model in a wide energy
range.
\end{abstract}


\maketitle

\section{\label{sec:intro} Introduction}

\begin{figure}
\begin{center}
\includegraphics[width=0.5\textwidth]{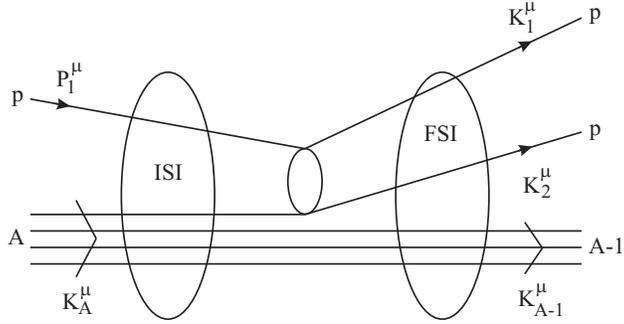}
\caption{Schematic representation of the $A(p,2p)$ reaction.  The
  incoming proton undergoes ``soft'' initial-state interactions with
  the target before knocking out a bound proton through the primary
  high-momentum-transfer $pp$ scattering.  Both the scattered and the
  ejected proton suffer final-state interactions while leaving the
  nucleus.  The scattered and the ejected proton are detected in
  coincidence, while the residual nucleus remains unobserved.}
\label{fig:schem_repres}
\end{center}
\end{figure}

Quasielastic nucleon knockout reactions have been extensively investigated
with the aim of obtaining precise information on nuclear structure.  The
present work focuses on exclusive proton-induced $A(p,2p)$ and
$A(p,pn)$ processes, whereby the residual $\mbox{A-1}$ nucleus is left
in the discrete part of its energy spectrum.  A sketch of the
$A(p,2p)$ reaction is given in Fig.~\ref{fig:schem_repres}.  For a
historical overview of the research into proton-induced nucleon
emission off nuclear targets the reader is referred to
Refs.~\cite{jacob66,jacob73,kitching85}.

In quasielastic $A(p,pN)$ scattering the projectile is elastically
scattered from a single bound nucleon in the target nucleus, resulting
in the struck nucleon being knocked out of the target nucleus.  This
relatively simple reaction mechanism of one ``hard'' nucleon-nucleon
collision is obscured by the ``soft'' initial- and final-state interactions
(IFSI) of the incident and two outgoing nucleons with the nuclear medium.
Consequently, every model for $A(p,2p)$ and $A(p,pn)$ reactions
has two issues to address~: first, the description of the
``hard'' wide-angle scattering that leads to the ejection of the
struck nucleon and second, the distorting mechanisms of the
``soft'' small-angle IFSI.

Concerning the treatment of the ``hard'' $NN$ scattering part,
essentially two methods exist.  A so-called ``cross-section
factorized'' approximation \cite{jacob66,chant77} can be adopted so
that the nucleon-nucleon scattering cross section enters as a
multiplicative factor in the differential $A(p,pN)$ cross section.
Some results of exclusive $A(p,2p)$ measurements
interpreted with this cross-section factorized form can be found in
Refs.~\cite{kullander71,bhowmik74,bengtsson78,kitching80,oers82,carman99}.
The inclusion of spin-dependence in the description of the
IFSI, however, breaks this factorization scheme.  In that situation an
alternative technique can be used~: the amplitude factorized form of the
cross section \cite{chant83}.  In this approach, the two-body $NN$
interaction can be approximated by the interpolation of phase shifts
\cite{SAID} from free elastic $NN$ scattering.  Various
phenomenological forms to fit the amplitudes have also been used in the
past.  Traditionally, the nucleon-nucleon scattering matrix has
been parametrized in terms of five Lorentz invariants
\cite{mcneil,horowitz,maxwell}, a method usually dubbed as the IA1
model or the SPVAT (scalar, pseudoscalar, vector, axial vector,
tensor) form of the $NN$ scattering matrix.  Differential cross
section calculations adopting these five-term representations have
been reported in
Refs~\cite{maxwellcooper,ikebata95,mano98,miller98,neveling02}.  It
should be noted, however, that although the SPVAT form gives
reasonable predictions of $A(p,pN)$ observables, it is, in principle,
not correct, as a five-term parametrization of the relativistic $NN$
scattering matrix is inherently ambiguous \cite{adams84}.  Tjon and
Wallace \cite{tjon85} have shown that a complete expansion of the $NN$
scattering matrix (commonly called the IA2 model) contains $44$
independent invariant amplitudes.  To date, the only calculations
employing this general Lorentz invariant representation have been
performed in the context of the relativistic plane wave impulse
approximation (RPWIA), i.e., a model which ignores all IFSI mechanisms
\cite{vanderventel04}.

The IFSI effects are typically computed by means of the distorted wave
impulse approximation (DWIA) theoretical framework
\cite{jackson65,chant77,chant83,kitching85}.  Generally, in a DWIA
approach the scattering wave functions of the incoming and two
outgoing nucleons are generated by solving the Schr\"odinger or Dirac
equation with complex optical potentials.  Parametrizations for these
optical potentials are usually not gained from basic grounds, but are
obtained by fitting elastic nucleon-nucleus scattering data.  Several
optical-potential parameter sets
\cite{nadasen81,oers82,schwandt82,madland87,cooper87,cooper88,hama90,cooper93}
have been used in the description of quasifree proton scattering off
nuclei.  In the past, both nonrelativistic and relativistic DWIA
versions
\cite{kullander71,bhowmik74,kitching80,antonuk81,oers82,samanta86,cowley91,%
cowley95,ikebata95,miller98,cowley98,mano98,carman99,neveling02,maxwellcooper}
have proven successful in predicting $A(p,pN)$ cross sections over a
wide energy range ($76$--$600$~MeV) and for a whole scope of target
nuclei.

Most of the available calculations for the exclusive $A(p,pN)$ process
addressed incident proton kinetic energies of a few hundred MeV.  In this
work we aim at extending the formalism to scattering in the GeV energy
regime.  As a matter of fact, the majority of DWIA frameworks rely on
partial-wave expansions of the exact solution to the scattering
problem, an approach which becomes increasingly cumbersome at higher
energies.  In this energy range the eikonal approximation
\cite{mccauley58,joachain75}, which belongs to the class of
high-energy semiclassical methods, offers a valid alternative for
describing the IFSI.  Nonrelativistic eikonal studies of the $A(p,2p)$
reaction used in combination with optical potentials can be found in
Refs.~\cite{jacob66,kullander71,bengtsson78}.

In this paper we propose a relativistic and cross-section factorized
formalism based on the eikonal approximation for computing exclusive
$A(p,pN)$ cross sections at incident proton energies in the few
hundred MeV to GeV range.  The eikonal formalism is implemented
relativistically in combination with optical potentials \cite{cooper93}, as
well as with Glauber theory \cite{glauber70,wallace75,yennie71}, which is a
multiple-scattering extension of the eikonal approximation.  The two
frameworks only differ in the way they treat the IFSI.  The main focus will
be on the optical-potential approach, as this method turns out to be the most
suitable for the description of the IFSI for the kinematical settings discussed
in this work.

The paper is organized as follows.  In
Secs.~\ref{sec:cross_sect_matr_elt} and \ref{sec:RPWIA} the factorized
cross section is derived in the RPWIA formalism.  Thereafter, the
different methods to deal with the IFSI are developed in
Sec.~\ref{sec:IFSI}.  Sec.~\ref{sec:IFSI_factor} is devoted to a
presentation of optical-potential and Glauber results for the
IFSI factor.  This is a function which accounts for all IFSI effects
when computing the $A(p,pN)$ observables.  The optical-potential predictions
of our model are compared with cross section data that have been collected at
PNPI and TRIUMF in Sec.~\ref{sec:cross_sect_results}.  First, we
present our calculations for the $^{12}$C, $^{16}$O, and
$^{40}$Ca$(p,2p)$ and $(p,pn)$ PNPI data for $1$~GeV incoming proton energies
\cite{belostotsky87}.  Second, the cross sections for $^{4}$He$(p,2p)$
scattering at $250$~MeV are compared to the TRIUMF data by van Oers
\textit{et al.} \cite{oers82}.  Finally, Sec.~\ref{sec:concl} states
our conclusions.

\section{\label{sec:formalism} $A(p,2p)$ formalism}

In this section the formalism for the description of $A(p,2p)$
reactions is outlined.  The generalization to $A(p,pn)$ reactions is
straightforward.  We conform to the conventions of Bjorken and Drell
\cite{bjorken64} for the $\gamma$ matrices and the Dirac spinors, and
take $\hbar = c = 1$.

\subsection{\label{sec:cross_sect_matr_elt} $A(p,2p)$ differential
  cross section and matrix element}

The four-momenta of the incident and scattered proton are denoted as
$P^{\mu}_{1} (E_{p1}, \vec{p}_{1})$ and $K^{\mu}_{1} (E_{k1},
\vec{k}_{1})$.  The proton momenta $\vec{p}_{1}$ and $\vec{k}_{1}$
define the scattering plane.  The four-momentum transfer is given by
$(\omega, \vec{q}) \equiv q^{\mu} = P^{\mu}_{1} - K^{\mu}_{1} =
K_{A-1}^{\mu} + K_{2}^{\mu} - K_{A}^{\mu}$, where $K_{A}^{\mu} (E_{A},
\vec{k}_{A})$, $K_{A-1}^{\mu} (E_{A-1}, \vec{k}_{A-1})$, and
$K_{2}^{\mu} (E_{k2}, \vec{k}_{2})$ are the four-momenta of the target
nucleus, residual nucleus, and the ejected proton.  The standard
convention $Q^{2} \equiv - q_{\mu} q^{\mu} = \left| \vec{q} \right|^2
- \omega^2 \geq 0$ is followed for the four-momentum transfer. 

In the laboratory frame, the fivefold differential cross section can
be written as
\begin{equation}
\left(\frac{d^{5} \sigma}{dE_{k1} d\Omega_{1} d\Omega_{2}}\right) = 
\frac{M_{p}^{3} M_{A-1}}{(2\pi)^{5} M_{A}} \frac{k_{1} k_{2}}{p_{1}}
f_{rec}^{-1} \:
\overline{\sum_{if}} \left| {\cal M}_{fi}^{(p,2p)} \right|^{2}
\; .
\label{eq:fivefold_diff_cross_section}
\end{equation}
Here, ${\cal M}_{fi}^{(p,2p)}$ is the invariant matrix element which
reflects the transition between the initial and final states.  The
hadronic recoil factor is given by
\begin{eqnarray}
f_{rec} = \frac{E_{A-1}}{E_{A}}
\left| 1 + \frac{E_{k2}}{E_{A-1}}
\left( 1 - \frac{\vec{q} \cdot \vec{k}_{2}}{k_{2}^{2}} \right) \right|
= \left| 1 + \frac{\omega k_2 - q E_{k2} \cos \theta_{k_{2}q}}{M_A k_2}
\right| \; ,
\label{eq:recoil_factor}
\end{eqnarray}
with the energy transfer $\omega = E_{p1} - E_{k1} = E_{A-1} + E_{k2}
- E_{A}$, the three-momentum transfer $\vec{q} = \vec{p}_{1} -
\vec{k}_{1} = \vec{k}_{A-1} + \vec{k}_{2} - \vec{k}_{A}$, and
$\theta_{k_{2}q}$ the angle between $\vec{k}_{2}$ and $\vec{q}$.

The $A(p,2p)$ matrix element is given by
\begin{equation}
{\cal M}_{fi}^{(p,2p)} =
\left< K^{\mu}_{1} \: m_{s1f}, \; K^{\mu}_{2} \: m_{s2f}, \;
\mbox{A-1} \: (K_{A-1}^{\mu}, \: J_R \: M_R) \right|
\widehat{\mathcal{O}}^{(2)} \left|
P^{\mu}_{1} \: m_{s1i}, \; A \: (K_{A}^{\mu}, \; 0^+, \; g.s.) \right>   
\; ,
\end{equation}
where 
\begin{equation}
\widehat{\mathcal{O}}^{(2)} =
\sum_{i < j = 0}^{A} O \left( \vec{r}_i, \vec{r}_j \right)
\label{eq:O_operator}
\end{equation}
is the unknown two-body operator describing the high-momentum transfer
``hard'' $pp$ scattering, $\left| A \: (K_{A}^{\mu},  \; 0^+, \; g.s.)
\right>$ the ground state of the even-even target nucleus and $\left|
\mbox{A-1} \: (K_{A-1}^{\mu}, \: J_R \: M_R) \right>$ the discrete state
in which the residual nucleus is left.  In coordinate space the matrix
element takes on the form
\begin{eqnarray}
{\cal M}_{fi}^{(p,2p)} & = &
\int d \vec{r}_0 \int d \vec{r}_1 \int d \vec{r}_2 \ldots
\int d \vec{r}_A
\left( \Psi^{\vec{k}_{1}, m_{s1f}, \vec{k}_{2}, m_{s2f}}_{A + 1}
\left( \vec{r}_0, \vec{r}_1, \vec{r}_2, \ldots, \vec{r}_A \right)
\right)^{\dagger}
\widehat{\mathcal{O}}^{(2)}
\nonumber \\ & & \times \:
\Psi^{\vec{p}_{1}, m_{s1i}, gs}_{A + 1}
\left( \vec{r}_0, \vec{r}_1, \vec{r}_2, \ldots, \vec{r}_A \right)
\; .
\label{eq:matrix_elt}
\end{eqnarray}
For the sake of brevity of the notations, only the spatial coordinates
are explicitly written.

\subsection{\label{sec:RPWIA} Relativistic plane wave impulse
  approximation}

In this section, the $A(p,2p)$ matrix element of
Eq.~(\ref{eq:matrix_elt}) will be analyzed in the RPWIA.  In this
approach, only one hard collision between the projectile and a bound nucleon is
assumed to occur, knocking the bound nucleon out of the target
nucleus.  The modelling of the ``soft'' IFSI processes, which affect
both the incoming and outgoing protons, will be considered in
Sec.~\ref{sec:IFSI}.

In evaluating the $A(p,2p)$ matrix element of
Eq.~(\ref{eq:matrix_elt}), a mean-field approximation for the nuclear
wave functions is adopted.  We also assume factorization between the
``hard'' $NN$ coupling and the nuclear dynamics.  For reasons of
conciseness, the forthcoming derivations are explained for the $A = 3$
case.  The generalization to arbitrary mass number $A$ is
rather straightforward.

The antisymmetrized $\mbox{(A+1)}$-body wave function in the initial
state is of the Slater determinant form 
\begin{equation}
\Psi^{\vec{p}_{1}, m_{s1i}, gs}_{A + 1}
\left( \vec{r}_0, \vec{r}_1, \vec{r}_2, \vec{r}_3 \right)
= \frac {1} {\sqrt{(A+1)!}}
\left| \begin{array}{cccc}
\phi_{\vec{p}_{1} m_{s1i}} \left( \vec{r}_0 \right) &
\phi_{\alpha_{1}} \left( \vec{r}_0 \right) &
\phi_{\alpha_{2}} \left( \vec{r}_0 \right) &
\phi_{\alpha_{3}} \left( \vec{r}_0 \right) \\
\phi_{\vec{p}_{1} m_{s1i}} \left( \vec{r}_1 \right) &
\phi_{\alpha_{1}} \left( \vec{r}_1 \right) &
\phi_{\alpha_{2}} \left( \vec{r}_1 \right) &
\phi_{\alpha_{3}} \left( \vec{r}_1 \right) \\
\phi_{\vec{p}_{1} m_{s1i}} \left( \vec{r}_2 \right) &
\phi_{\alpha_{1}} \left( \vec{r}_2 \right) &
\phi_{\alpha_{2}} \left( \vec{r}_2 \right) &
\phi_{\alpha_{3}} \left( \vec{r}_2 \right) \\
\phi_{\vec{p}_{1} m_{s1i}} \left( \vec{r}_3 \right) &
\phi_{\alpha_{1}} \left( \vec{r}_3 \right) &
\phi_{\alpha_{2}} \left( \vec{r}_3 \right) &
\phi_{\alpha_{3}} \left( \vec{r}_3 \right) 
\end{array} \right| \; .
\label{eq:ini_wave_fctn_Slater} 
\end{equation}
Details on the bound-state single-particle wave functions
$\phi_{\alpha_{i}} \left( \vec{r}, \vec{\sigma} \right)$ entering this
mean-field $\mbox{(A+1)}$-body wave function can be found in
Appendix~\ref{app:bound}.  The wave function of the incoming proton is
given by a relativistic plane wave
\begin{eqnarray}
\phi_{\vec{k} m_{s}} \left( \vec{r} \right) \equiv
\sqrt{\frac{E+M}{2M}} \left[
\begin{array}{c}
1 \\
\frac{1}{E+M} \vec{\sigma} \cdot \hat{\vec{p}}
\end{array} 
\right] e^{i \vec{k} \cdot \vec{r}} \chi_{\frac{1}{2}m_{s}}
= e^{i \vec{k} \cdot \vec{r}} \: u ( \vec{k}, m_{s} ) \; .
\label{eq:relat_plane_wave}
\end{eqnarray}
The $\mbox{(A+1)}$-body wave function in the final state reads
\begin{equation}
\Psi^{\vec{k}_{1}, m_{s1f}, \vec{k}_{2}, m_{s2f}}_{A + 1}
\left( \vec{r}_0, \vec{r}_1, \vec{r}_2, \vec{r}_3 \right)
= \frac {1} {\sqrt{(A+1)!}}
\left| \begin{array}{cccc}
\phi_{\vec{k}_{1} m_{s1f}} \left( \vec{r}_0 \right) &
\phi_{\vec{k}_{2} m_{s2f}} \left( \vec{r}_0 \right) &
\phi_{\alpha_{2}} \left( \vec{r}_0 \right) &
\phi_{\alpha_{3}} \left( \vec{r}_0 \right) \\
\phi_{\vec{k}_{1} m_{s1f}} \left( \vec{r}_1 \right) &
\phi_{\vec{k}_{2} m_{s2f}} \left( \vec{r}_1 \right) &
\phi_{\alpha_{2}} \left( \vec{r}_1 \right) &
\phi_{\alpha_{3}} \left( \vec{r}_1 \right) \\
\phi_{\vec{k}_{1} m_{s1f}} \left( \vec{r}_2 \right) &
\phi_{\vec{k}_{2} m_{s2f}} \left( \vec{r}_2 \right) &
\phi_{\alpha_{2}} \left( \vec{r}_2 \right) &
\phi_{\alpha_{3}} \left( \vec{r}_2 \right) \\
\phi_{\vec{k}_{1} m_{s1f}} \left( \vec{r}_3 \right) &
\phi_{\vec{k}_{2} m_{s2f}} \left( \vec{r}_3 \right) &
\phi_{\alpha_{2}} \left( \vec{r}_3 \right) &
\phi_{\alpha_{3}} \left( \vec{r}_3 \right) 
\end{array} \right|
\; .
\label{eq:fin_wave_fctn_Slater} 
\end{equation}
Relative to the target nucleus ground state written in
Eq.~(\ref{eq:ini_wave_fctn_Slater}), the wave function of
Eq.~(\ref{eq:fin_wave_fctn_Slater}) refers to the situation whereby
the struck proton resides in a state ``$\alpha_{1}$'', leaving the
residual $\mbox{A-1}$ nucleus as a hole state in that particular
single-particle level.  The outgoing protons are represented by
relativistic plane waves.

Since both the initial and the final wave functions are fully
antisymmetrized, one can choose the operator
$\widehat{\mathcal{O}}^{(2)}$ to act on two particular coordinates
($\vec{r}_0$ and $\vec{r}_1$).  Without any loss of generality, the
$A(p,2p)$ matrix element of Eq.~(\ref{eq:matrix_elt}) can be written
as
\begin{eqnarray}
{\cal M}_{fi}^{(p,2p)} & = &
\frac{A (A + 1)}{2} \frac {1} {(A + 1)!}
\int d \vec{r}_0 \int d \vec{r}_1 \int d \vec{r}_2 \int d \vec{r}_3
\nonumber \\ & & \times \:
\sum_{ k,l \in \left\{ \vec{k}_{1} m_{s1f}, \vec{k}_{2} m_{s2f}
  \right\} }
\; \; \sum_{ m,n \in \left\{ \alpha_2, \alpha _3 \right\} } 
\; \; \sum_{ o,p \in \left\{ \vec{p}_{1} m_{s1i}, \alpha_1 \right\} }
\; \; \sum_{ q,r \in \left\{ \alpha_2, \alpha _3 \right\} } 
\nonumber \\ & & \times \:
\epsilon_{klmn} \epsilon_{opqr}
\phi_{k} ^{\dagger} \left( \vec{r}_0 \right)
\phi_{l} ^{\dagger} \left( \vec{r}_1 \right)
\phi_{m} ^{\dagger} \left( \vec{r}_2 \right)
\phi_{n} ^{\dagger} \left( \vec{r}_3 \right)
\nonumber \\ & & \times \:
O \left( \vec{r}_0, \vec{r}_1 \right)
\phi_{o} \left( \vec{r}_0 \right)
\phi_{p} \left( \vec{r}_1 \right)
\phi_{q} \left( \vec{r}_2 \right)
\phi_{r} \left( \vec{r}_3 \right) \; ,
\label{eq:matrix_elt_part_coord}
\end{eqnarray}
with $\epsilon_{ijkl}$ the Levi-Civita tensor.  In the RPWIA,
\begin{eqnarray}
\int d \vec{r}_0 \int d \vec{r}_1 \int d \vec{r}_2 
\phi_{k} ^{\dagger} \left( \vec{r}_0 \right)
\phi_{l} ^{\dagger} \left( \vec{r}_1 \right)
\phi_{m} ^{\dagger} \left( \vec{r}_2 \right)
O \left( \vec{r}_0, \vec{r}_1 \right)
\phi_{o} \left( \vec{r}_0 \right)
\phi_{p} \left( \vec{r}_1 \right)
\phi_{q} \left( \vec{r}_2 \right)
\nonumber \\ 
= \delta_{mq} \int d \vec{r}_0 \int d \vec{r}_1 \int d \vec{r}_2
\phi_{k} ^{\dagger} \left( \vec{r}_0 \right)
\phi_{l} ^{\dagger} \left( \vec{r}_1 \right)
O \left( \vec{r}_0, \vec{r}_1 \right)
\phi_{o} \left( \vec{r}_0 \right)
\phi_{p} \left( \vec{r}_1 \right)
\left| \phi_{q} \left( \vec{r}_2 \right) \right|^2
\; .
\label{eq:delta_mq}
\end{eqnarray}
Inserting this expression in Eq.~(\ref{eq:matrix_elt_part_coord}) one
obtains
\begin{eqnarray}
{\cal M}_{fi}^{(p,2p)} & = & \frac{A (A + 1)}{2} \frac {1} {(A + 1)!}
\int d \vec{r}_0 \int d \vec{r}_1 \int d \vec{r}_2 \int d \vec{r}_3 
\nonumber \\ & & \times \:
\sum_{ k,l \in \left\{ \vec{k}_{1} m_{s1f}, \vec{k}_{2} m_{s2f}
  \right\} }
\; \; \sum_{ o,p \in \left\{ \vec{p}_{1} m_{s1i}, \alpha_1 \right\} }
\; \; \sum_{ m,n \in \left\{ \alpha_2, \alpha _3 \right\} } 
\nonumber \\ & & \times \:
\epsilon_{klmn} \epsilon_{opmn}
\phi_{k} ^{\dagger} \left( \vec{r}_0 \right)
\phi_{l} ^{\dagger} \left( \vec{r}_1 \right)
\left| \phi_{m} \left( \vec{r}_2 \right) \right|^2
\left| \phi_{n} \left( \vec{r}_3 \right) \right|^2
O \left( \vec{r}_0, \vec{r}_1 \right)
\phi_{o} \left( \vec{r}_0 \right)
\phi_{p} \left( \vec{r}_1 \right)
\; .
\end{eqnarray}
There are $\left(A-1\right)!$ possible choices (permutations) for the
indices $m$, $n$, \ldots, all giving the same contribution to the
matrix element.  Accordingly, the above expression can be rewritten as
\begin{eqnarray}
{\cal M}_{fi}^{(p,2p)} & = & \frac{1}{2}
\int d \vec{r}_0 \int d \vec{r}_1 \int d \vec{r}_2 \int d \vec{r}_3
\sum_{ k,l \in \left\{ \vec{k}_{1} m_{s1f}, \vec{k}_{2} m_{s2f}
  \right\} }
\; \; \sum_{ o,p \in \left\{ \vec{p}_{1} m_{s1i}, \alpha_1 \right\} }
\nonumber \\ & & \times \:
\epsilon_{kl\alpha_2\alpha_3} \epsilon_{op\alpha_2\alpha_3}
\phi_{k} ^{\dagger} \left( \vec{r}_0 \right)
\phi_{l} ^{\dagger} \left( \vec{r}_1 \right)
\left| \phi_{\alpha_2} \left( \vec{r}_2 \right) \right|^2
\left| \phi_{\alpha_3} \left( \vec{r}_3 \right) \right|^2
O \left( \vec{r}_0, \vec{r}_1 \right)
\phi_{o} \left( \vec{r}_0 \right)
\phi_{p} \left( \vec{r}_1 \right)
\; .
\end{eqnarray}  
Because the bound-state wave functions are normalized to unity ($\int
d \vec{r} \left| \phi_{\alpha} \left( \vec{r} \right) \right|^2 = 1$)
and $O \left( \vec{r}_0, \vec{r}_1 \right) = O \left( \vec{r}_1,
\vec{r}_0 \right)$, the matrix element can be further simplified to
\begin{eqnarray}
{\cal M}_{fi}^{(p,2p)} & = & 
\int d \vec{r}_0 \int d \vec{r}_1
\left( \phi_{\vec{k}_{1} m_{s1f}} ^{\dagger} \left( \vec{r}_0 \right)
\phi_{\vec{k}_{2} m_{s2f}} ^{\dagger} \left( \vec{r}_1 \right) -
\phi_{\vec{k}_{2} m_{s2f}} ^{\dagger} \left( \vec{r}_0 \right)
\phi_{\vec{k}_{1} m_{s1f}} ^{\dagger} \left( \vec{r}_1 \right) \right)
\nonumber \\ & & \times \:
O \left( \vec{r}_0, \vec{r}_1 \right)
\phi_{\vec{p}{1} m_{s1i}} \left( \vec{r}_0 \right)
\phi_{\alpha_1} \left( \vec{r}_1 \right) 
\; ,
\label{eq:rpwia_dir_exch_matrix_elt}
\end{eqnarray}
including a direct and an exchange term.

Substitution of the general form of the scattering operator $O \left(
\vec{r}_0, \vec{r}_1 \right) = \int \frac{d \vec{p}}{(2\pi)^{3}} e^{i
  \vec{p} \cdot (\vec{r}_1 - \vec{r}_0)} \widehat{F}$, with
$\widehat{F}$ the $NN$ scattering amplitude in momentum space, in the
above expression (\ref{eq:rpwia_dir_exch_matrix_elt}) leads to 
\begin{eqnarray}
& \int \frac{d \vec{p}}{(2\pi)^{3}} \int d \vec{r}_0 \int d \vec{r}_1
e^{- i \vec{k}_{1} \cdot \vec{r}_{0}}
u^{\dagger} ( \vec{k}_{1}, m_{s1f} )
e^{- i \vec{k}_{2} \cdot \vec{r}_{1}}
u^{\dagger} ( \vec{k}_{2}, m_{s2f} )
e^{i \vec{p} \cdot (\vec{r}_1 - \vec{r}_0)} \widehat{F} \:
e^{i \vec{p}_{1} \cdot \vec{r}_{0}}
u ( \vec{p}_{1}, m_{s1i} )
\phi_{\alpha_1} \left( \vec{r}_1 \right) 
\nonumber \\ & =
u^{\dagger} ( \vec{k}_{1}, m_{s1f} )
u^{\dagger} ( \vec{k}_{2}, m_{s2f} ) \:
\widehat{F} \:
u ( \vec{p}_{1}, m_{s1i} )
\phi_{\alpha_1} \left( \vec{p}_{m} \right)
\end{eqnarray}
for the direct term and a similar expression for the exchange term.
Here $\phi_{\alpha} \left( \vec{p} \right)$ is the relativistic wave
function for the bound nucleon in momentum space (for details see
Appendix~\ref{app:bound}) and $\vec{p}_{m} = \vec{k}_{1} + \vec{k}_{2}
- \vec{p}_{1}$ is the missing momentum.  In order to arrive at a
cross-section factorized expression for
Eq.~(\ref{eq:fivefold_diff_cross_section}), the quasielastic off-shell
proton-proton scattering matrix element will be related to the free
on-shell proton-proton cross section.  For this purpose, we insert the
completeness relation
\begin{equation}
\sum_{s} \left[ u ( \vec{p}_{m}, s ) \bar{u} ( \vec{p}_{m}, s ) - 
v ( \vec{p}_{m}, s ) \bar{v} ( \vec{p}_{m}, s ) \right] = 1
\label{eq:completeness}
\end{equation}
in
\begin{equation}
{\cal M}_{fi}^{(p,2p)} =
u^{\dagger} ( \vec{k}_{1}, m_{s1f} )
u^{\dagger} ( \vec{k}_{2}, m_{s2f} ) \:
\widehat{F} \:
u ( \vec{p}_{1}, m_{s1i} )
\phi_{\alpha_1} \left( \vec{p}_{m} \right)
\: - \: 
\left( \vec{k}_{1} m_{s1f} \leftrightarrow \vec{k}_{2} m_{s2f} \right) 
\; ,
\end{equation}
and obtain the following expression for the matrix element
\begin{eqnarray}
{\cal M}_{fi}^{(p,2p)} & = & \sum_{s}
\left( {\cal M}_{fi}^{pp} \right)_{m_{s1i},s,m_{s1f},m_{s2f}} \:
\bar{u} ( \vec{p}_{m}, s ) \phi_{\alpha_1} \left( \vec{p}_{m} \right)
\nonumber \\ & &
- \: \textrm{negative-energy projection term} \; .
\end{eqnarray}
Here, ${\cal M}_{fi}^{pp}$ is the matrix element for free $pp$
scattering
\begin{eqnarray}
\left( {\cal M}_{fi}^{pp} \right)_{m_{s1i},m_{s2i},m_{s1f},m_{s2f}} & = &
u^{\dagger} ( \vec{k}_{1}, m_{s1f} )
u^{\dagger} ( \vec{k}_{2}, m_{s2f} ) \:
\widehat{F} \:
u ( \vec{p}_{1}, m_{s1i} )
u ( \vec{p}_{2}, m_{s2i} )
\nonumber \\ & & - \:
\left( \vec{k}_{1} m_{s1f} \leftrightarrow \vec{k}_{2} m_{s2f} \right) 
\; .
\end{eqnarray}
Factorization breaks down, even when IFSI are disregarded, owing to
the negative-energy projection term.  To recover factorization, the
negative-energy projection term is neglected in the remainder of
this work
\begin{equation}
{\cal M}_{fi}^{(p,2p)} \approx \sum_{s}
\left( {\cal M}_{fi}^{pp} \right)_{m_{s1i}, s, m_{s1f}, m_{s2f}} \:
\bar{u} ( \vec{p}_{m}, s ) \phi_{\alpha_1} \left( \vec{p}_{m} \right)
\; .
\label{eq:pos_energy_proj_matr_elt}
\end{equation}

Using the expression of the relativistic bound-nucleon wave
function in momentum space given in Appendix~\ref{app:bound}, the
$\bar{u} \: \phi_{\alpha}$ contraction in
Eq.~(\ref{eq:pos_energy_proj_matr_elt}) reduces to \cite{caballero98}
\begin{equation}
\bar{u} ( \vec{p}, s ) \: \phi_{\alpha} \left( \vec{p} \right) =
\left( - i \right)^l (2\pi)^{3/2} \sqrt{\frac{\bar{E} + M_p}{2 M_p}}
\: 
\alpha_{n \kappa} \left( p \right)
\:
\chi_{\frac{1}{2} s}^{\dagger} {\cal Y}_{\kappa m} (\Omega_p)
\; ,
\label{eq:ubar_phialpha_contract}
\end{equation}
where $\bar{E} = \sqrt{p^2 + M_p^2}$, $\chi_{\frac{1}{2} s}^{\dagger}
{\cal Y}_{\kappa m}$ indicates the spin projection of the spin
spherical harmonic ${\cal Y}_{\kappa m} (\Omega_p)$ on a spin state
$\chi_{\frac{1}{2} s}$, and the radial function in momentum space
$\alpha_{n \kappa}$ is given by 
\begin{equation}
\alpha_{n \kappa} \left( p \right) =
g_{n \kappa} ( p ) -
\frac{p}{\bar{E} + M_p} S_{\kappa} \; f_{n \kappa} ( p )
\; ,
\label{eq:alpha_nkappa}
\end{equation}
with $g_{n \kappa}$ and $f_{n \kappa}$ the Bessel transforms of the
standard upper and lower radial functions of the bound-nucleon wave
function in coordinate space (see Appendix~\ref{app:bound} for
details) and $S_{\kappa} = \kappa / \left| \kappa \right|$.

Upon squaring Eq.~(\ref{eq:pos_energy_proj_matr_elt}), the $pp$ and
nuclear bound-state parts get coupled by the summation over the
intermediate spins $s$ and $s'$~: 
\begin{eqnarray}
\left| {\cal M}_{fi}^{(p,2p)} \right|^{2} & \approx &
\sum_{s, s'}
\left( {\cal M}_{fi}^{pp} \right)_{m_{s1i}, s, m_{s1f}, m_{s2f}}^{*}
\: 
\left( {\cal M}_{fi}^{pp} \right)_{m_{s1i}, s', m_{s1f}, m_{s2f}}
\nonumber \\ & & \times \:
\left( \bar{u} ( \vec{p}_{m}, s )
\phi_{\alpha_1} \left( \vec{p}_{m} \right) \right)^{*}
\bar{u} ( \vec{p}_{m}, s' )
\phi_{\alpha_1} \left( \vec{p}_{m} \right)
\; .
\label{eq:squared_matrix_elt}
\end{eqnarray}
After summation over $m$, the struck nucleon's generalized
angular momentum quantum number, the square of $\bar{u} ( \vec{p}_{m},
s ) \phi_{\alpha_1} \left( \vec{p}_{m} \right)$ yields a $\delta_{s
s'}$, i.e., becomes diagonal in $s$.  Thereby, use is made of the
following identity
\begin{equation}
\sum_{m}
\left( \chi_{\frac{1}{2} s}^{\dagger} {\cal Y}_{\kappa m} \right)^{*}
\chi_{\frac{1}{2} s'}^{\dagger} {\cal Y}_{\kappa m}
= \frac{2 j + 1}{8 \pi} \delta_{s s'}
\; .
\end{equation}
This leads to the decoupling between the $pp$ scattering and the
bound-state part in the matrix element~:
\begin{eqnarray}
\overline{\sum_{if}} \left| {\cal M}_{fi}^{(p,2p)} \right|^{2}
& \approx & (2\pi)^{3} \frac{2 j + 1}{8 \pi}
\left| \tilde \alpha_{n \kappa} \left( p_m \right) \right|^{2}
\nonumber \\ & & \times \:
\overline{\sum}_{m_{s1i}, m_{s1f}, m_{s2f}} \sum_{s}
\left|
\left( {\cal M}_{fi}^{pp} \right)_{m_{s1i}, s, m_{s1f}, m_{s2f}}
\right|^{2}
\; ,
\label{eq:decoupled_squared_matrix_elt}
\end{eqnarray}
with
\begin{equation}
\tilde \alpha_{n \kappa} \left( p \right) =
\sqrt{\frac{\bar{E} + M_p}{2 M_p}}
\alpha_{n \kappa} \left( p \right) \; .
\label{eq:alpha_tilde}
\end{equation}

The last factor in Eq.~(\ref{eq:decoupled_squared_matrix_elt}) can be
related to the free $pp$ scattering center-of-mass cross section
\begin{equation}
\left( \frac{d \sigma^{pp}}{d \Omega} \right)_{\textrm{c.m.}} = 
\frac{M_{p}^{4}}{(2\pi)^{2} s} 
\frac{1}{2} \overline{\sum}_{m_{s1i}, m_{s1f}, m_{s2f}} \sum_{s}
\left| \left( {\cal M}_{fi}^{pp} \right)_{m_{s1i}, s, m_{s1f}, m_{s2f}}
\right|^{2}
\; ,
\end{equation}
so that the RPWIA differential $A(p,2p)$ cross section of
Eq.~(\ref{eq:fivefold_diff_cross_section}) can be written in the
cross-section factorized form
\begin{equation}
\left(
\frac{d^{5}\sigma}{dE_{k1} d\Omega_{1} d\Omega_{2}}
\right)^{\textrm{RPWIA}} 
\approx 
\frac{s M_{A-1}}{M_{p} M_{A}} \frac{k_{1} k_{2}}{p_{1}}
f_{rec}^{-1} \:
\frac{2 j + 1}{4 \pi}
\left| \tilde \alpha_{n \kappa} \left( p_m \right) \right|^{2}
\left(\frac{d \sigma^{pp}}{d\Omega}\right)_{\textrm{c.m.}}
\; .
\label{eq:factorized_rpwia_cross_sect}
\end{equation}
Here, $s$ is the Mandelstam variable for the $pp$ scattering, not to
be confused with the intermediate spin from the preceding
Eqs.~(\ref{eq:completeness})--(\ref{eq:decoupled_squared_matrix_elt}).
In the numerical calculations which will be presented in the
forthcoming sections, the free proton-proton cross section
$\left( \frac{d \sigma^{pp}}{d \Omega} \right)_{\textrm{c.m.}}$ is
obtained from the SAID code \cite{SAID}. 

\subsection{\label{sec:IFSI} Treatment of the IFSI}

It is well known that the factorized RPWIA result of
Eq.~(\ref{eq:factorized_rpwia_cross_sect}) adopts an
oversimplified description of the reaction mechanism.  The momentum
distribution $\frac{2 j + 1}{4 \pi} \left| \tilde \alpha_{n \kappa}
\left( p_m \right) \right|^{2}$, which represents the probability of
finding a proton in the target nucleus with missing momentum
$\vec{p}_m$, will be modified by the scatterings of the incoming and
outgoing protons in the nucleus.  Therefore it is necessary to
incorporate the effects of these IFSI in the model.

First, in Sec.~\ref{sec:dist_factoriz}, the differential $A(p,2p)$
cross section is written in a factorized form taking IFSI effects into
account.  Next, the relativistic eikonal methods used for dealing with
the IFSI effects in this work, are discussed in depth.  Two methods will
be used.  The relativistic optical model eikonal approximation (ROMEA)
is the subject of Sec.~\ref{sec:ROMEA}, whereas the relativistic
multiple-scattering Glauber approximation (RMSGA) is discussed in
Sec.~\ref{sec:RMSGA}.

\subsubsection{\label{sec:dist_factoriz} Factorization assumption and
  the distorted momentum distribution}

In both versions of the relativistic eikonal framework for $A(p,pN)$
reactions presented here (ROMEA and RMSGA), the antisymmetrized
initial- and final-state $\mbox{(A+1)}$-body wave functions,
\begin{equation}
\Psi^{\vec{p}_{1}, m_{s1i}, gs}_{A + 1}
\left( \vec{r}_0, \vec{r}_1, \ldots, \vec{r}_A \right)
= {\widehat{\mathcal{A}}}
\Biggl[
{\widehat{\mathcal{S}}_{p1}
\left( \vec{r}_0, \vec{r}_2, \ldots, \vec{r}_A \right) }
\:
e^{i \vec{p}_1 \cdot \vec{r}_0} \: u ( \vec{p}_1, m_{s1i} )
\:
\Psi^{gs}_{A}
\left( \vec{r}_1, \vec{r}_2, \ldots, \vec{r}_A \right)
\Biggr]
\label{eq:dist_ini_wave_fctn}
\end{equation}
and
\begin{eqnarray}
\Psi^{\vec{k}_{1}, m_{s1f}, \vec{k}_{2}, m_{s2f}}_{A + 1}
\left( \vec{r}_0, \vec{r}_1, \ldots, \vec{r}_A \right)
= {\widehat{\mathcal{A}}}
\Biggl[
{\widehat{\mathcal{S}}_{k1}^{\dagger}
\left( \vec{r}_0, \vec{r}_2, \ldots, \vec{r}_A \right) } 
\:
e^{i \vec{k}_1 \cdot \vec{r}_0} \: u ( \vec{k}_1, m_{s1f} )
\nonumber \\ \times \:
{\widehat{\mathcal{S}}_{k2}^{\dagger}
\left( \vec{r}_1, \vec{r}_2, \ldots, \vec{r}_A \right) }
\:
e^{i \vec{k}_2 \cdot \vec{r}_1} \: u ( \vec{k}_2, m_{s2f} )
\:
\Psi_{\mbox{A-1}}^{J_R \; M_R}
\left( \vec{r}_2, \ldots, \vec{r}_A \right)
\Biggr] \; ,
\label{eq:dist_fin_wave_fctn}
\end{eqnarray}
differ from their respective RPWIA expressions of
Eqs.~(\ref{eq:ini_wave_fctn_Slater}) and
(\ref{eq:fin_wave_fctn_Slater}) through the presence of the operators
$\widehat{\mathcal{S}}_{p1}$, $\widehat{\mathcal{S}}_{k1}$, and
$\widehat{\mathcal{S}}_{k2}$.  These define the accumulated effect of
all interactions that the incoming and emerging protons undergo in
their way into and out of the target nucleus.

Since the IFSI violate factorization, some additional approximations
are in order.  First, only central IFSI are
considered, i.e., spin-orbit contributions are omitted.  Further, the
zero-range approximation is adopted for the ``hard'' $NN$ interaction,
allowing one to replace the coordinates of
the two interacting protons ($\vec{r}_0$ and $\vec{r}_1$) by one single
collision point in the
distorting functions $\widehat{\mathcal{S}}_{p1}$,
$\widehat{\mathcal{S}}_{k1}$, and $\widehat{\mathcal{S}}_{k2}$.  This
leads to the distorted momentum-space wave function
\begin{equation}
\phi_{\alpha_1}^{D} \left( \vec{p}_{m} \right)
= \int d \vec{r}
e^{- i \vec{p}_{m} \cdot \vec{r}}
\phi_{\alpha_1} \left( \vec{r} \right)
\mathcal{S}_{IFSI} \left( \vec{r} \right)
\; ,
\label{eq:dist_mom_space_wave_fctn}
\end{equation}
similar to Eq.~(\ref{eq:Fourier_bound_wave}), but with the additional
IFSI factor
\begin{eqnarray}
\mathcal{S}_{IFSI} (\vec{r}) & = &
\int d \vec{r}_2 \ldots \int d \vec{r}_A
\left| \phi_{\alpha_2} \left( \vec{r}_2 \right) \right|^2 \ldots 
\left| \phi_{\alpha_A} \left( \vec{r}_A \right) \right|^2 
{\widehat{\mathcal{S}}_{k1}
\left( \vec{r}, \vec{r}_2, \ldots, \vec{r}_A \right) }
\nonumber \\ & & \times \:
{\widehat{\mathcal{S}}_{k2}
\left( \vec{r}, \vec{r}_2, \ldots, \vec{r}_A \right) } \: 
{\widehat{\mathcal{S}}_{p1}
\left( \vec{r}, \vec{r}_2, \ldots, \vec{r}_A \right) } 
\label{eq:IFSI_factor}
\end{eqnarray}
accounting for the soft IFSI effects.

Now, along the lines of \cite{vignote04}, it is natural to define a
distorted wave amplitude 
\begin{equation}
\psi^{D} \left( \vec{p}_{m} \right) =
\bar{u} ( \vec{p}_{m}, s )
\phi_{\alpha_1}^{D} \left( \vec{p}_{m} \right) \; ,
\label{eq:dist_wave_ampl}
\end{equation}
so that the distorted momentum distribution is given by the square of this
amplitude,
\begin{equation}
\rho^{D} \left( \vec{p}_{m} \right)
= \frac{1}{(2\pi)^3} \sum_{m} \sum_{s}
\left| \psi^{D} \left( \vec{p}_{m} \right) \right|^{2} 
\; . 
\label{eq:dist_mom_dist}
\end{equation}
This distorted momentum distribution has the following properties.  First, it
takes into account the distortions for the incoming
and outgoing protons.  Second, it reduces to the plane wave momentum
distribution $\frac{2 j + 1}{4 \pi} \left| \tilde \alpha_{n \kappa}
\left( p_m \right) \right|^{2}$ in the plane wave limit when
assuming that $\phi_{\alpha_1} \left( \vec{p}_{m} \right)$ satisfies the
relation
\begin{equation}
\frac{\vec{\sigma} \cdot \vec{p}}{\bar{E} + M_p} \phi^{u} = \phi^{d}
\label{eq:free_relation}
\end{equation}
between the upper and lower components.

Using the ansatz (\ref{eq:dist_mom_dist}) for the distorted momentum
distribution, the differential $A(p,2p)$ cross section can be cast in the form
\begin{equation}
\left(
\frac{d^{5}\sigma}{dE_{k1} d\Omega_{1} d\Omega_{2}}
\right)^{D}
\approx 
\frac{s M_{A-1}}{M_{p} M_{A}} \frac{k_{1} k_{2}}{p_{1}}
f_{rec}^{-1} \:
\rho^{D} \left( \vec{p}_{m} \right)
\left(\frac{d \sigma^{pp}}{d\Omega}\right)_{\textrm{c.m.}}
\; .
\label{eq:factorized_ifsi_cross_sect}
\end{equation}
It differs from the RPWIA expression
(\ref{eq:factorized_rpwia_cross_sect}) through the introduction
of a ``distorted'' momentum distribution $\rho^{D}$.

\subsubsection{\label{sec:ROMEA} Relativistic optical model eikonal
  approximation}

As shown for example in Refs.~\cite{amado83,debruyne00}, in the
relativistic eikonal limit the scattering wave function of a nucleon
with energy $E = \sqrt{k^{2} + M^{2}}$ and spin state $\left|
\frac{1}{2} m_s \right>$ subject to a scalar ($V_{s}$) and a vector
potential ($V_{v}$) takes on the form
\begin{equation}
\psi_{\vec{k}, m_{s}}^{(+)} = \sqrt { \frac{E+M}{2M} }
\left[
\begin{array}{c}
1 \\
\frac{1}{E+M+V_{s}-V_{v}} \vec{\sigma} \cdot \hat{\vec{p}}
\end{array}
\right]
e^{i \vec{k} \cdot \vec{r}} e^{i S(\vec{r})}
\chi_{\frac{1}{2} m_{s}}
\; ,
\label{eq:eikonal_scatt_wave}
\end{equation}
where the eikonal phase $S(\vec{b},z)$ reads
\begin{equation}
i S (\vec{b}, z) = - i \frac{M}{K} \int_{-\infty}^{z} dz \, ' \,
\biggl[ V_{c} (\vec{b},z \, ') +
V_{so} (\vec{b},z \, ')
[ \vec{\sigma} \cdot (\vec{b} \times \vec{K} )- i Kz \, ']
\biggr]
\; ,
\label{eq:eikonal_phase}
\end{equation}
with $\vec{r} \equiv (\vec{b}, z)$ and the average momentum $\vec{K}$
pointing along the $z$-axis.  The central and spin-orbit potentials
$V_{c}$ and $V_{so}$ in the above expression are determined by $V_{s}$
and $V_{v}$ and their derivatives.  In general, a fraction of the strength
from the incident beam is removed from the elastic channel into the
inelastic ones.  These inelasticities are commonly implemented by means
of the imaginary part of the optical potential.

In evaluating the IFSI effects, three approximations are introduced.
First, the dynamical enhancement of the lower component of the
scattering wave function (\ref{eq:eikonal_scatt_wave}), which is due
to the combination of the scalar and vector potentials, is neglected.  
Second, the impulse operator $\hat{\vec{p}}$ is replaced by the
asymptotic momentum $\vec{k}$ of the nucleon.  As mentioned before, the
spin-orbit potential $V_{so}$ is also omitted.

\begin{figure}
\begin{center}
\includegraphics[width=0.5\textwidth]{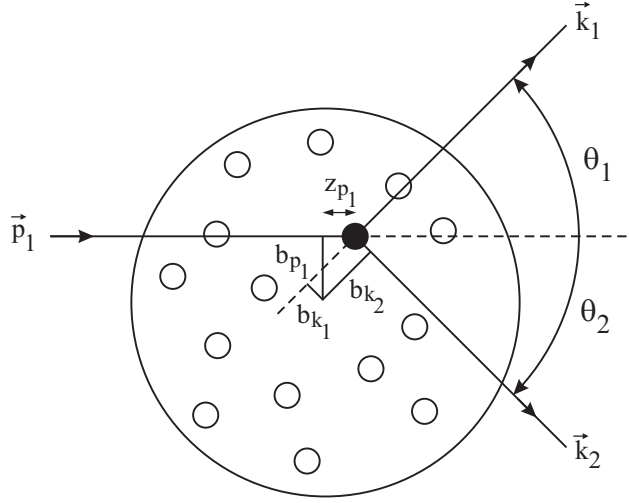}
\caption{Geometry of the scattering process.  The vectors
  $\vec{b}_{p_{1}}$, $\vec{b}_{k_{1}}$, and $\vec{b}_{k_{2}}$ are the
  impact parameters for each of the three paths for a collision
  occurring at $\vec{r}$.  $z_{p_{1}}$, $z_{k_{1}}$, and $z_{k_{2}}$
  are the $z$ coordinates of the collision point in the respective
  coordinate systems.  $\theta_{1}$ and $\theta_{2}$ are the angles of
  the outgoing nucleons relative to the incoming proton direction.}
\label{fig:geometry}
\end{center}
\end{figure}

As a result, the effects of the interactions of the incoming and
outgoing protons with the residual nucleus are implemented in the
distorted momentum-space wave function of
Eq.~(\ref{eq:dist_mom_space_wave_fctn}) through the following phase
factors
\begin{subequations}
\begin{equation}
\widehat{\mathcal{S}}_{p1} \left( \vec{r} \right) =
e^{- i \frac{M_{p}}{p_1} \int_{- \infty}^{z_{p_{1}}} dz \,
V_{c} (\vec{b}_{p_{1}}, z)}
\; ,
\end{equation}
\begin{equation}
\widehat{\mathcal{S}}_{k1} \left( \vec{r} \right) =
e^{- i \frac{M_{p}}{k_1} \int_{z_{k_{1}}}^{+ \infty} dz \, ' \,
V_{c} (\vec{b}_{k_{1}}, z \, ')}
\; ,
\end{equation}
\begin{equation}
\widehat{\mathcal{S}}_{k2} \left( \vec{r} \right) =
e^{- i \frac{M_{p}}{k_2} \int_{z_{k_{2}}}^{+ \infty} dz \, '' \,
V_{c} (\vec{b}_{k_{2}}, z \, '')}
\; ,
\end{equation}
\label{eq:eikonal_IFSI_operators}
\end{subequations}
with the $z$-axes of the different coordinate systems lying along the
trajectories of the respective particles ($z$ along the direction of
the incoming proton $\vec{p}_{1}$, $z \, '$ along the trajectory of
the scattered proton $\vec{k}_{1}$, and $z \, ''$ along the path of
the ejected nucleon $\vec{k}_{2}$) and $(\vec{b}_{p_{1}}, z_{p_{1}})$,
$(\vec{b}_{k_{1}}, z_{k_{1}})$, and $(\vec{b}_{k_{2}}, z_{k_{2}})$
the coordinates of the collision point $\vec{r}$ in the
respective coordinate systems.  The geometry of the scattering process
is illustrated in Fig.~\ref{fig:geometry}.  The integration limits
guarantee that the incoming proton only undergoes ISI up to the point
where the ``hard'' $NN$ collision occurs and the outgoing protons are
only subject to FSI after this ``hard'' collision.

It is worth remarking that the eikonal IFSI operators of
Eq.~(\ref{eq:eikonal_IFSI_operators}) are one-body operators,
i.e., they do not depend on the coordinates $\left( \vec{r}_2,
\vec{r}_3, \ldots, \vec{r}_A \right)$ of the residual nucleons.  The
normalization of the bound-state wave functions simplifies the
IFSI factor (\ref{eq:IFSI_factor}) considerably to $\mathcal{S}_{IFSI}
(\vec{r}) = {\widehat{\mathcal{S}}_{k1} \left( \vec{r} \right) } \:
{\widehat{\mathcal{S}}_{k2} \left( \vec{r} \right) } \:
{\widehat{\mathcal{S}}_{p1} \left( \vec{r} \right) }$ in the ROMEA
case.

In the numerical calculations, we have employed the global $S-V$
parametrizations of Cooper \textit{et al.} \cite{cooper93} and the
optical potential of van Oers \textit{et al.} \cite{oers82} to
describe the PNPI and TRIUMF data, respectively.  Hereafter, the $A(p,2p)$
calculations which adopt Eq.~(\ref{eq:eikonal_IFSI_operators}) as a
starting basis are labeled the relativistic optical model
eikonal approximation (ROMEA).

\subsubsection{\label{sec:RMSGA} Relativistic multiple-scattering
  Glauber approximation}

In the ROMEA approach, all the IFSI effects are parametrized in terms of
mean-field like optical potentials, i.e., the IFSI are seen as a scattering
of the nucleon with the residual nucleus as a whole.  As the energy increases,
shorter distances are probed and the scattering with the individual nucleons
becomes more relevant.  For proton kinetic energies $T_{p} \geq 1$~GeV, the
highly inelastic and diffractive character of the underlying elementary
proton-nucleon scattering cross sections makes the Glauber approach
\cite{glauber70,wallace75,yennie71} more natural.  This method reestablishes
the link between proton-nucleus interactions and the elementary proton-proton
and proton-neutron scattering.  It essentially relies on the eikonal, or,
equivalently, the small-angle approximation and the assumption of consecutive
cumulative scatterings of a fast nucleon on a composite target containing
``frozen'' point scatterers (nucleons).

In relativistic Glauber theory, the scattering wave function of a
nucleon with energy $E = \sqrt{k^{2} + M^{2}}$ and spin state $\left|
\frac{1}{2} m_s \right>$ reads \cite{debruyne02,ryckebusch03}
\begin{equation}
\psi_{\vec{k}, m_{s}}^{(+)} = \sqrt { \frac{E+M}{2M} }
{\widehat{\mathcal{S}}}
\left[
\begin{array}{c}
1 \\
\frac{1}{E+M} \vec{\sigma} \cdot \hat{\vec{p}}
\end{array}
\right]
e^{i \vec{k} \cdot \vec{r}} \chi_{\frac{1}{2}m_{s}}
\; .
\label{eq:Glauber_scatt_wave}
\end{equation}
The operator ${\widehat{\mathcal{S}}}$ implements the subsequent
elastic or ``mildly inelastic'' collisions of the fast nucleon with
the ``frozen'' spectator nucleons
\begin{equation}
{\widehat{\mathcal{S}}
\left( \vec{r}, \vec{r}_2, \vec{r}_3, \ldots, \vec{r}_A \right) }
\equiv \prod_{j=2}^{A}
\left[ 1 - \Gamma \left( \vec{b} - \vec{b_j} \right)
\theta \left( z - z_j \right) \right]
\; ,
\label{eq:Glauber_operator}
\end{equation}
where $\theta \left( z - z_j \right)$ ensures that the nucleon only
interacts with other nucleons if they are localized in its forward
propagation path.  Given the diffractive nature of $pN$ collisions at
GeV energies, the profile function $\Gamma \left( \vec{b} \right)$ for
central elastic $pN$ scattering is parametrized in a functional form
of the type
\begin{equation}
\Gamma \left( \vec{b} \right)
= \frac{{\sigma^{tot}_{pN}} \left(1 - i {\epsilon_{pN}} \right)}
{4\pi \beta_{pN}^{2}} \:
exp \left( - \frac{\vec{b}^{2}}{2\beta_{pN}^{2}}  \right)
\; .
\label{eq:profile_fctn}
\end{equation}
At lower energies that part of the profile function proportional to
$\epsilon_{pN}$ is non-Gaussian and makes significant contributions
to nuclear scattering.  Rather than Eq.~(\ref{eq:profile_fctn}) a
parametrization in terms of the Arndt $NN$ phases \cite{SAID} is
appropriate at lower energies.  For the calculations presented here,
which address higher energies, the Gaussian-like real part of
$\Gamma \left( \vec{b} \right)$ is the dominant contributor and the
use of Eq.~(\ref{eq:profile_fctn}) is justified.  The parameters in
Eq.~(\ref{eq:profile_fctn}) can be determined directly from
elementary nucleon-nucleon scattering experiments and include the
total $pN$ cross sections $\sigma^{tot}_{pN}$, the slope parameters
$\beta_{pN}$, and the ratios of the real to the imaginary part of
the scattering amplitude $\epsilon_{pN}$.  We obtained these Glauber
parameters through interpolation of the data base available from the
Particle Data Group \cite{pdg} (for more details, see
Ref.~\cite{ryckebusch03}).  As in the ROMEA framework, only the central
spin-independent contribution is retained and the impulse operator
is replaced by the nucleon momentum.

The Glauber operators in Eq.~(\ref{eq:IFSI_factor}) take the following
form
\begin{subequations}
\begin{equation}
\widehat{\mathcal{S}}_{p1} 
\left( \vec{r}, \vec{r}_2, \vec{r}_3, \ldots, \vec{r}_A \right)
= \prod_{j=2}^{A} \left[ 1 - \Gamma \left( \vec{b} - \vec{b_j} 
\right) \theta \left( z - z_j \right) \right]
\; ,
\end{equation}
\begin{equation}
\widehat{\mathcal{S}}_{k1} 
\left( \vec{r}, \vec{r}_2, \vec{r}_3, \ldots, \vec{r}_A \right)
= \prod_{j=2}^{A}
\left[ 1 - \Gamma \left( \vec{b} \, ' - \vec{b_j} \, ' 
\right) \theta \left( z_j \, ' - z \, ' \right) \right] 
\; ,
\end{equation}
\begin{equation}
\widehat{\mathcal{S}}_{k2} 
\left( \vec{r}, \vec{r}_2, \vec{r}_3, \ldots, \vec{r}_A \right)
= \prod_{j=2}^{A}
\left[ 1 - \Gamma \left( \vec{b} \, '' - \vec{b_j} \, '' 
\right) \theta \left( z_j \, '' - z \, '' \right) \right]
\; ,
\end{equation}
\label{eq:Glauber_IFSI_operators}
\end{subequations}
where $\vec{r}$ denotes the collision point and $\left( \vec{r}_2,
\vec{r}_3, \ldots, \vec{r}_A \right)$ are the positions of the frozen
spectator protons and neutrons in the target.  The $(\vec{b}, z)$,
$(\vec{b} \, ', z \, ')$, and $(\vec{b} \, '', z \, '')$ coordinate
systems are defined as in the previous section.  The step functions
make sure that the incoming proton can only interact with the
spectator nucleons which it encounters before the ``hard'' collision
and the outgoing protons can only interact with the spectator nucleons
which they find in their forward propagation paths.

Contrary to ROMEA, the Glauber IFSI operators of
Eq.~(\ref{eq:Glauber_IFSI_operators}) are genuine $A$-body operators,
so the integration over the coordinates of the spectator nucleons in
Eq.~(\ref{eq:IFSI_factor}) has to be carried out explicitly.  This
makes the numerical evaluation of the Glauber IFSI factor very
challenging.  Standard numerical integration techniques were adopted
to evaluate the IFSI factor and no additional approximations, such as
the commonly used thickness-function approximation, were introduced.

Henceforth, we refer to calculations based on
Eq.~(\ref{eq:Glauber_IFSI_operators}) as the relativistic
multiple-scattering Glauber approximation (RMSGA).

\section{\label{sec:IFSI_factor} Numerical results for the IFSI factor}

In this section, results for the IFSI factor (\ref{eq:IFSI_factor})
are given for the knockout of nucleons from the Fermi level in
$^{12}$C, $^{16}$O, and $^{40}$Ca, at an incident energy $T_{p1} =
1$~GeV and a scattered proton kinetic energy $T_{k1} = 870$~MeV.  Thereby,
we adopt coplanar scattering angles $(\theta_{1}, \theta_{2}) = (13.4^{\circ},
67^{\circ})$ on opposite sides of the incident beam, i.e., kinematics
corresponding with the PNPI experiment of Ref.~\cite{belostotsky87}.
All IFSI effects are included in the IFSI factor $\mathcal{S}_{IFSI}
(\vec{r})$.  Note that in the absence of initial- and final-state
interactions the real part of the IFSI factor equals one, whereas the
imaginary part vanishes identically.

The $A(p,2p)$ IFSI factor is a function of three independent variables
$(r, \theta, \phi)$.  The $z$-axis is chosen along the direction of
the incoming beam $\vec{p}_{1}$, the $y$-axis lies along $\vec{p}_{1}
\times \vec{k}_{1}$ and the $x$-axis lies in the scattering plane
defined by the proton momenta $\vec{p}_{1}$ and $\vec{k}_{1}$.  $\theta$
and $\phi$ denote the polar and azimuthal angles with respect to the
$z$-axis and the $x$-axis, respectively.  The radial coordinate $r$
represents the distance relative to the center of the target nucleus.

\subsection{\label{sec:theta_depend_IFSI} $\theta$ dependence}

\begin{figure*}
\begin{center}
\resizebox{0.95 \textwidth}{!}
{\includegraphics{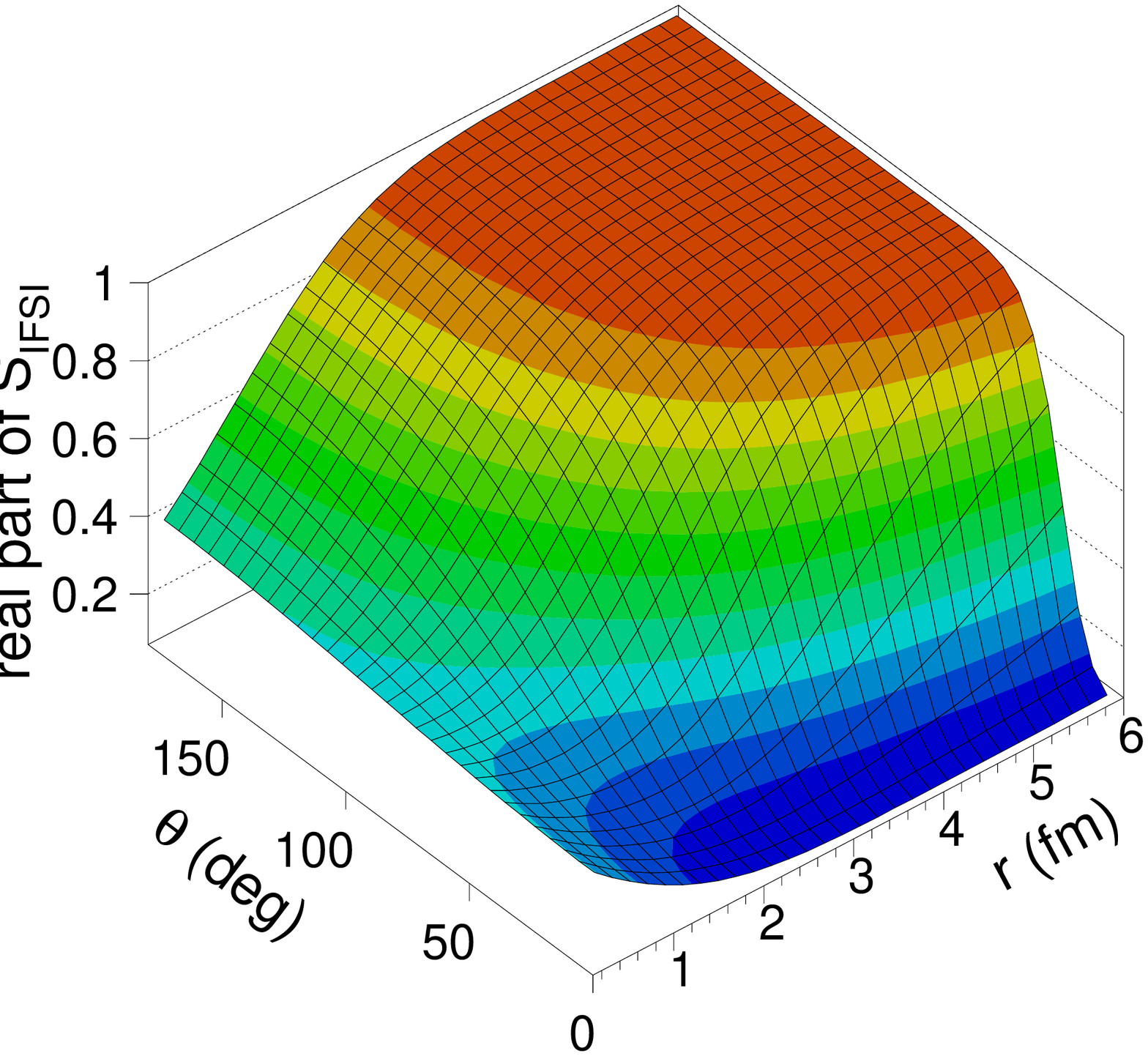}
  \includegraphics{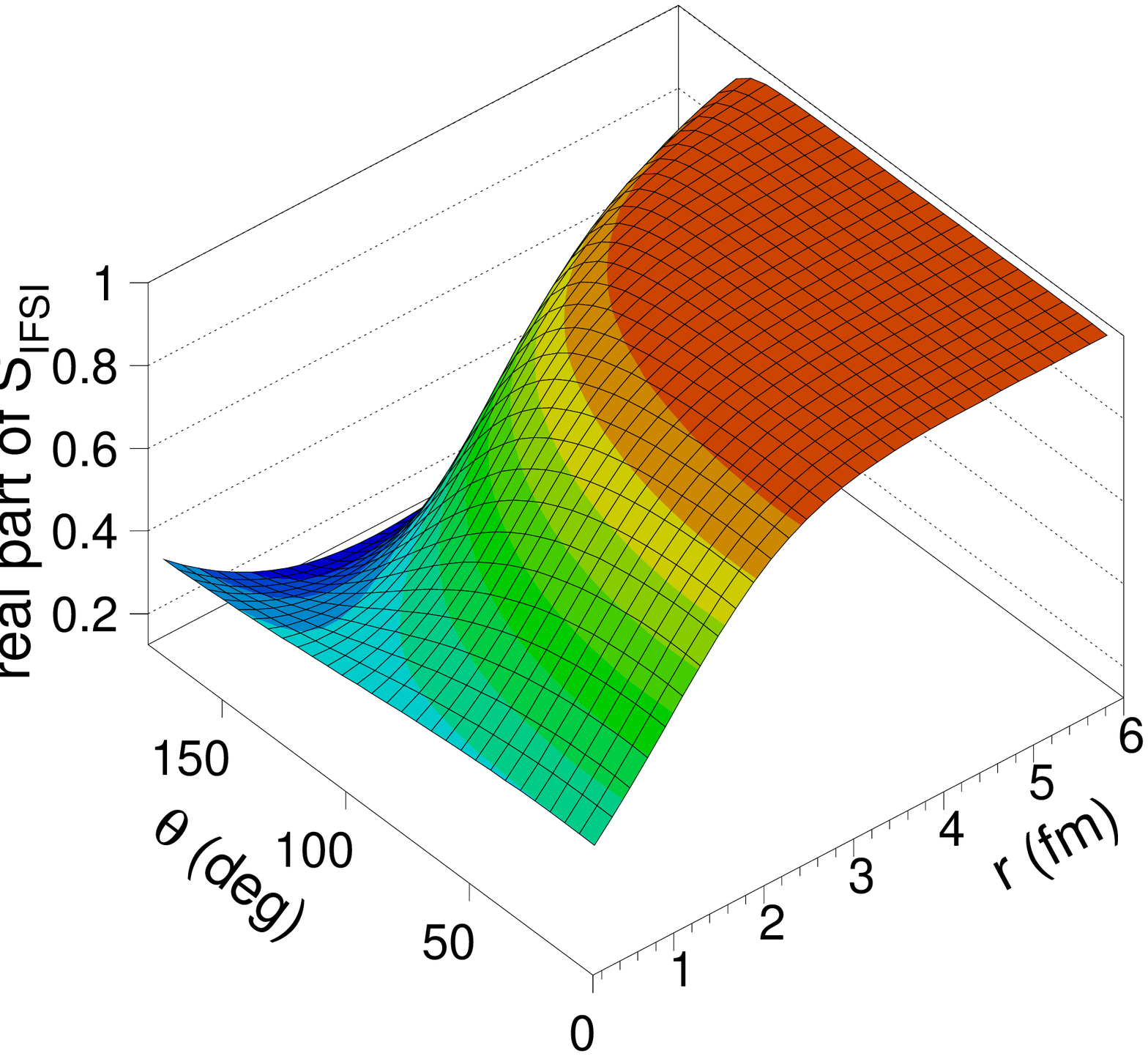}}
\resizebox{0.95 \textwidth}{!}
{\includegraphics{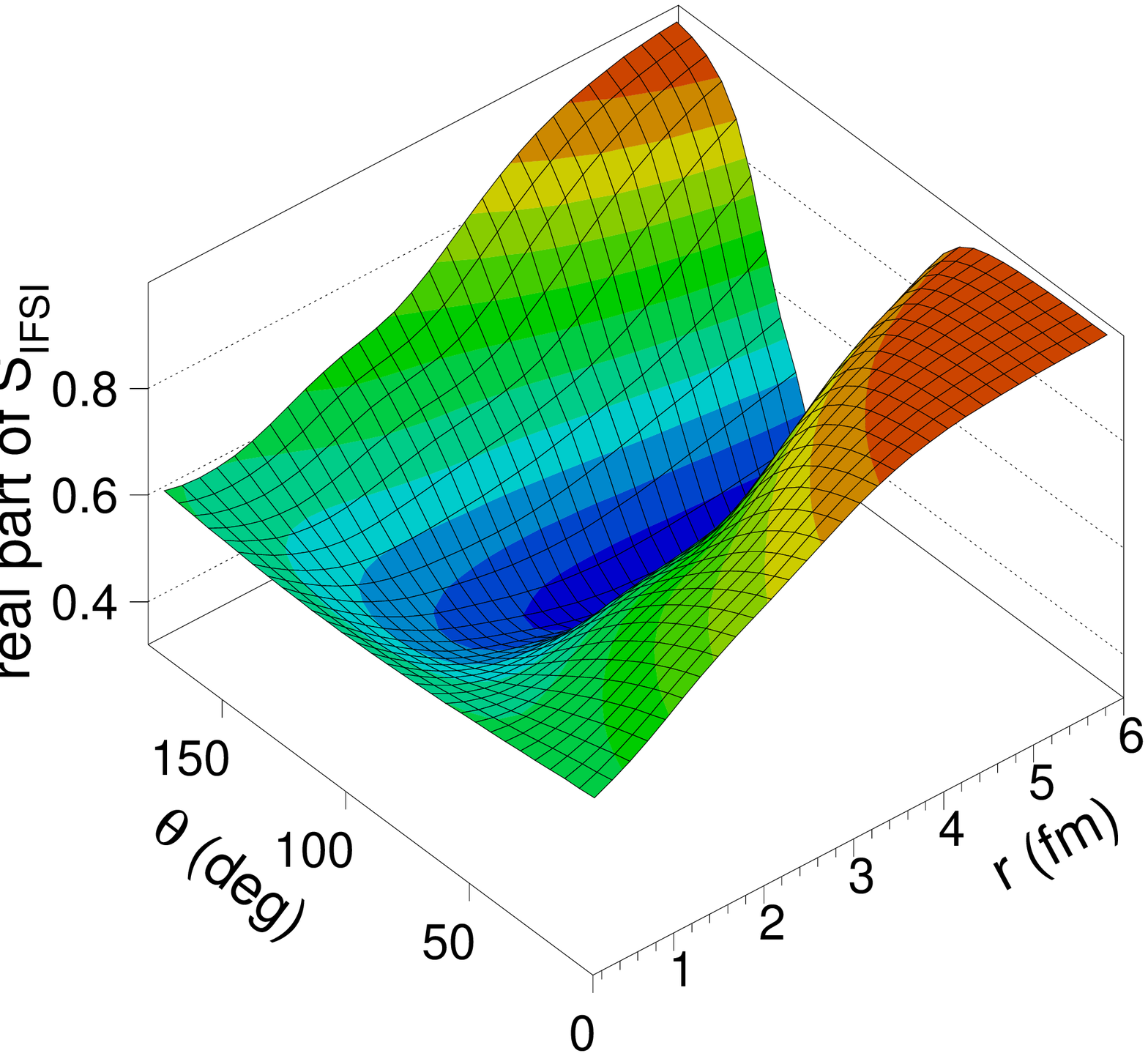}
  \includegraphics{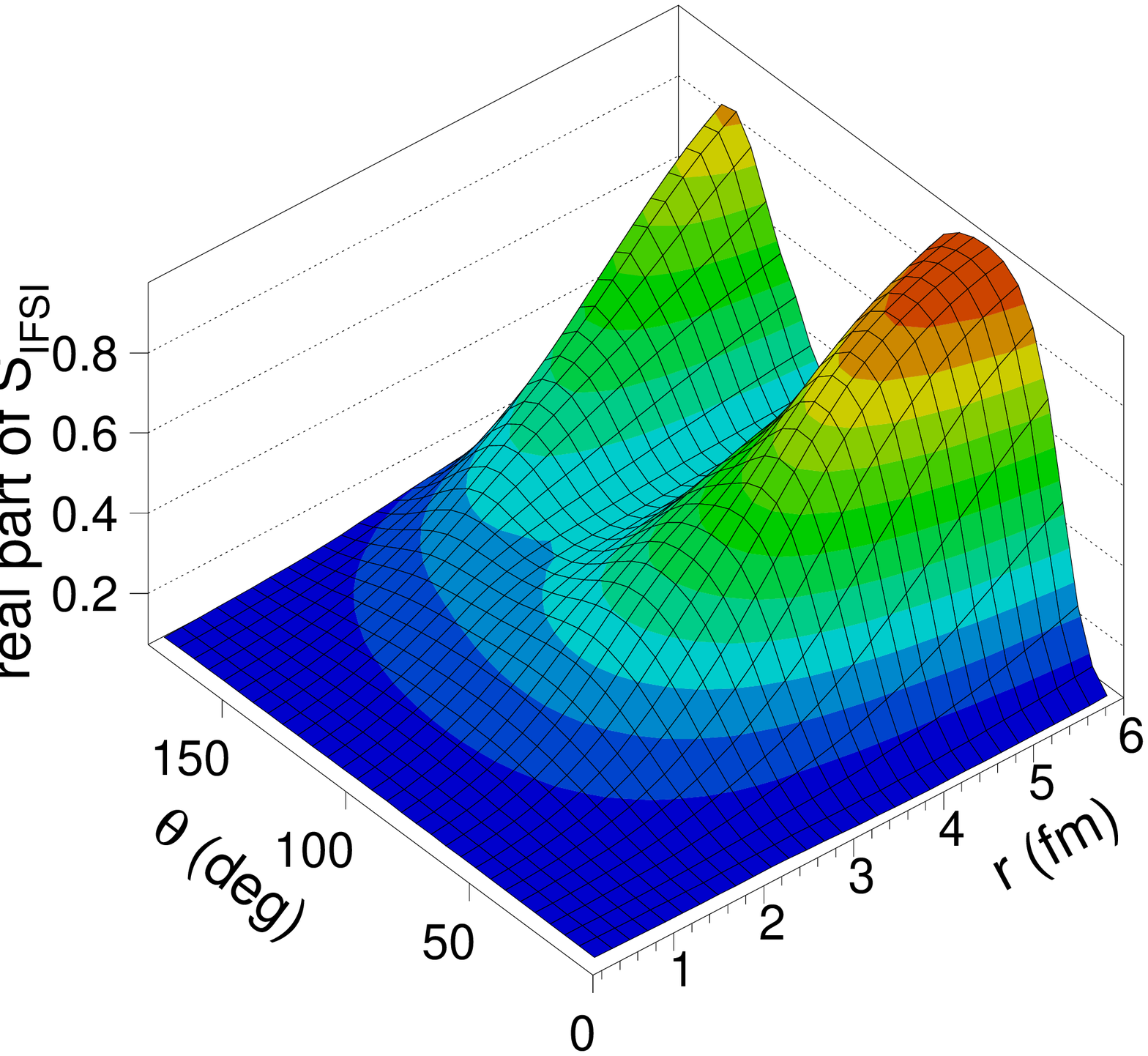}}
\caption{(Color online) The radial and polar-angle dependence of the real part
  of the IFSI factor $\mathcal{S}_{IFSI}$ in the scattering plane ($\phi =
  0^{\circ}$) for proton knockout from the Fermi level in $^{12}$C.
  The upper left panel is the contribution from the impinging proton
  ($\widehat{\mathcal{S}}_{p1}$), while the
  upper right panel shows the effect of the FSI of the scattered
  proton ($\widehat{\mathcal{S}}_{k1}$).  In the bottom left figure,
  the effect of the FSI of the ejected proton
  ($\widehat{\mathcal{S}}_{k2}$) is presented and the bottom right
  figure shows the complete IFSI factor ($\widehat{\mathcal{S}}_{k1}
  \: \widehat{\mathcal{S}}_{k2} \: \widehat{\mathcal{S}}_{p1}$).  The
  kinematics was $T_{p1} = 1$~GeV, $T_{k1} = 870$~MeV, $\theta_{1} =
  13.4^{\circ}$, and $\theta_{2} = 67^{\circ}$.} 
\label{fig:re.12c.p.edad.phi0.3D}
\end{center}
\end{figure*}

\begin{figure*}
\begin{center}
\resizebox{0.95 \textwidth}{!}
{\includegraphics{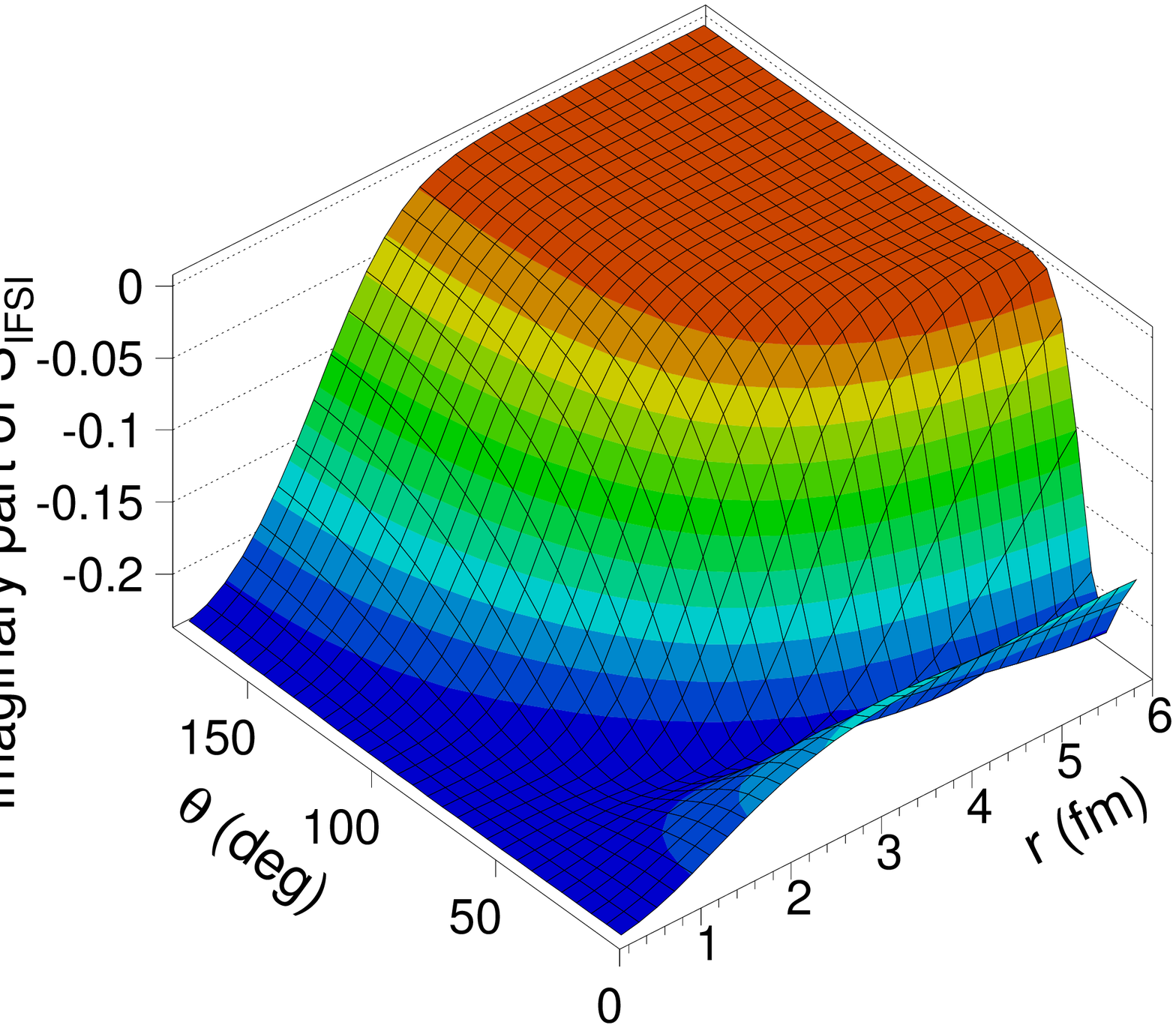}
  \includegraphics{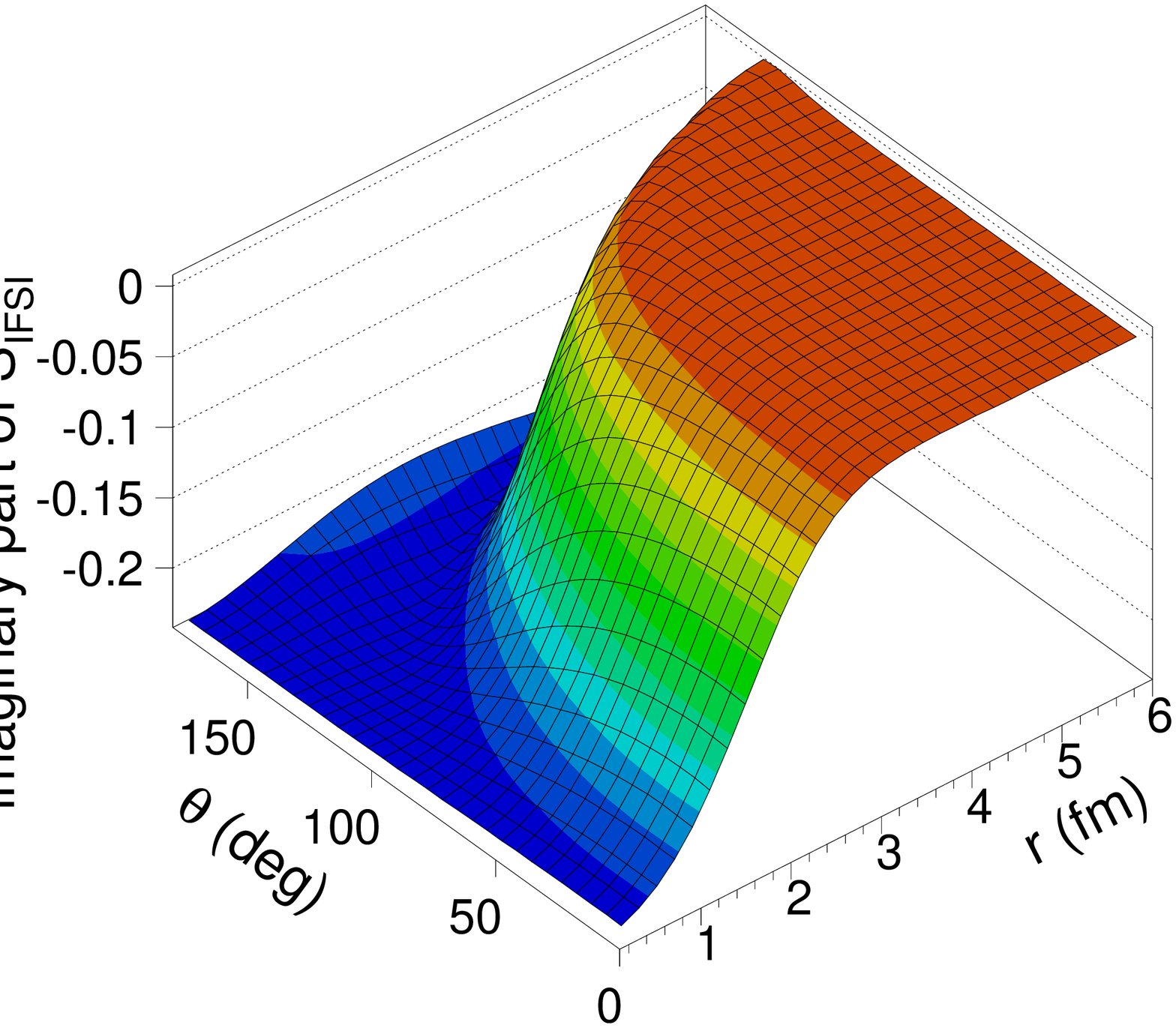}}
\resizebox{0.95 \textwidth}{!}
{\includegraphics{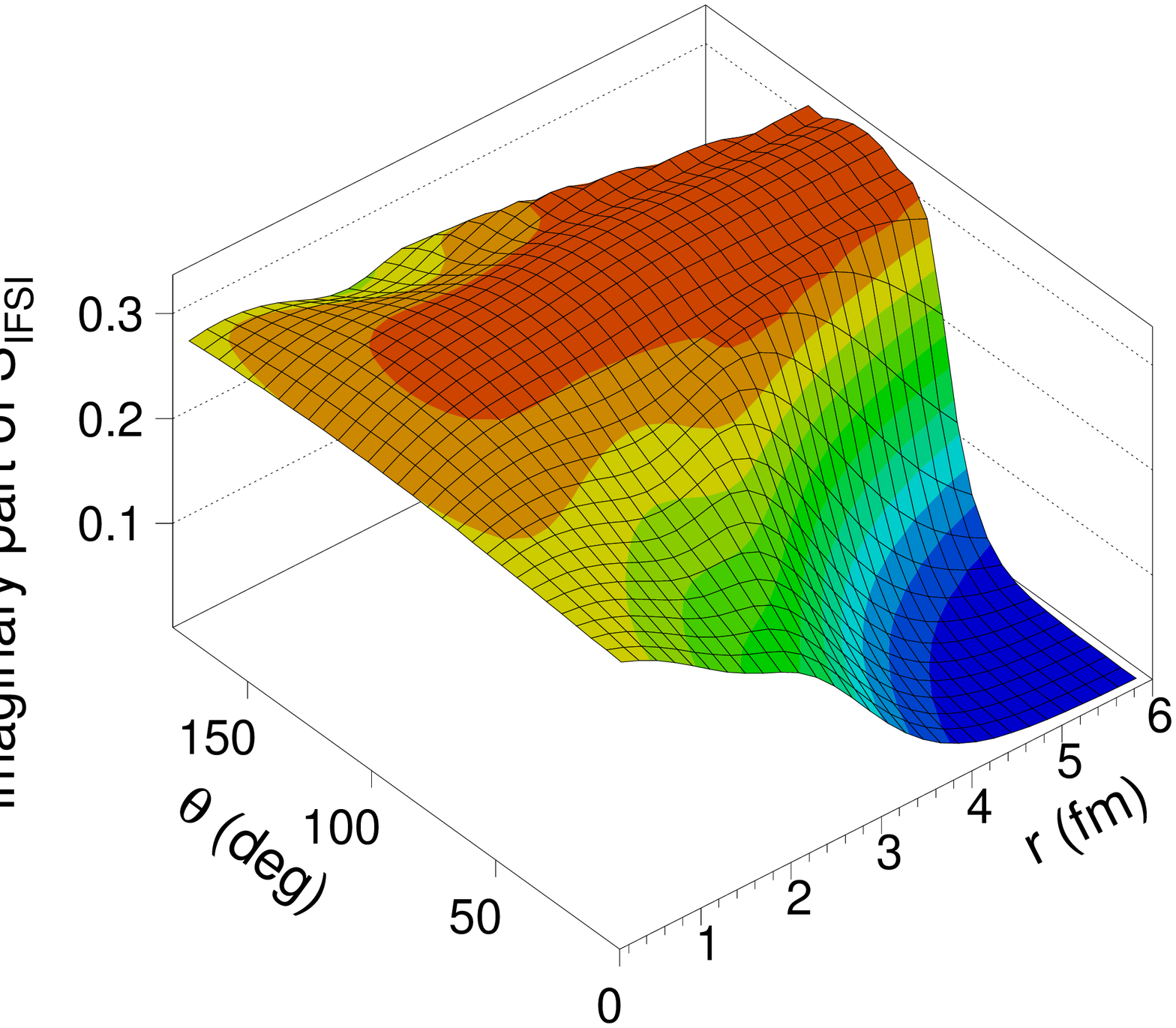}
  \includegraphics{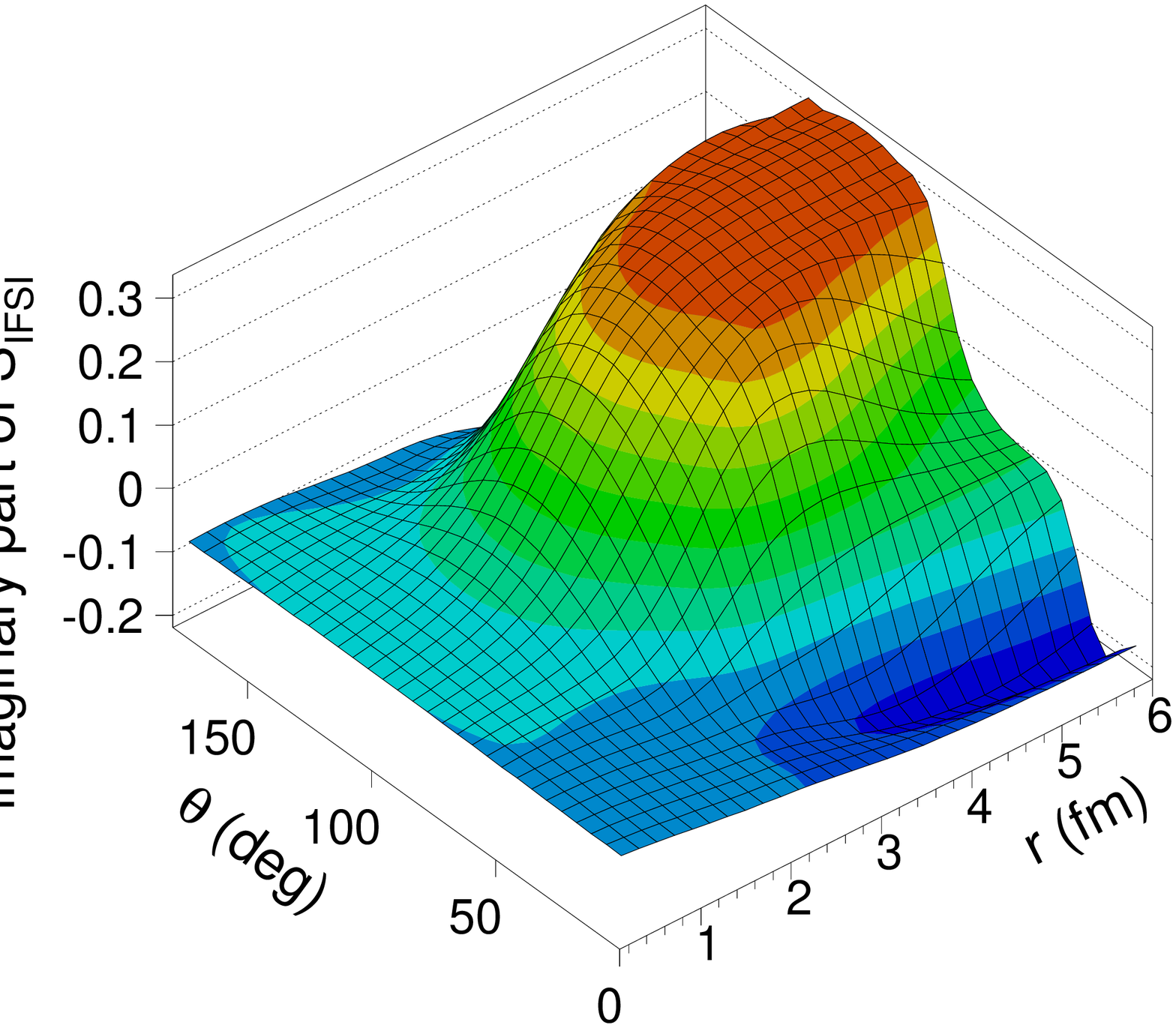}}
\caption{(Color online) As in Fig.~\ref{fig:re.12c.p.edad.phi0.3D} but now
  for the imaginary part of the IFSI factor.}
\label{fig:im.12c.p.edad.phi0.3D}
\end{center}
\end{figure*}

To gain a better insight into the dependence of the IFSI factor on
$r$, $\theta$, and $\phi$, we calculated the contribution of the three
distorting functions $\widehat{\mathcal{S}}_{p1}$,
$\widehat{\mathcal{S}}_{k1}$, and $\widehat{\mathcal{S}}_{k2}$ to the
IFSI factor.  In
Figs.~\ref{fig:re.12c.p.edad.phi0.3D} and
\ref{fig:im.12c.p.edad.phi0.3D} results are displayed for the computed
real and imaginary part of $\mathcal{S}_{IFSI} (r, \theta, \phi = 0)$,
for proton emission from the Fermi level in $^{12}$C.  The results were
computed within the ROMEA framework, using the EDAD1 optical-potential
parametrization of \cite{cooper93}.

The $\theta$ dependence can be interpreted as follows.  For a given
$r$, the distance that the incoming proton travels through the target
nucleus before colliding ``hard'' with a target nucleon, decreases with
increasing angle
$\theta$.  As a consequence, small values of $\theta$ induce the largest
ISI.  For the FSI of the scattered proton, the opposite holds true and
$\theta = 180^{\circ}$ for a large $r$ value corresponds to
an event whereby the ``hard'' collision transpires at the outskirts
of the nucleus and the scattered proton has to travel through the
whole nucleus before it becomes asymptotically free, thus giving rise
to the smallest (largest) values for the real (imaginary) part of the
IFSI factor.  Unlike the scattered proton, which moves almost
collinear to the $z$-axis, the ejected nucleon leaves the
nucleus under a large scattering angle $\theta_{2}$.  Hence, the FSI
are minimal for $\theta$ close to $0^{\circ}$ or $180^{\circ}$ and
maximal for $\theta$ around $180^{\circ} - \theta_{2}$.  Finally, the
$\theta$ dependence of the complete IFSI factor is the result of the
interplay between the three distorting effects, with the strongest
scattering and absorption observed at $\theta$ close to $0^{\circ}$,
$180^{\circ} - \theta_{2}$, and $180^{\circ}$.

\subsection{\label{sec:r_depend_IFSI} $r$ dependence}

\begin{figure*}
\begin{center}
\resizebox{0.95 \textwidth}{!}
{\includegraphics{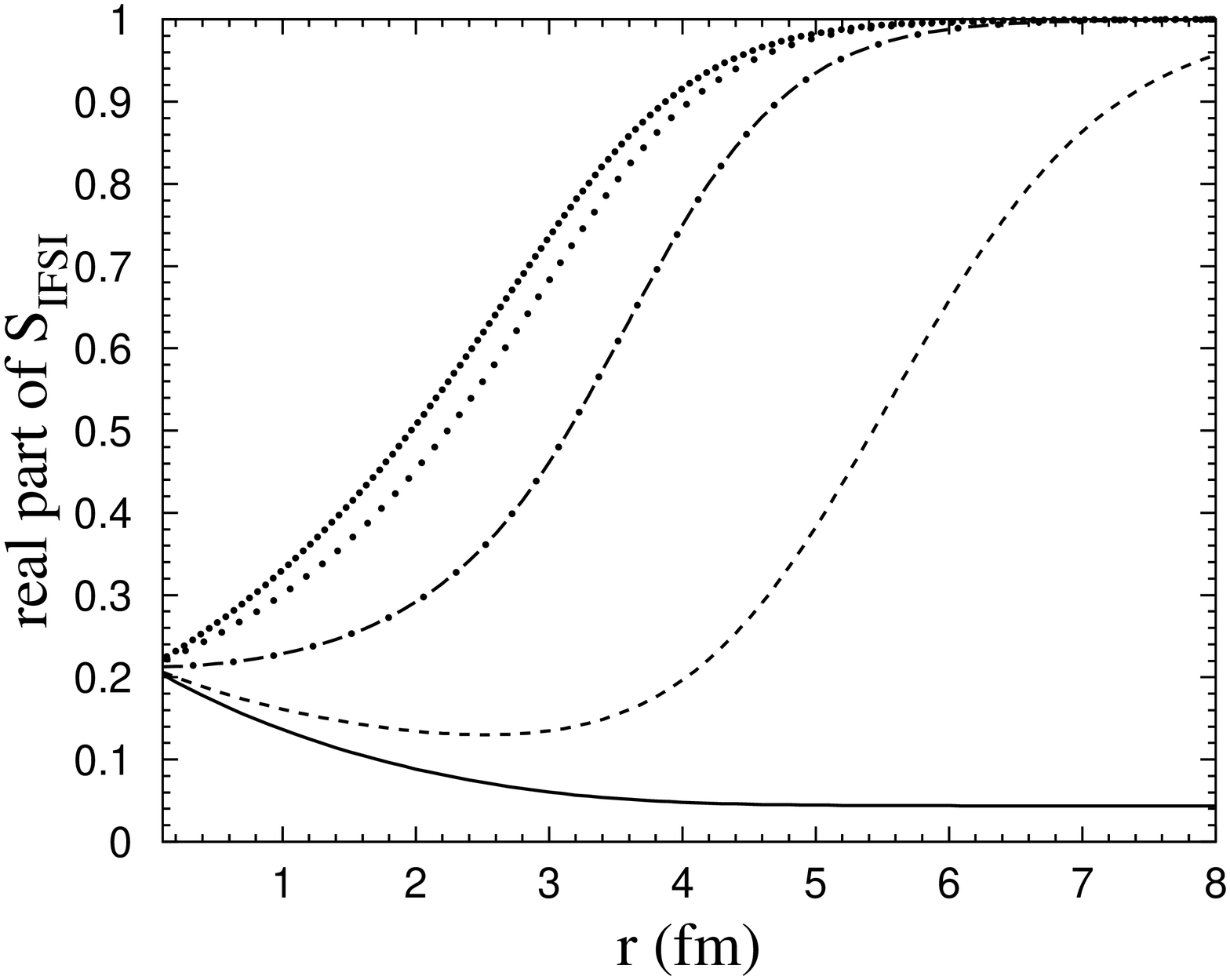}
  \includegraphics{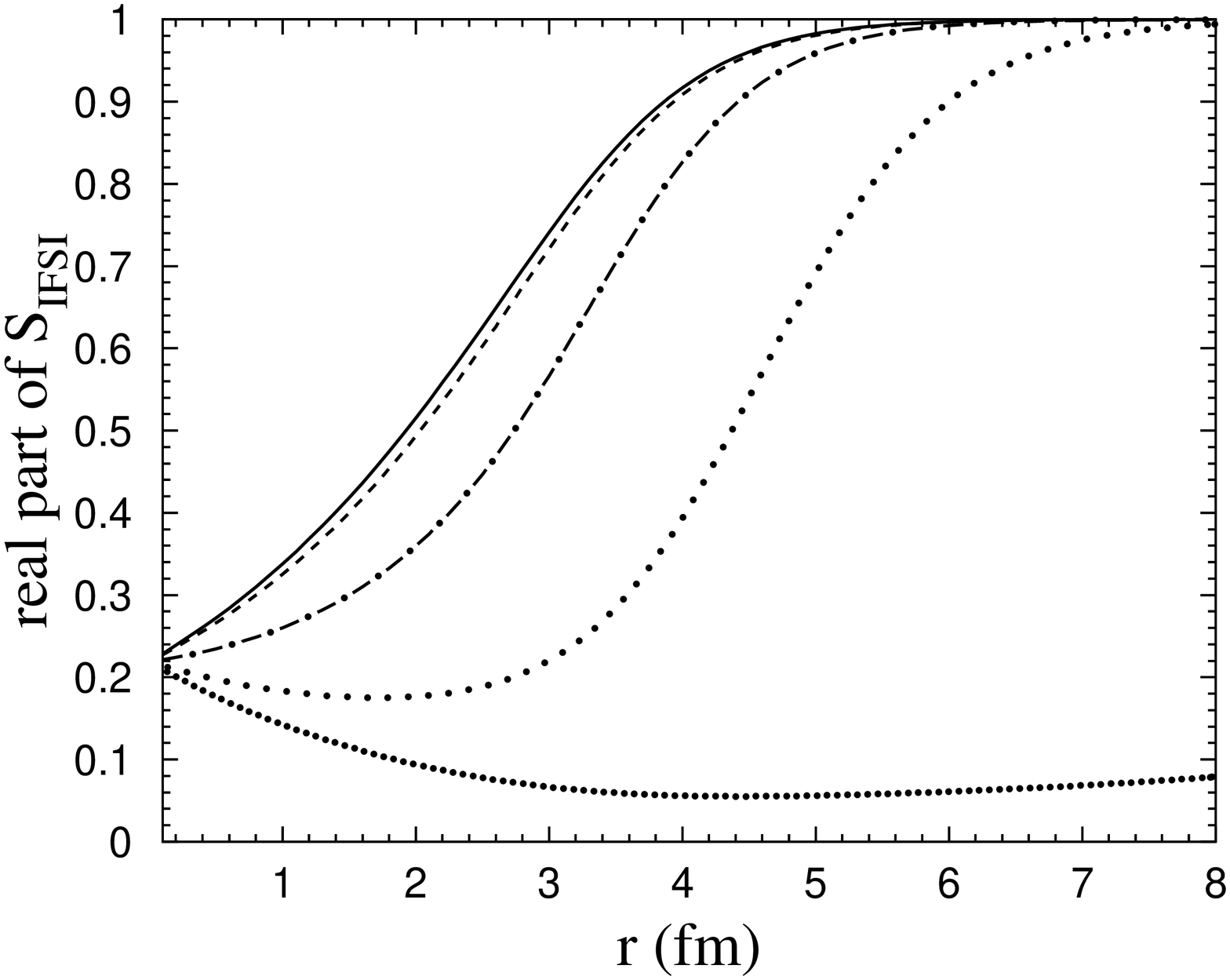}}
\resizebox{0.95 \textwidth}{!}
{\includegraphics{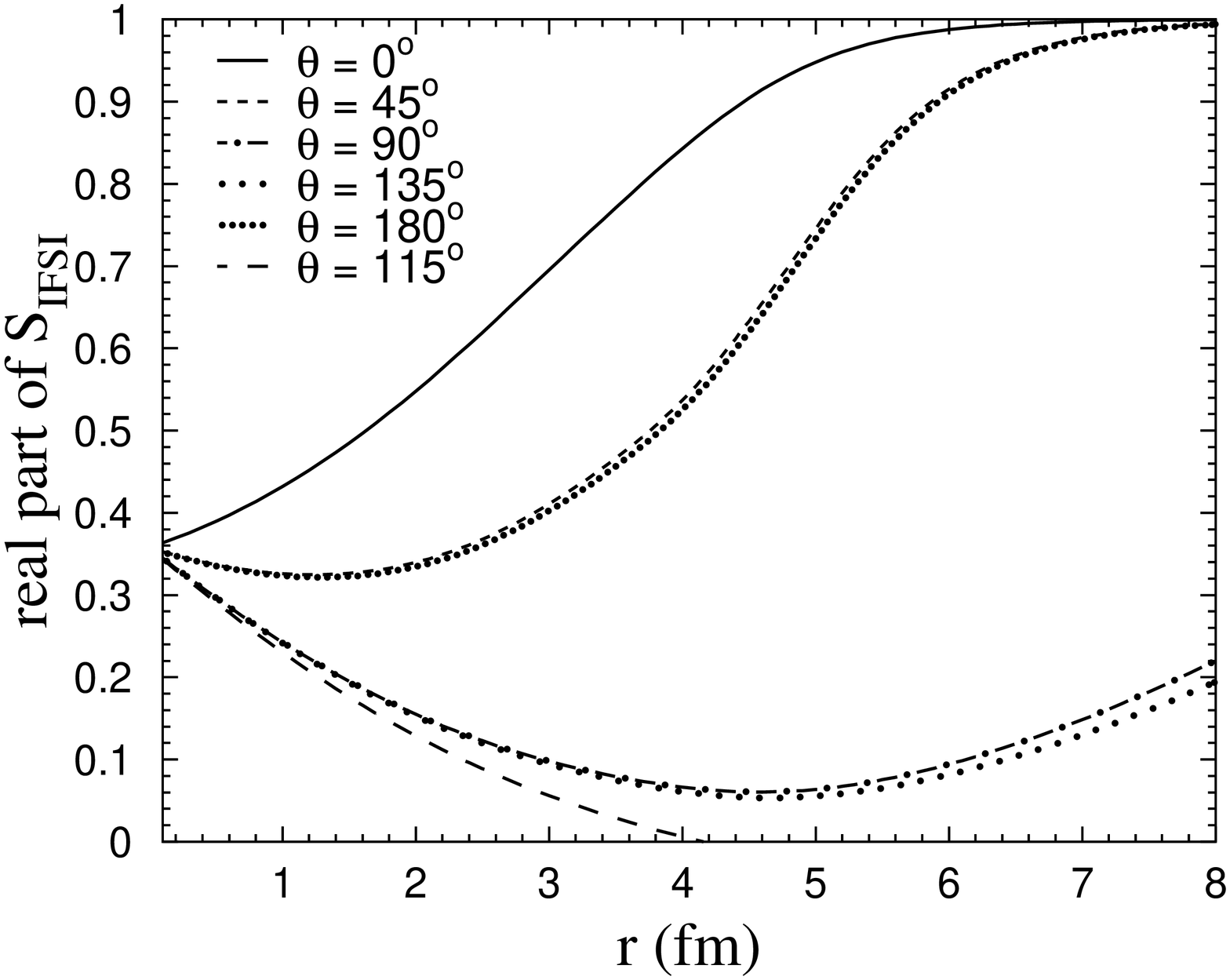}
  \includegraphics{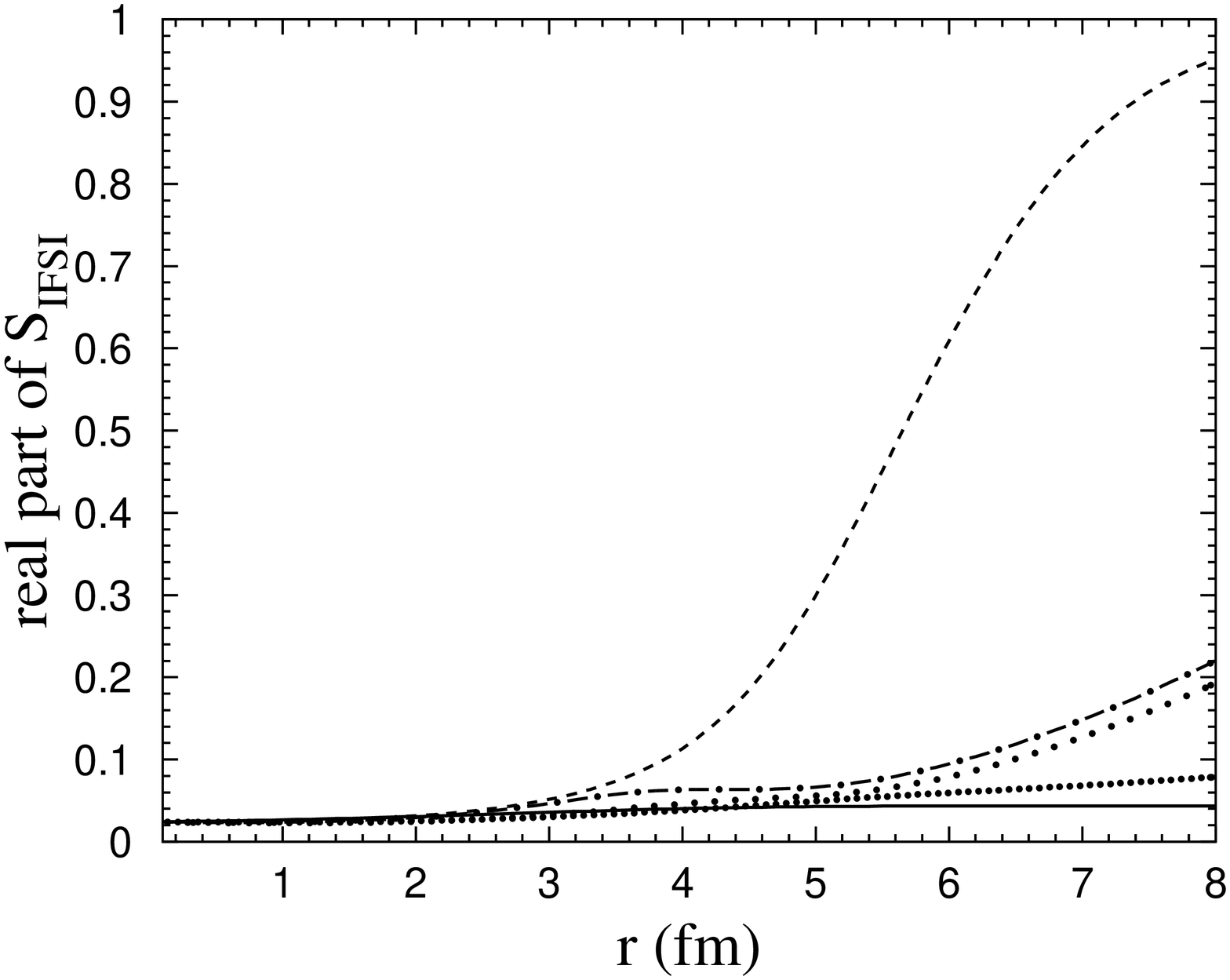}}
\caption{The radial dependence of the real part of the IFSI factor
  $\mathcal{S}_{IFSI}$ in the scattering plane ($\phi = 0^{\circ}$)
  for neutron knockout from the Fermi level in $^{40}$Ca.  The upper
  left, upper right, and bottom left panels display
  $\widehat{\mathcal{S}}_{p1}$, $\widehat{\mathcal{S}}_{k1}$, and
  $\widehat{\mathcal{S}}_{k2}$, respectively, while the complete IFSI
  factor is shown in the bottom right picture.}
\label{fig:re.40ca.n.edai.phi0.radial}
\end{center}
\end{figure*}

Figure~\ref{fig:re.40ca.n.edai.phi0.radial} displays the real part of
the IFSI factor as a function of $r$ at various values of $\theta$.
The ROMEA calculations were performed for the same kinematics as in
Figs.~\ref{fig:re.12c.p.edad.phi0.3D} and
\ref{fig:im.12c.p.edad.phi0.3D}, and use the EDAI optical-potential
fit of \cite{cooper93}. 

Turning first to the upper left panel, it can be inferred from the
picture that the ISI effects increase with growing $r$ for $\theta =
0^{\circ}$.  For $\theta = 45^{\circ}$ and increasing $r$, initially,
the growing distance the proton has to travel through the nucleus
leads to a decrease of the real part of $\widehat{\mathcal{S}}_{p1}$.
This is followed by an increase for larger $r$ up to
$\widehat{\mathcal{S}}_{p1} = 1$.  This reduction in ISI effects with
increasing $r$ is brought about by the incoming proton's path through
the nucleus moving away from the nuclear interior and closer to the
less dense nuclear surface.  The other curves of the upper left figure
illustrate a general trend for $90^{\circ} \leq \theta \leq
180^{\circ}$~: as $r$ increases, the real part of the IFSI factor
grows correspondingly.  As can be appreciated from
Fig.~\ref{fig:re.40ca.n.edai.phi0.radial} as well as from the previous
figures, the global behavior of the $\widehat{\mathcal{S}}_{k1}$
factor describing the scattered proton's FSI can be related to that of
the ISI factor $\widehat{\mathcal{S}}_{p1}$ through the substitution
$\theta \rightarrow 180^{\circ} - \theta$.  This approximate symmetry
can be attributed to the small scattering angle $\theta_{1}$, i.e.,
the scattered proton leaves the nucleus almost
parallel to the incoming proton's direction.  In the bottom left
panel, the additional curve ($\theta = 115^{\circ}$, i.e., close to
$180^{\circ} - \theta_{2}$) represents the situation of maximal FSI of
the ejected nucleon.  For this $\theta$ value, the path of the ejected
nucleon passes through the center of the nucleus and the distance
travelled through the nucleus increases with $r$.  Accordingly, the
real part of $\widehat{\mathcal{S}}_{k2}$ is a monotonously decreasing
function of $r$.  The other extreme is the $\theta = 0^{\circ}$ case
where increasing $r$ means less FSI.  For the other $\theta$ values,
the absorption reaches its maximum for some intermediate $r$ value.
Again, the combination of $\widehat{\mathcal{S}}_{p1}$,
$\widehat{\mathcal{S}}_{k1}$, and $\widehat{\mathcal{S}}_{k2}$
determines the total IFSI factor with the strongest attenuation
predicted in the nuclear interior.

\subsection{\label{sec:phi_depend_IFSI} $\phi$ dependence}

\begin{figure*}
\begin{center}
\resizebox{0.95 \textwidth}{!}
{\includegraphics{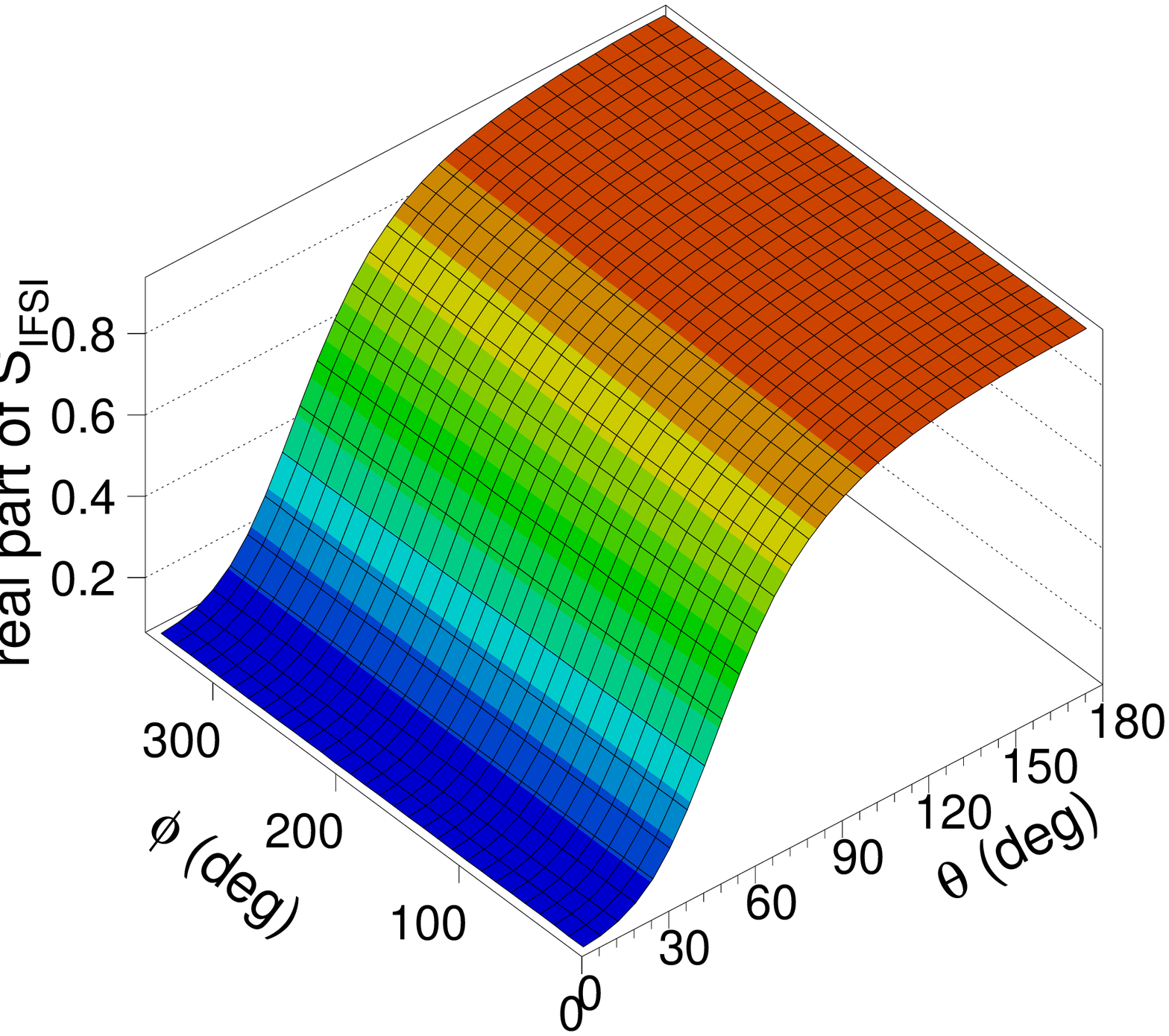}
  \includegraphics{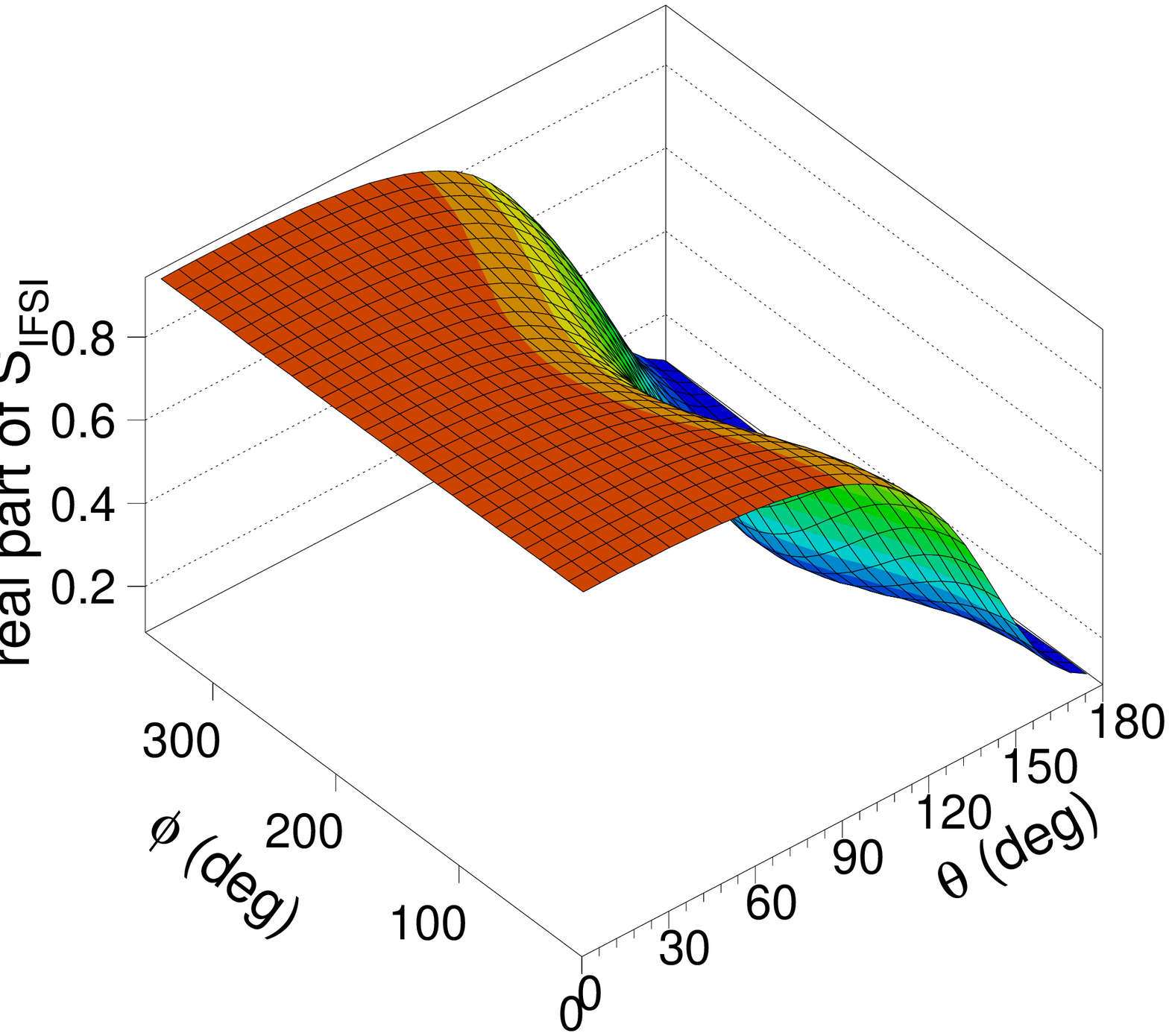}}
\resizebox{0.95 \textwidth}{!}
{\includegraphics{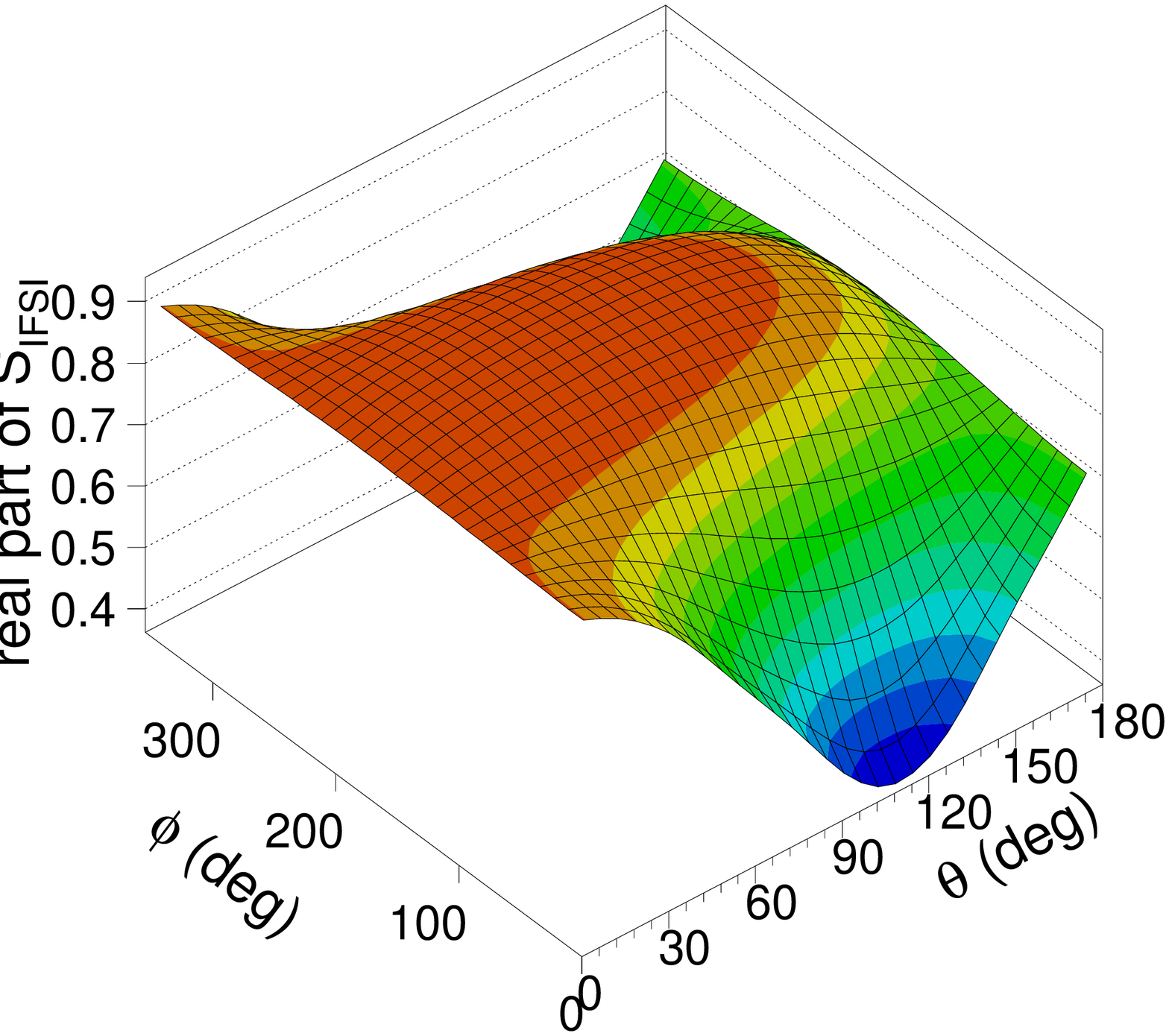}
  \includegraphics{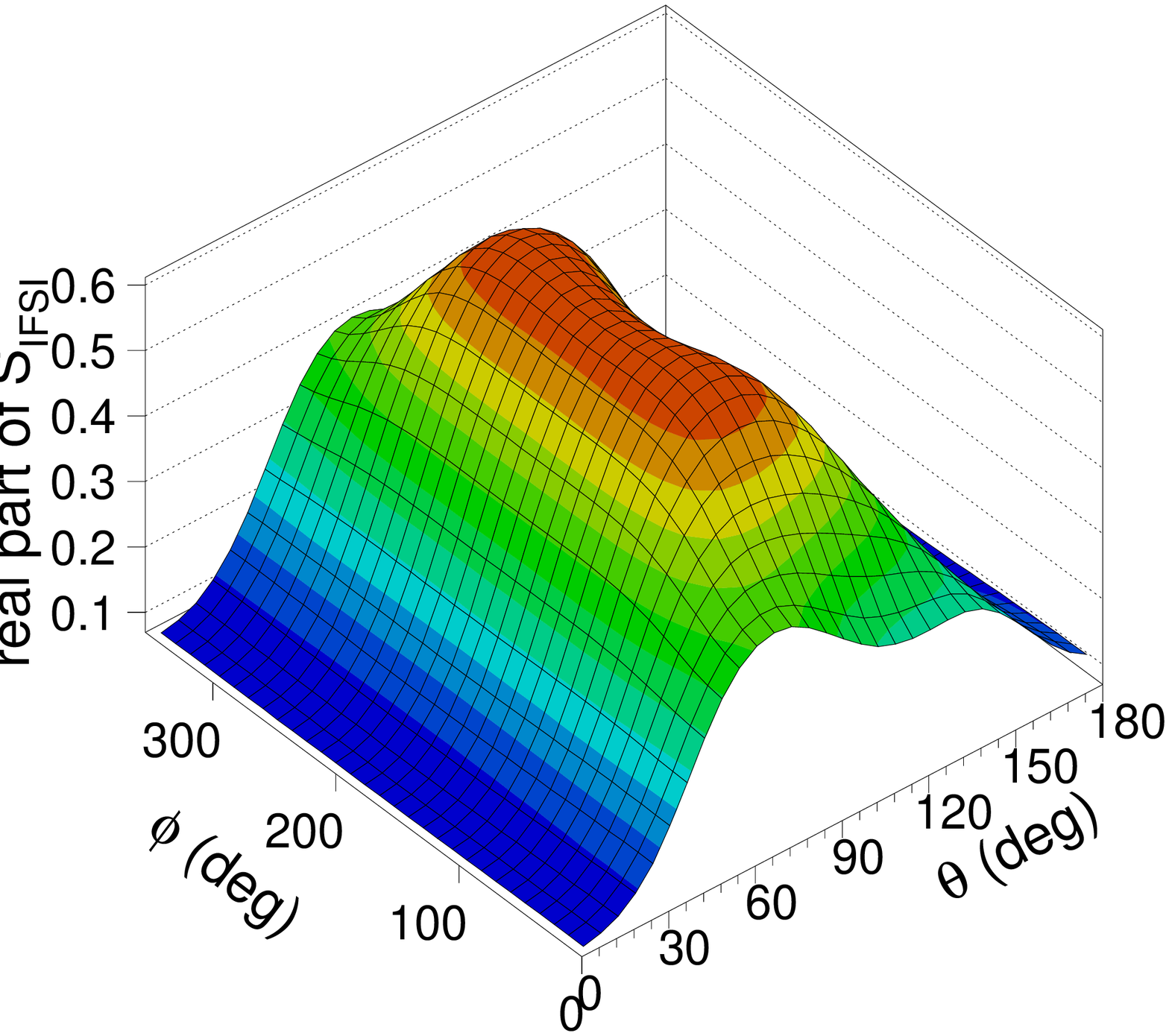}}
\caption{(Color online) The polar- and azimuthal-angle dependence of the
  real part of $\mathcal{S}_{IFSI} (r = 3~fm, \theta, \phi)$ for proton
  knockout from the Fermi level in $^{16}$O.  Kinematics as in
  Fig.~\ref{fig:re.12c.p.edad.phi0.3D}.  ROMEA calculation with the
  EDAD2 optical potential \cite{cooper93}.  As in the previous
  figures, the upper left, upper right, and bottom left panels
  represent the effect of the ISI of the incoming proton, the FSI of
  the scattered proton, and the FSI of the struck nucleon,
  respectively, whereas the bottom right figure displays the complete
  IFSI factor.}
\label{fig:re.16o.p.edad2.r3fm.3D}
\end{center}
\end{figure*}

The dependence of the IFSI factor on the azimuthal angle of the
collision point is quite straightforward.  One representative result
is displayed in Fig.~\ref{fig:re.16o.p.edad2.r3fm.3D}.  Here, $\cos
\phi \geq 0$ ($\cos \phi \leq 0$) refers to a situation where the
``hard'' $NN$ collision occurs in the upper (lower) hemisphere with
respect to the $yz$-plane (see Fig.~\ref{fig:geometry} for a collision
point located in the upper hemisphere).  Due to the cylindrical
symmetry about the $z$-axis, the factor describing the ISI of the
incoming proton is independent of $\phi$.  Regarding the scattered
proton, we observe the least FSI in the upper hemisphere, since the
proton then avoids passing through the highly absorbing nuclear
interior.  For the ejected nucleon, the contrary applies and the
strongest FSI effects are found for $\phi = 0^{\circ}$.  As the
$xz$-plane is defined as the scattering plane, the IFSI factor
possesses the symmetry $\mathcal{S}_{IFSI} (r, \theta, 2\pi - \phi) =
\mathcal{S}_{IFSI} (r, \theta, \phi)$ for coplanar scattering.

\subsection{\label{sec:level_A_depend} Level and $A$ dependence}

\begin{figure*}
\begin{center}
\resizebox{0.95 \textwidth}{!}
{\includegraphics{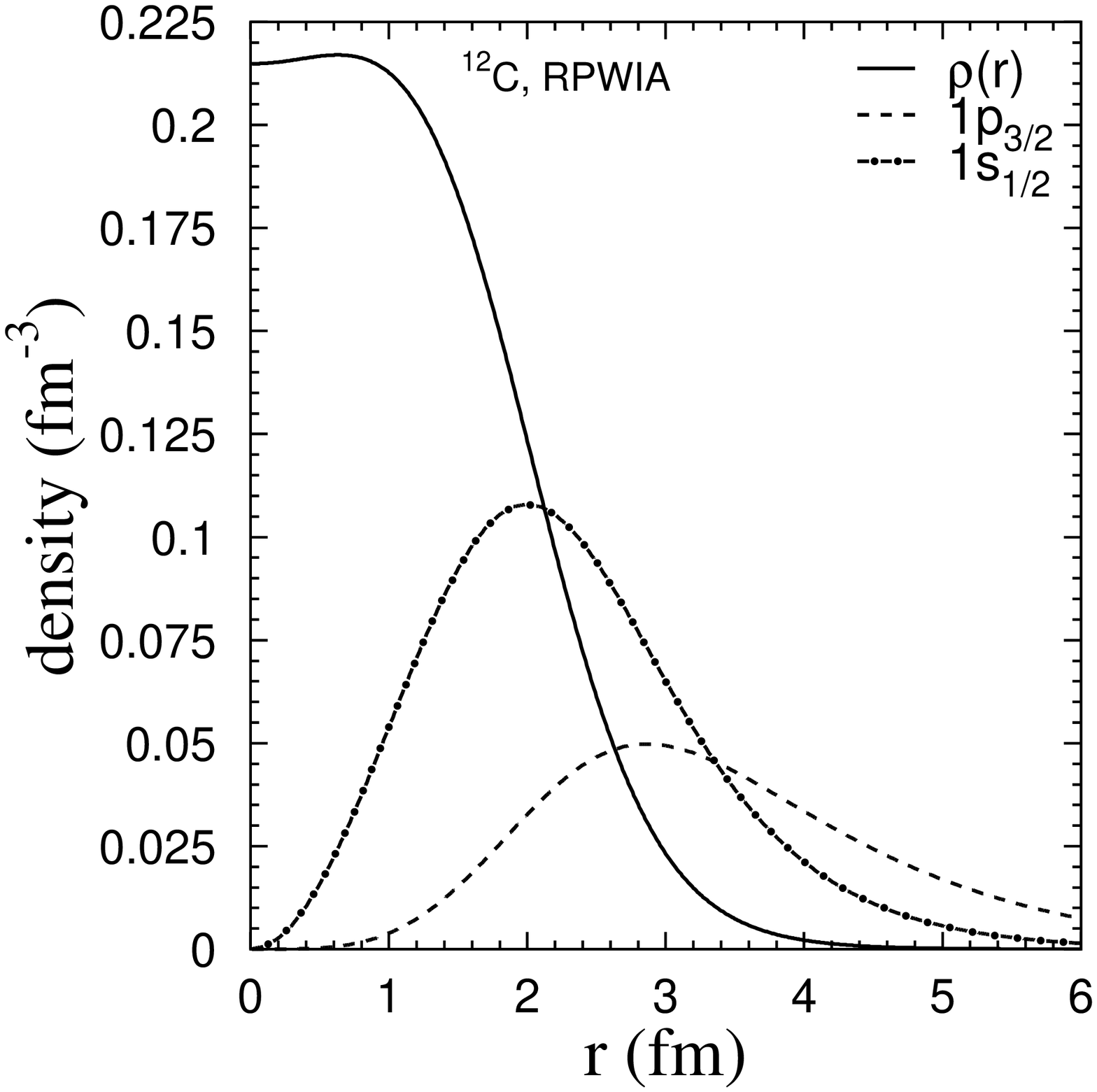}
 \includegraphics{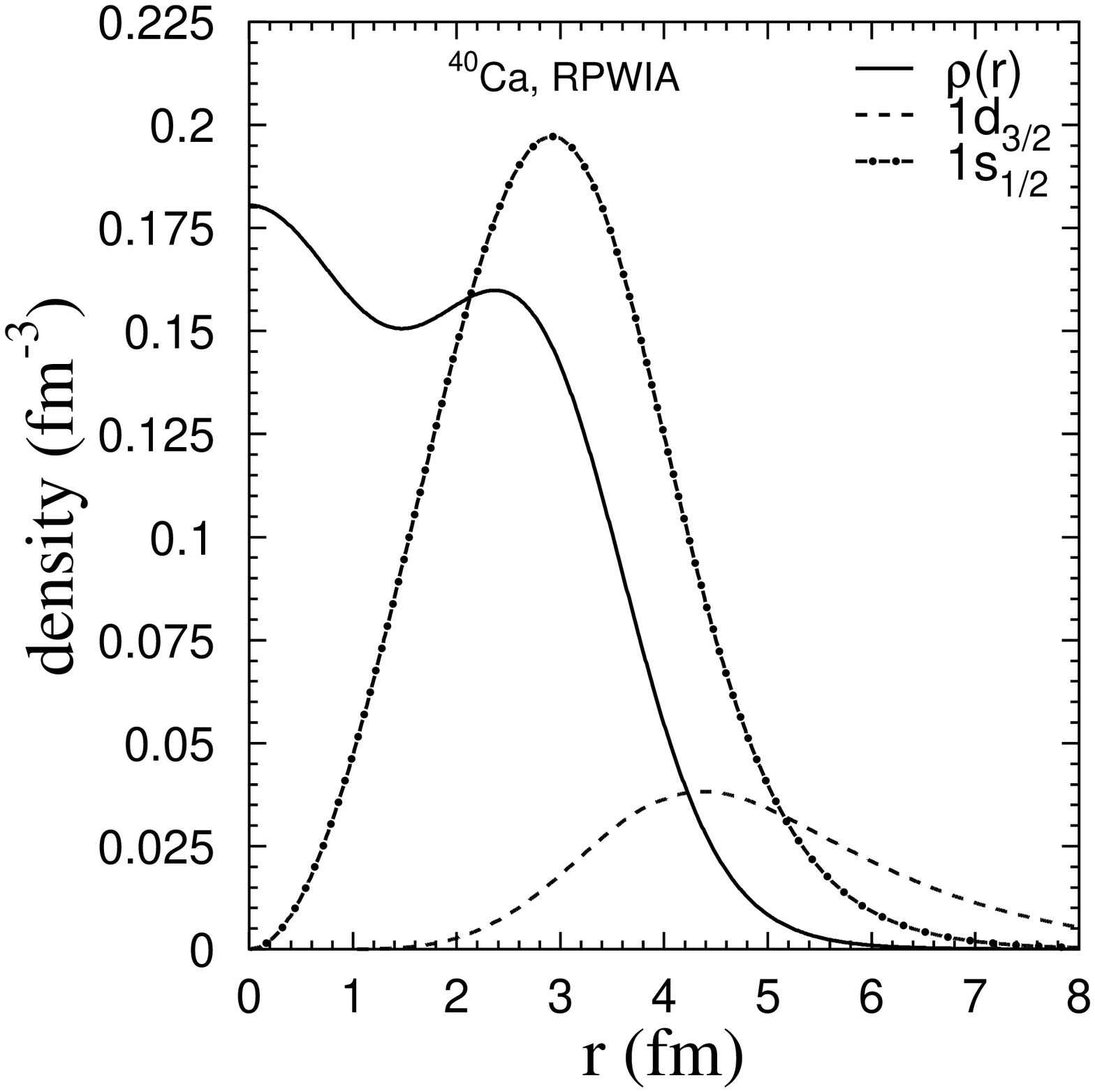}}
\resizebox{0.95 \textwidth}{!}
{\includegraphics{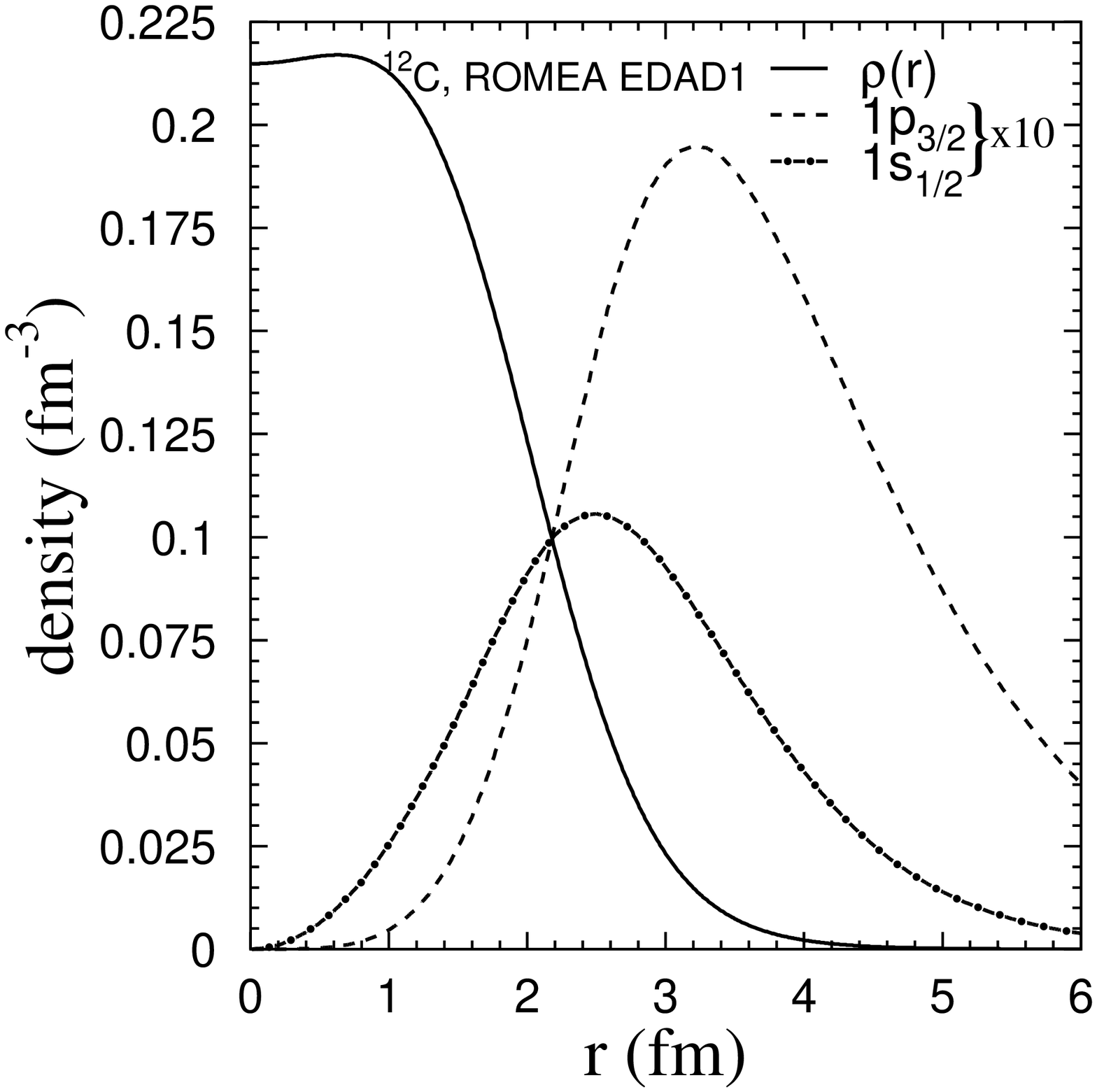}
 \includegraphics{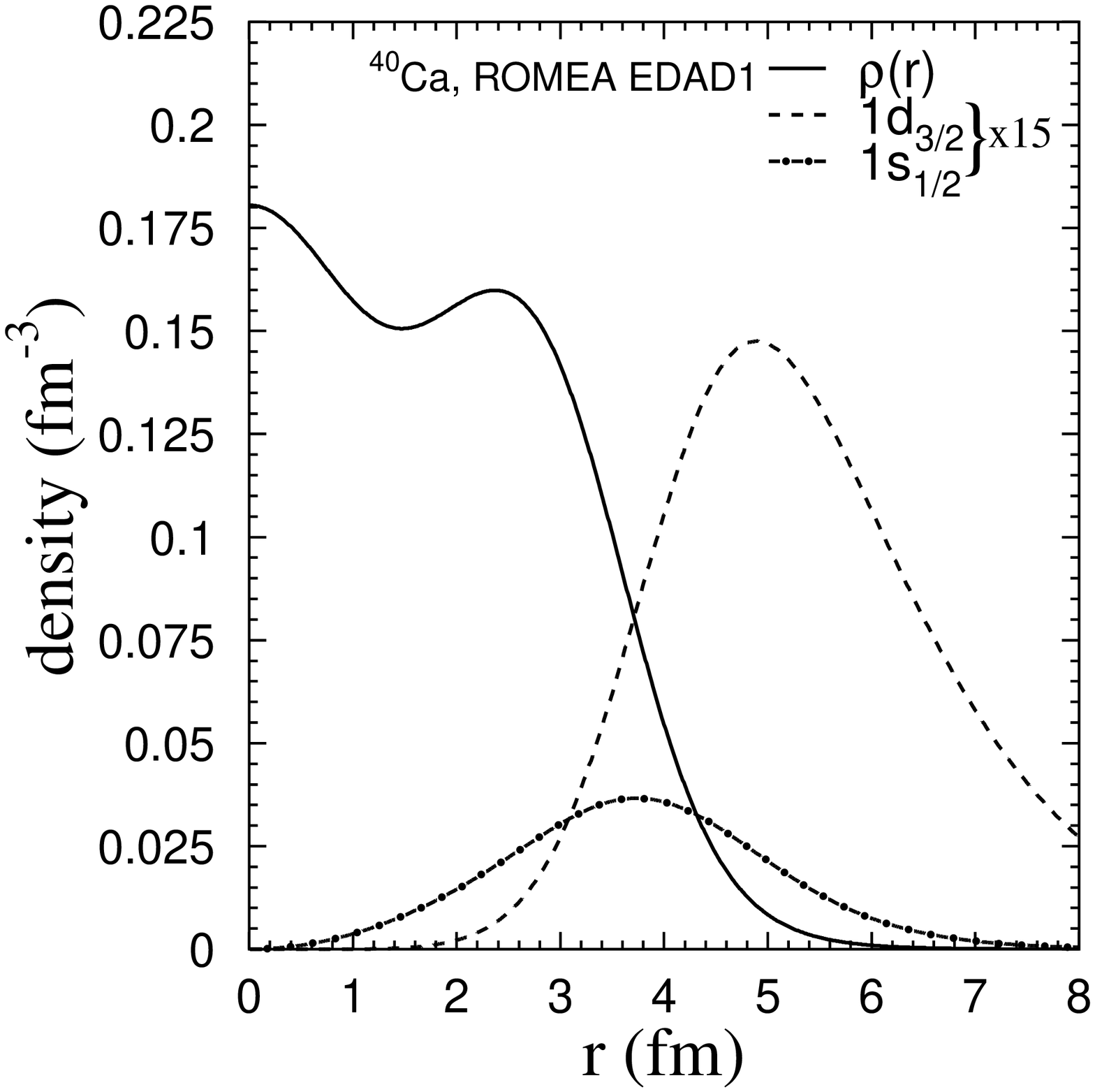}}
\caption{Contribution to the $A(p,2p)$ cross section $\delta \left( r
  \right)$ as a function of $r$.  The upper figures present the
  results obtained after setting the IFSI factor $\mathcal{S}_{IFSI}
  (\vec{r})$ equal to one in Eq.~(\ref{eq:delta_r}), whereas in the
  lower panels the ROMEA calculations using the EDAD1 optical
  potential are depicted.  The dashed (dot-dashed) curves show the
  result for emission from the Fermi (lowest-lying $1s_{1/2}$) level.
  The baryon density $\rho \left( r \right)$ is also shown (solid
  curve).  The ordinate is given for $\rho \left( r \right)$.  The
  $\delta \left( r \right)$ are plotted in units of fm$^2$ up to an
  arbitrary scaling factor.  The kinematics was $T_{p1} = 1$~GeV,
  $T_{k1} = 870$~MeV, $\theta_{1} = 13.4^{\circ}$, and $\theta_{2} =
  67^{\circ}$.}
\label{fig:delta_r}
\end{center}
\end{figure*}

Results for the emission from levels other than the Fermi level have
not been plotted here, as it turns out that the IFSI factors hardly
depend on the single-particle level in which the struck nucleon resides.  The
peculiar spatial characteristics of the different
single-particle orbits have an impact on the observables, though.
Indeed, the distorted momentum-space wave function $\phi_{\alpha_1}^{D}$ of
Eq.~(\ref{eq:dist_mom_space_wave_fctn}) is determined by the values of
the IFSI factor folded with a relativistic bound-state wave function
$\phi_{\alpha_1} \left( \vec{r} \right)$.
As the particles experience less IFSI close to the nuclear surface,
one obtains a stronger reduction of the quasifree cross section for
nucleon knockout from a level which has a larger fraction of its
density in the nuclear interior.  This will become apparent in
Fig.~\ref{fig:delta_r}, but even more so in Sec.~\ref{sec:PNPI}. 

Figure~\ref{fig:delta_r} shows a function $\delta \left( r \right)$
which represents the contribution of the nuclear region with radial
coordinate $r$ to the differential cross section.  The procedure to
calculate this function is similar to the method exposed in
Ref.~\cite{hatanaka97} and is developed in
Appendix~\ref{app:radial_contrib}.  Comparison of the upper and lower
panels illustrates that IFSI mechanisms make the $A(p,2p)$ cross
sections reflect surface mechanisms, unlike the $A(e,e'p)$ reaction
where the weakly
interacting electron probes the entire nuclear volume and only the
outgoing proton interacts with the residual system.  Apart from the
shift to higher $r$, the IFSI brings about a strong reduction in the
magnitude of the cross sections, whereby the Fermi level is least affected.
Even though $\delta \left( r \right)$ is concentrated in the surface
region, the average density seen through this reaction still amounts
to $0.069$~fm$^{-3}$ ($0.080$~fm$^{-3}$) or $32\%$ ($45\%$) of the
central density in the case of $1s_{1/2}$ knockout from $^{12}$C
($^{40}$Ca).  In the case of emission from the Fermi level, on the
other hand, the average density is only $12\%$ ($13\%$) of the central
density for a $^{12}$C ($^{40}$Ca) target.

Also, the IFSI factors for neutron emission are almost identical to
the corresponding IFSI factors for proton knockout and, as expected,
the overall effect of IFSI is more pronounced for heavier target
nuclei.

\subsection{\label{sec:ROMEA_vs_RMSGA} Comparison between ROMEA and
  RMSGA calculations}

\begin{figure}
\begin{center}
\includegraphics[width=0.5\textwidth]{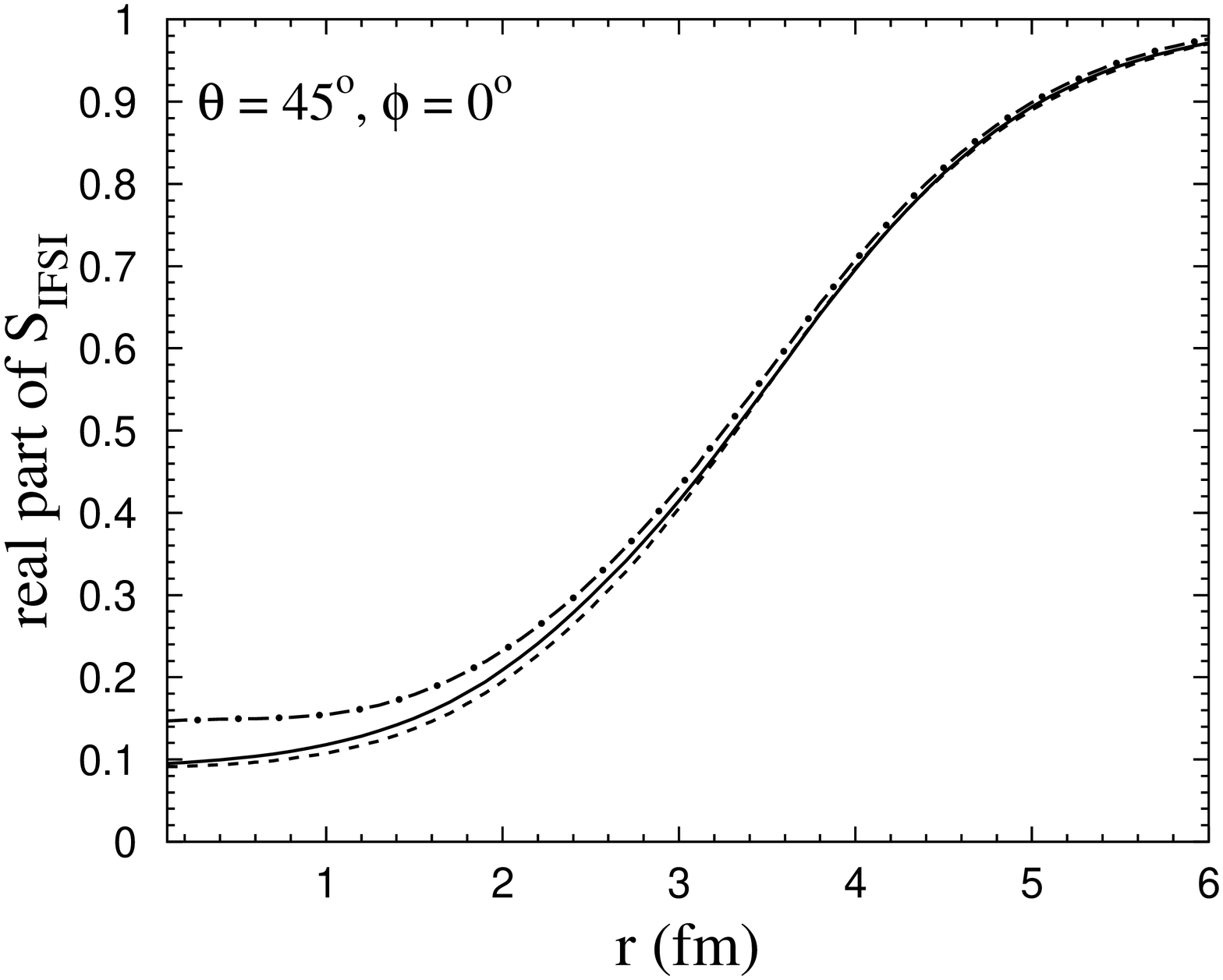}
\includegraphics[width=0.5\textwidth]{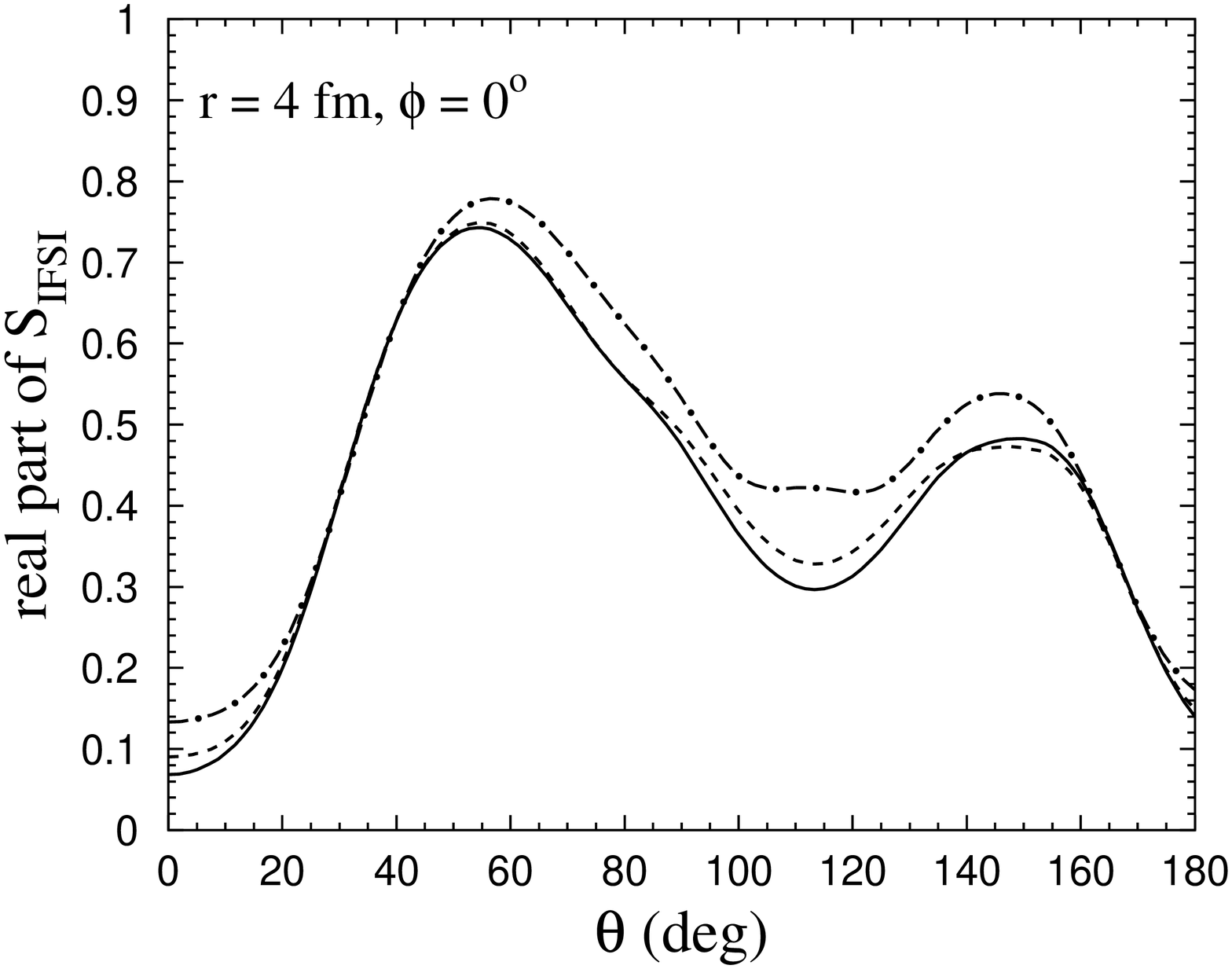}
\includegraphics[width=0.5\textwidth]{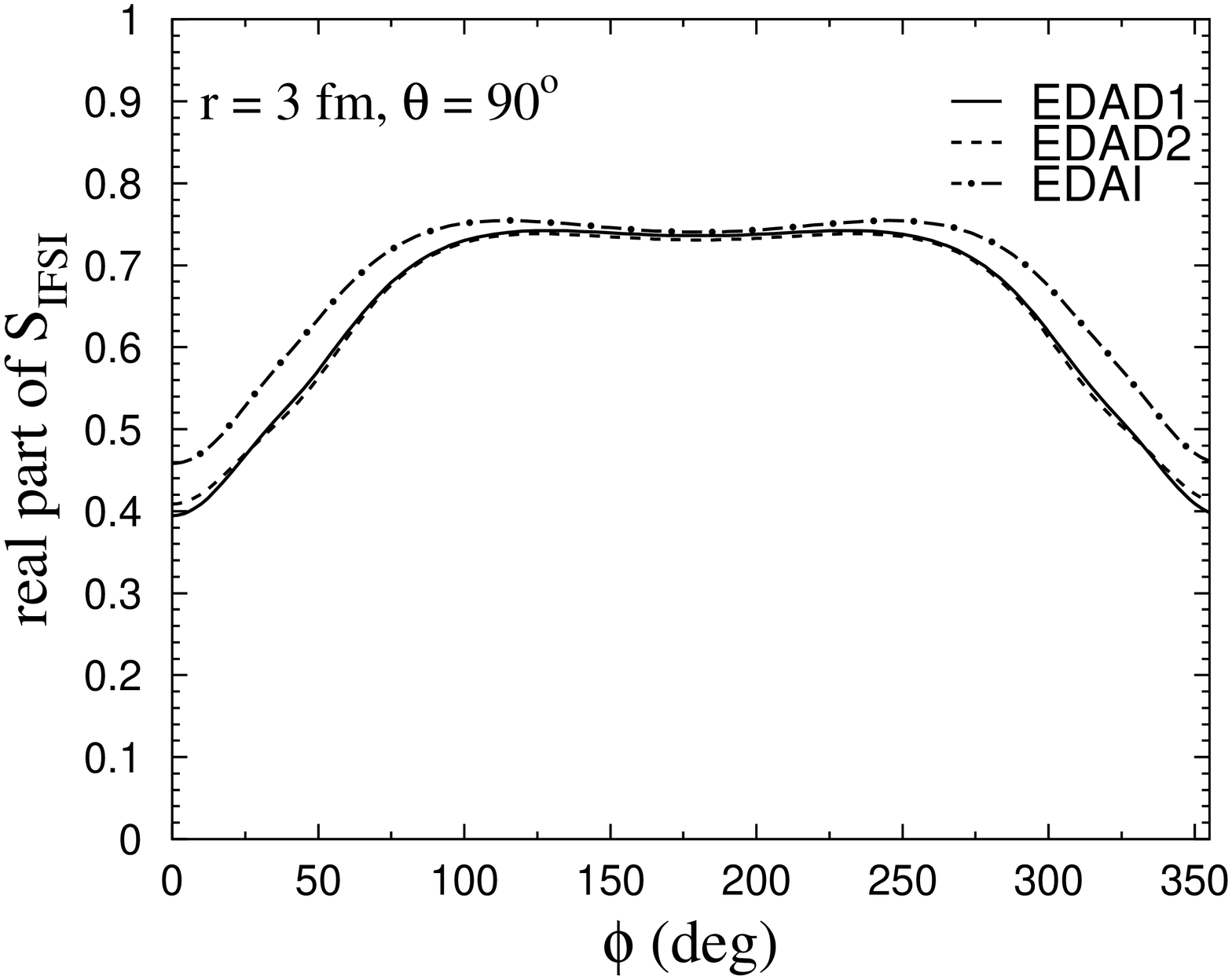}
\caption{The sensitivity of the real part of the complete IFSI factor
  for neutron knockout from $^{12}$C to the adopted choice for the
  parametrization of the optical potentials.  Results of ROMEA
  calculations with the EDAD1 (solid curve), EDAD2 (dashed curve), and
  EDAI (dot-dashed curve) optical potentials are shown.  Kinematics as
  in Fig.~\ref{fig:re.12c.p.edad.phi0.3D}.}
\label{fig:re.12c.n.romea_compar.radial_polar_azimuth}
\end{center}
\end{figure}

\begin{figure*}
\begin{center}
\resizebox{0.95 \textwidth}{!}
{\includegraphics{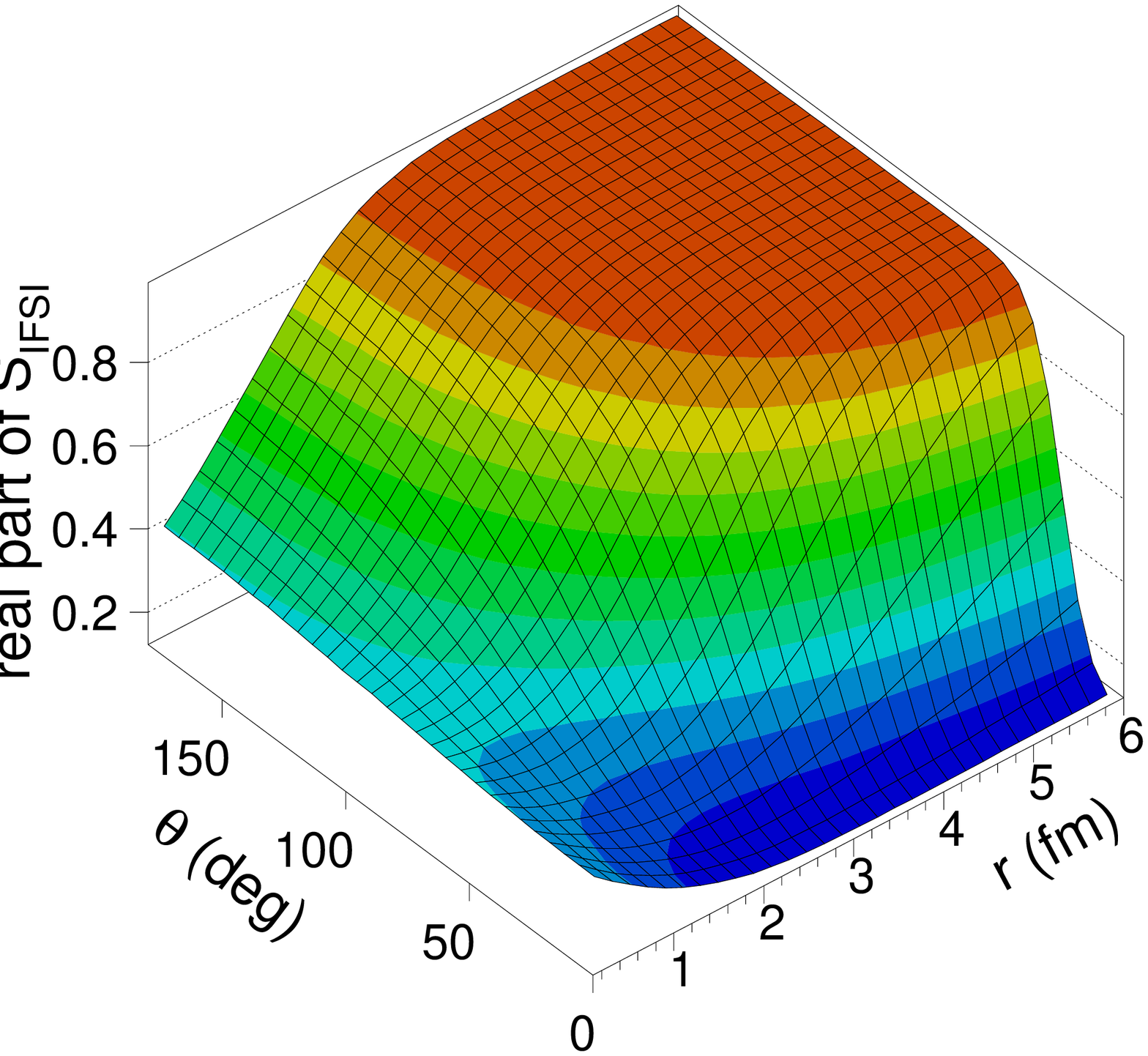}
  \includegraphics{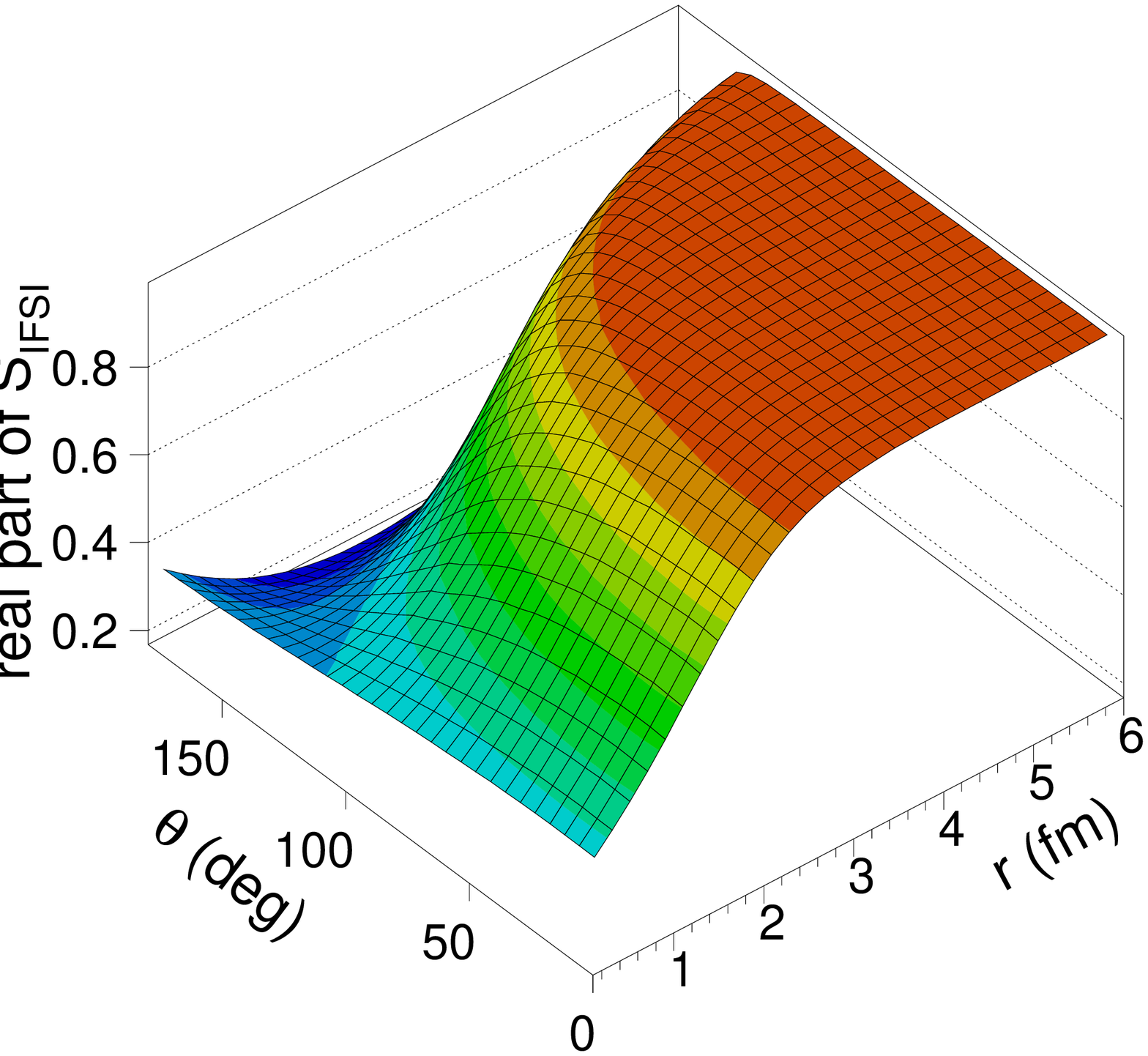}}
\resizebox{0.95 \textwidth}{!}
{\includegraphics{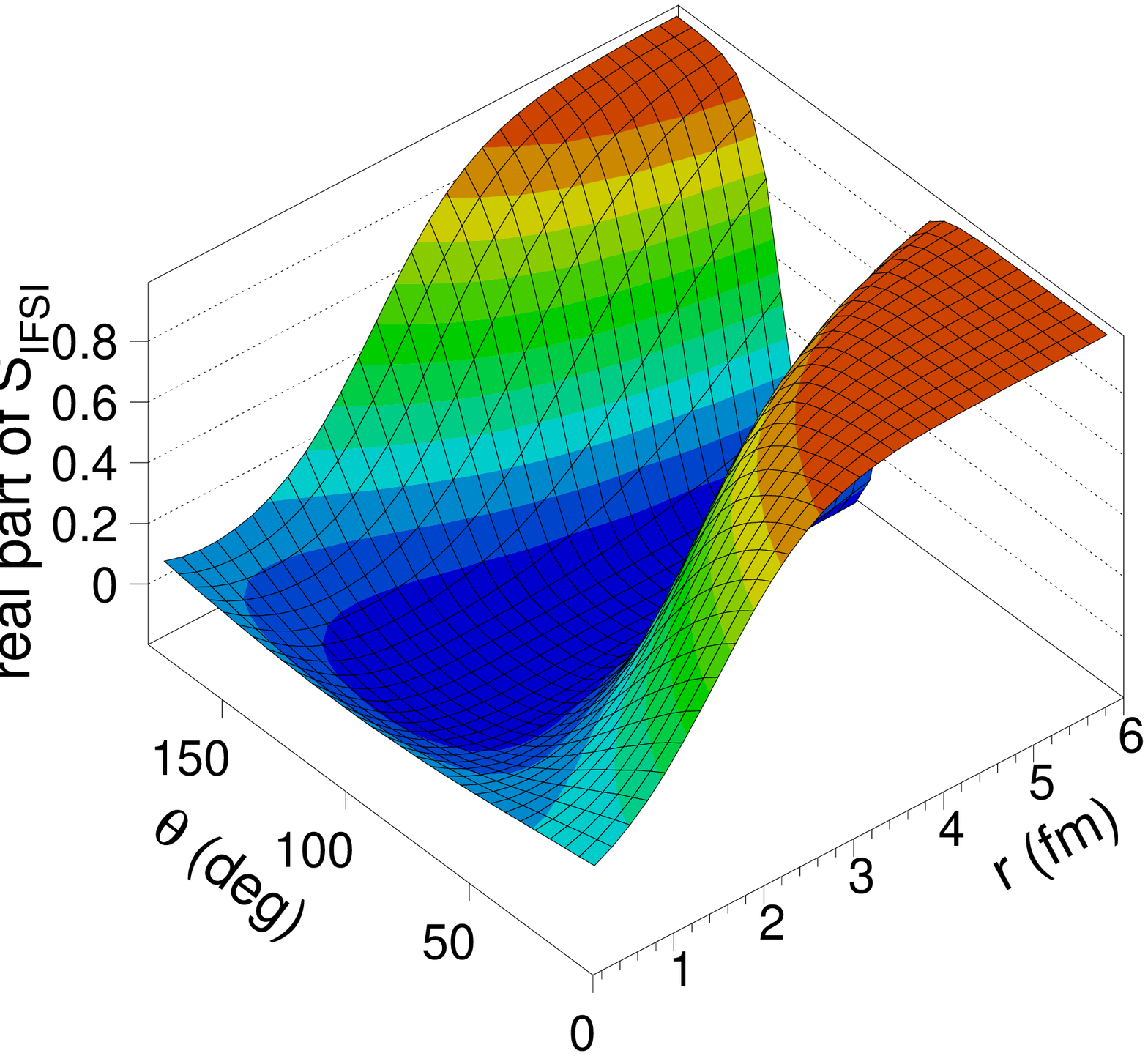}
  \includegraphics{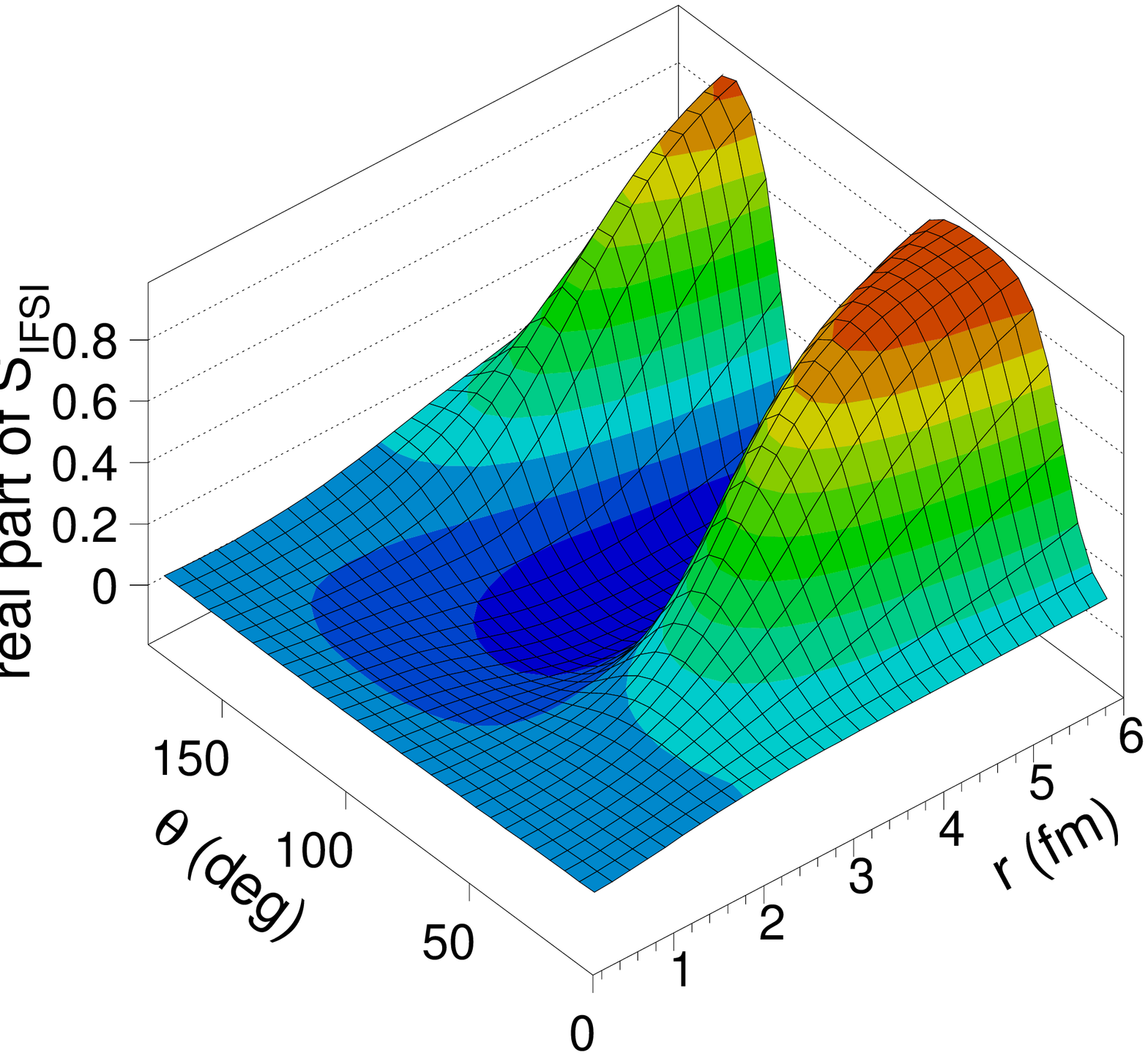}}
\caption{(Color online) As in Fig.~\ref{fig:re.12c.p.edad.phi0.3D} but
  now using the RMSGA method.}
\label{fig:re.12c.p.rmsga.phi0.3D}
\end{center}
\end{figure*}

In this subsection, we investigate the sensitivity of the computed IFSI factors
to the adopted parametrizations for the optical potentials and compare
the ROMEA results with the RMSGA predictions.  As can be seen in
Fig.~\ref{fig:re.12c.n.romea_compar.radial_polar_azimuth}, the IFSI
factor depends on whether $A$-dependent (EDAD1/EDAD2) or
$A$-independent (EDAI) fits for the potentials are selected, but the
global features are comparable.
Figure~\ref{fig:re.12c.p.rmsga.phi0.3D}, as contrasted to
Fig.~\ref{fig:re.12c.p.edad.phi0.3D}, demonstrates that the RMSGA
method adequately describes the ISI of the incoming proton and the FSI
of the scattered proton.  However, the discrepancies between ROMEA and
RMSGA become significant in the calculation of the FSI of the ejected
nucleon (note the different scales in the bottom left panels of
Figs.~\ref{fig:re.12c.p.edad.phi0.3D} and \ref{fig:re.12c.p.rmsga.phi0.3D}),
and therefore also in the complete IFSI factor.  The noted
difference is attributed to the low ejectile kinetic energy ($T_{k2}
\approx 114$~MeV for the specific case of
Fig.~\ref{fig:re.12c.p.rmsga.phi0.3D}, and comparable values for
knockout from other levels and other nuclei).  At such low energies,
the RMSGA predictions are not realistic due to the underlying
approximations, mostly the postulation of linear trajectories and
frozen spectator nucleons.  So, for the kinematics discussed here, the
ROMEA method is to be preferred over the RMSGA one, as the latter
overestimates the distortion for the low-energetic ejectile nucleon.

\section{\label{sec:cross_sect_results} Numerical results for
  $A(p,pN)$ differential cross sections} 

\subsection{\label{sec:PNPI} The PNPI experiment}

The PNPI experiment \cite{belostotsky87} was carried out with an
incident proton beam of energy $1$~GeV.  The scattered proton was
detected at $\theta_{1} = 13.4^{\circ}$ with a kinetic energy between
$800$ and $950$~MeV, while the knocked-out nucleon was observed at
$\theta_{2} = 67^{\circ}$ having a kinetic energy below $200$~MeV.

\begin{figure*}
\begin{center}
\resizebox{0.95 \textwidth}{!}
{\includegraphics{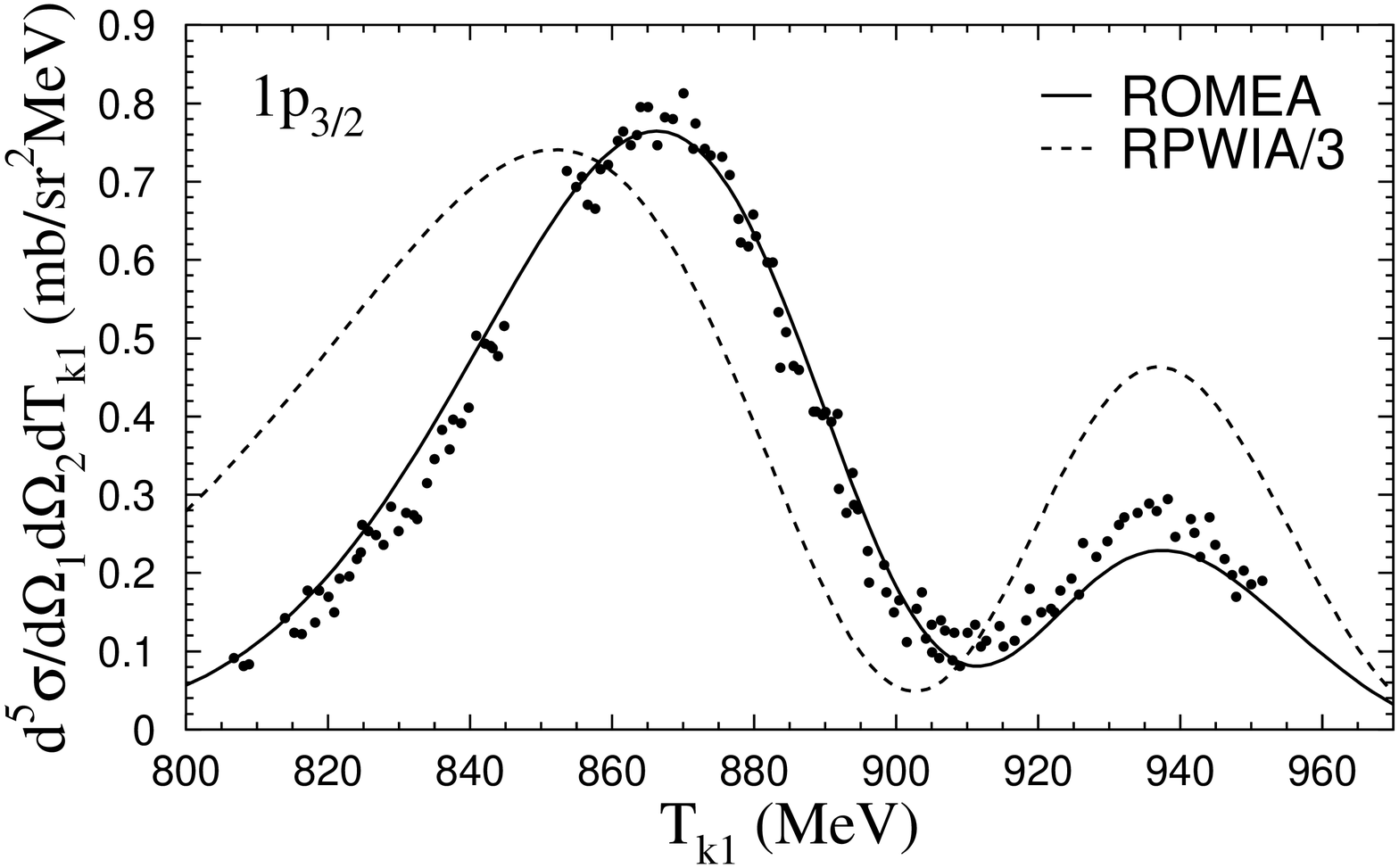}
  \includegraphics{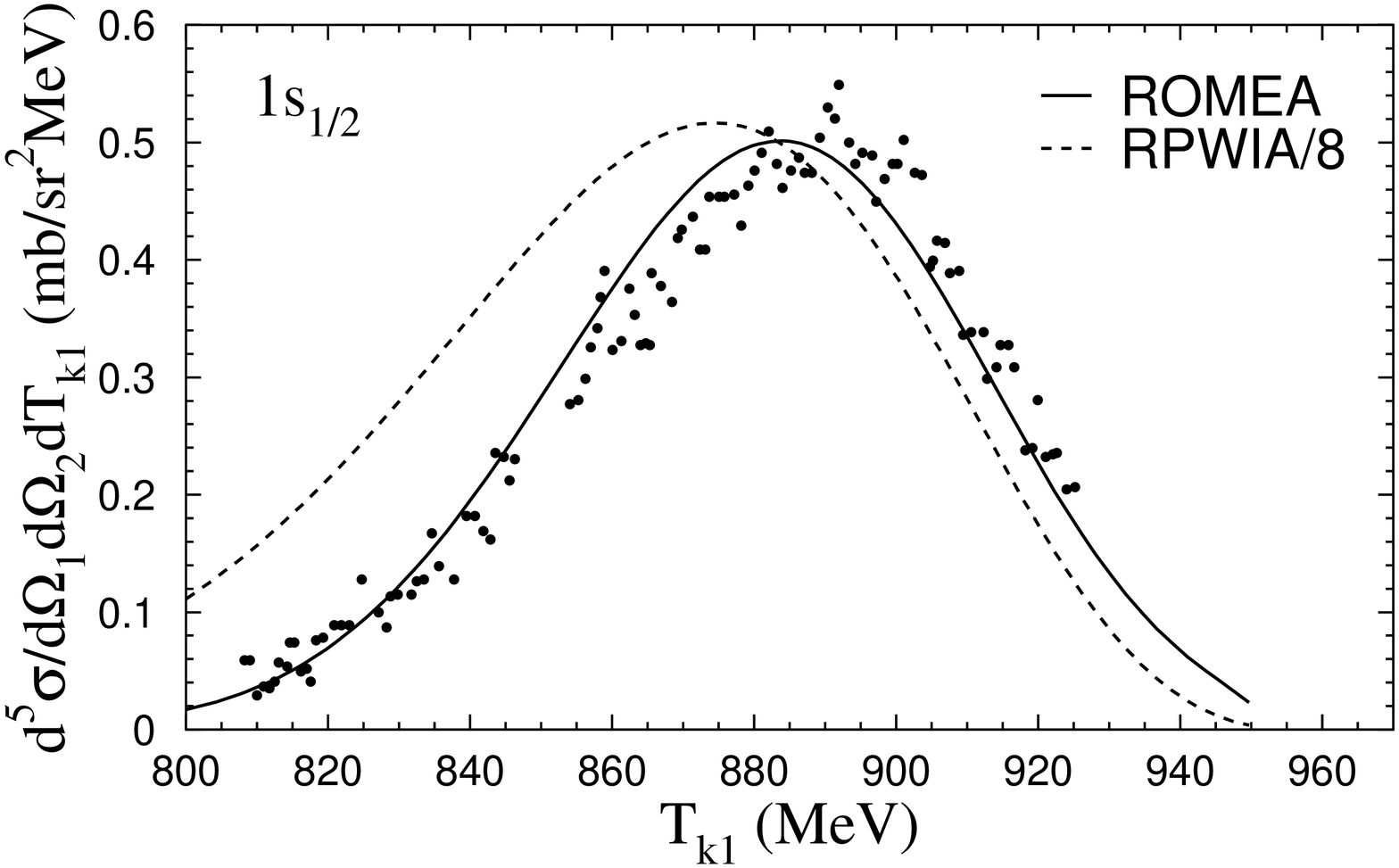}}
\caption{Differential cross section for the $^{12}$C$(p,2p)$ reaction.
  The solid curve represents the ROMEA calculation, whereas the dashed
  curve is the plane wave result reduced by the indicated factor.  The ROMEA
  results are normalized to the data.  Data points are from
  Ref.~\cite{belostotsky87}.  The magnitude of the experimental error bars is
  estimated to be of the order of $5$--$10\%$.}
\label{fig:diffcross.12c.p.romea.edai}
\end{center}
\end{figure*}

\begin{figure*}
\begin{center}
\resizebox{0.95 \textwidth}{!}
{\includegraphics{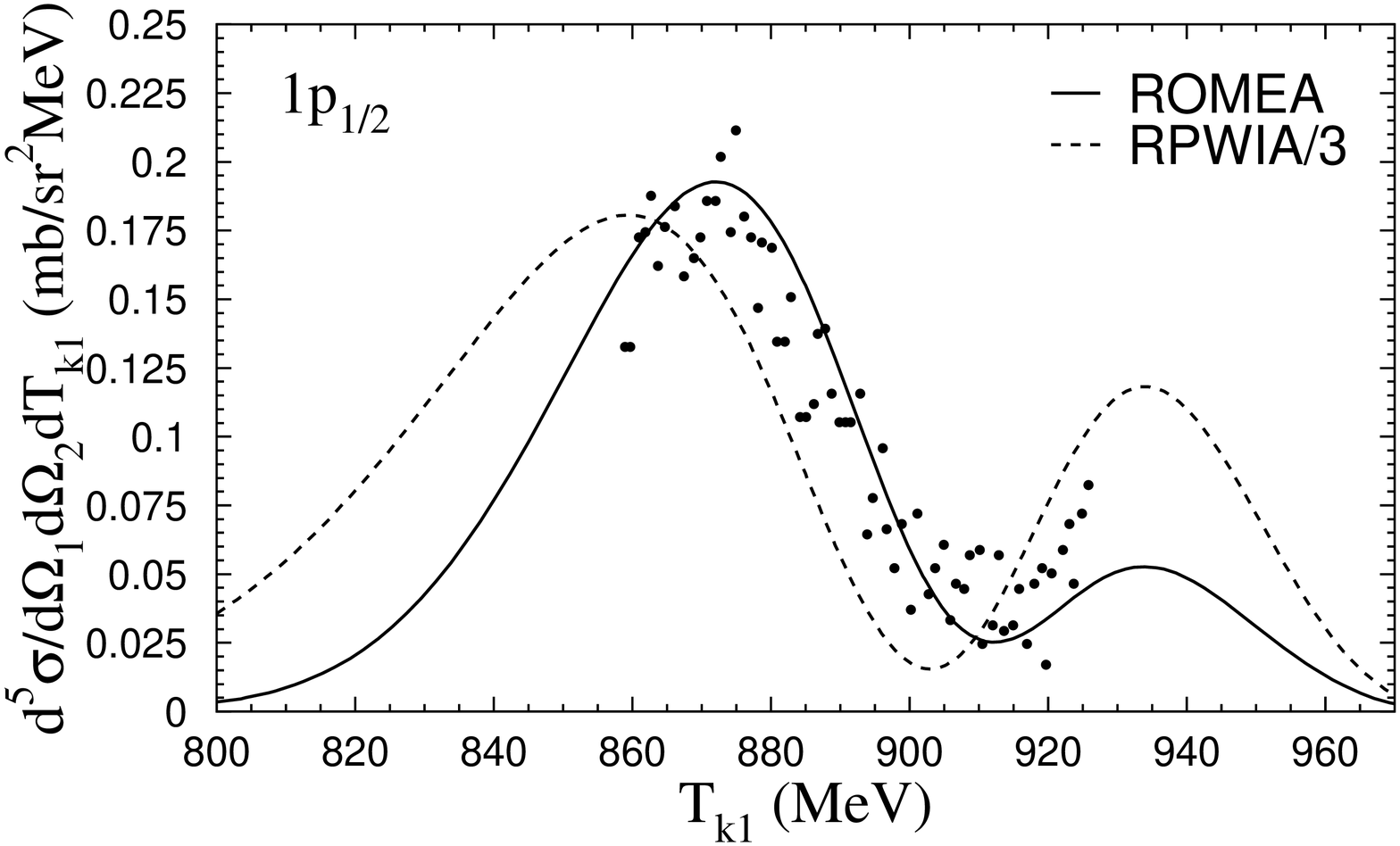}
  \includegraphics{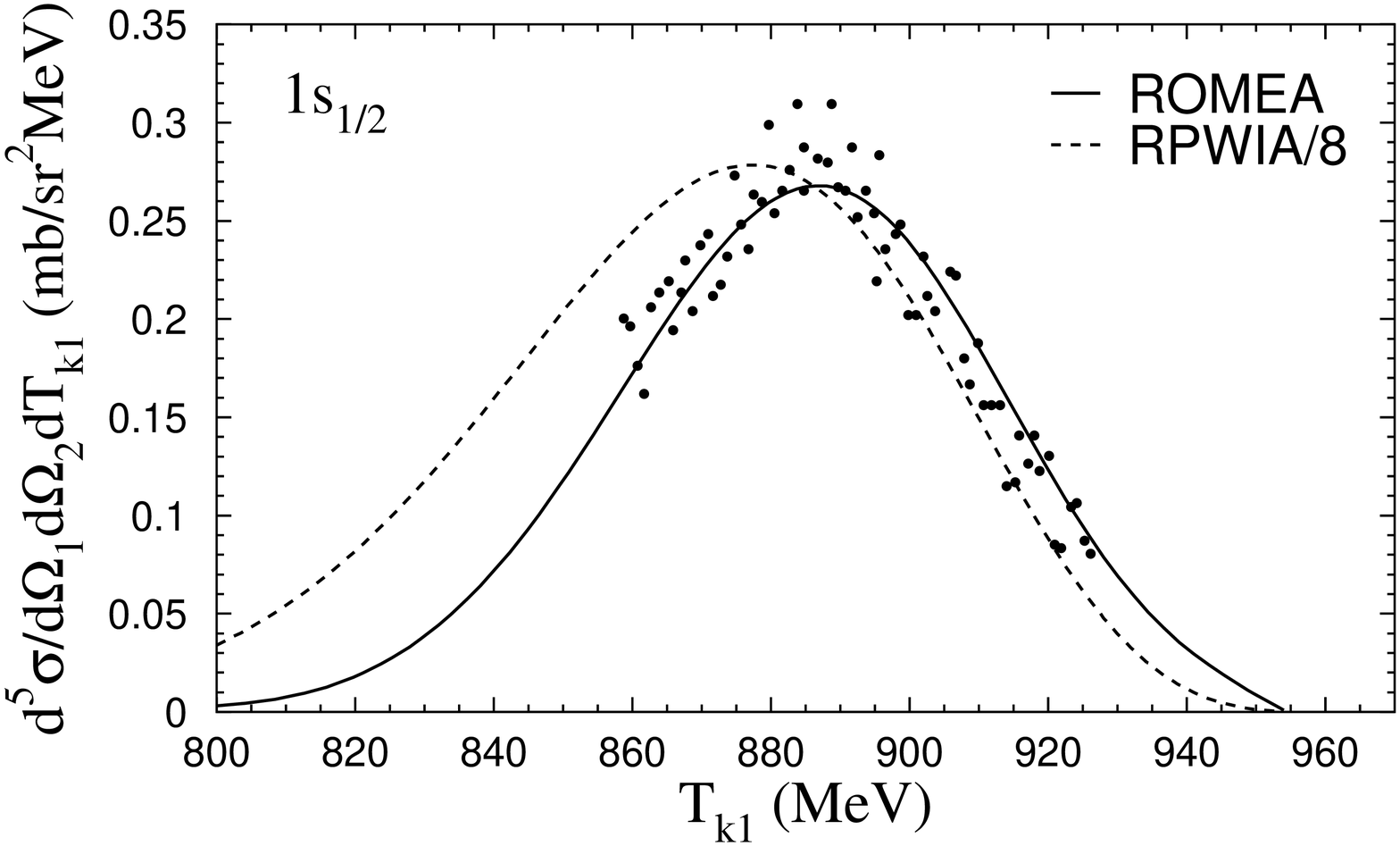}}
\caption{As in Fig.~\ref{fig:diffcross.12c.p.romea.edai} but for the
  $^{16}$O$(p,pn)$ reaction.}
\label{fig:diffcross.16o.n.romea.edai}
\end{center}
\end{figure*}

\begin{figure*}
\begin{center}
\resizebox{0.95 \textwidth}{!}
{\includegraphics{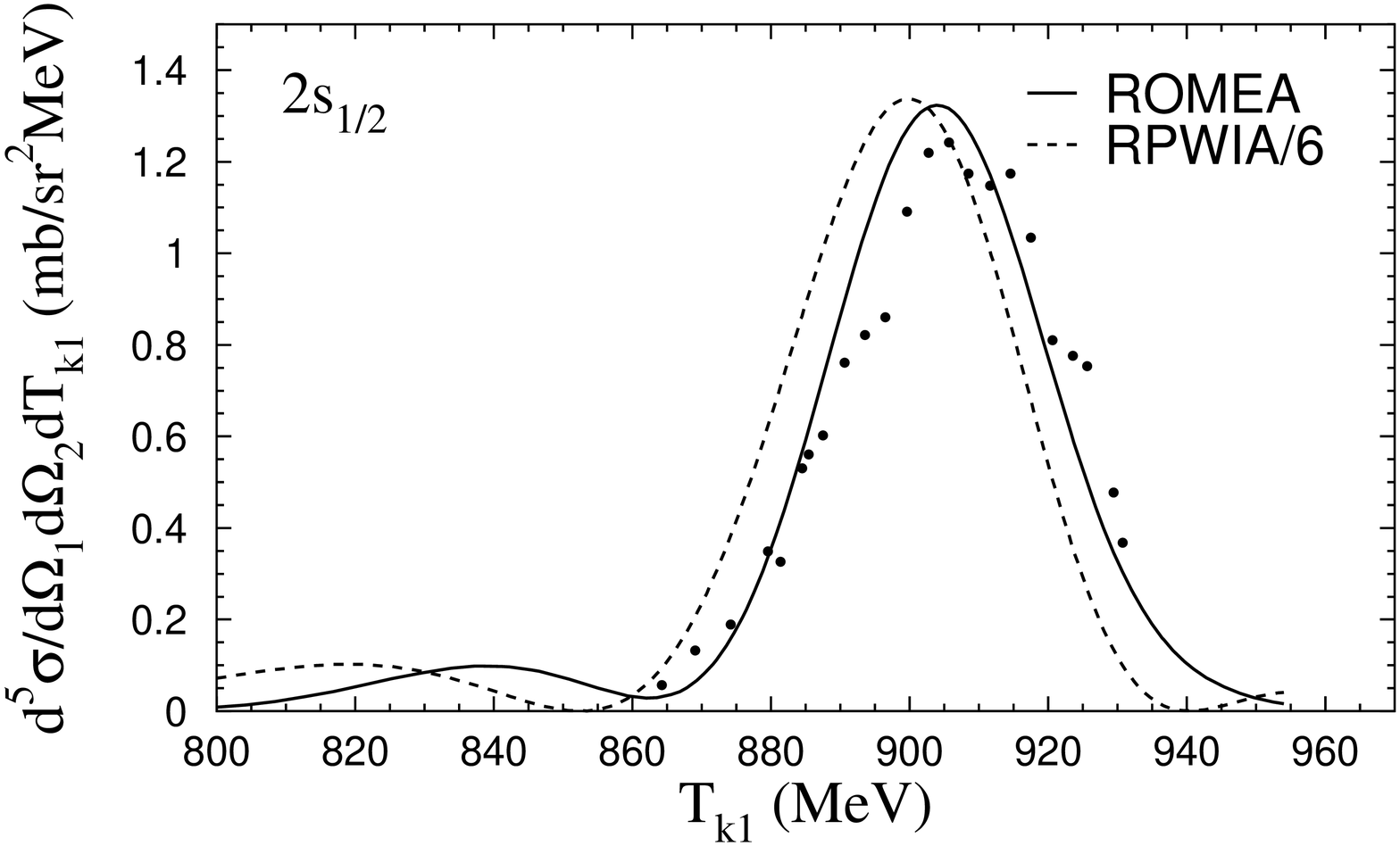}
  \includegraphics{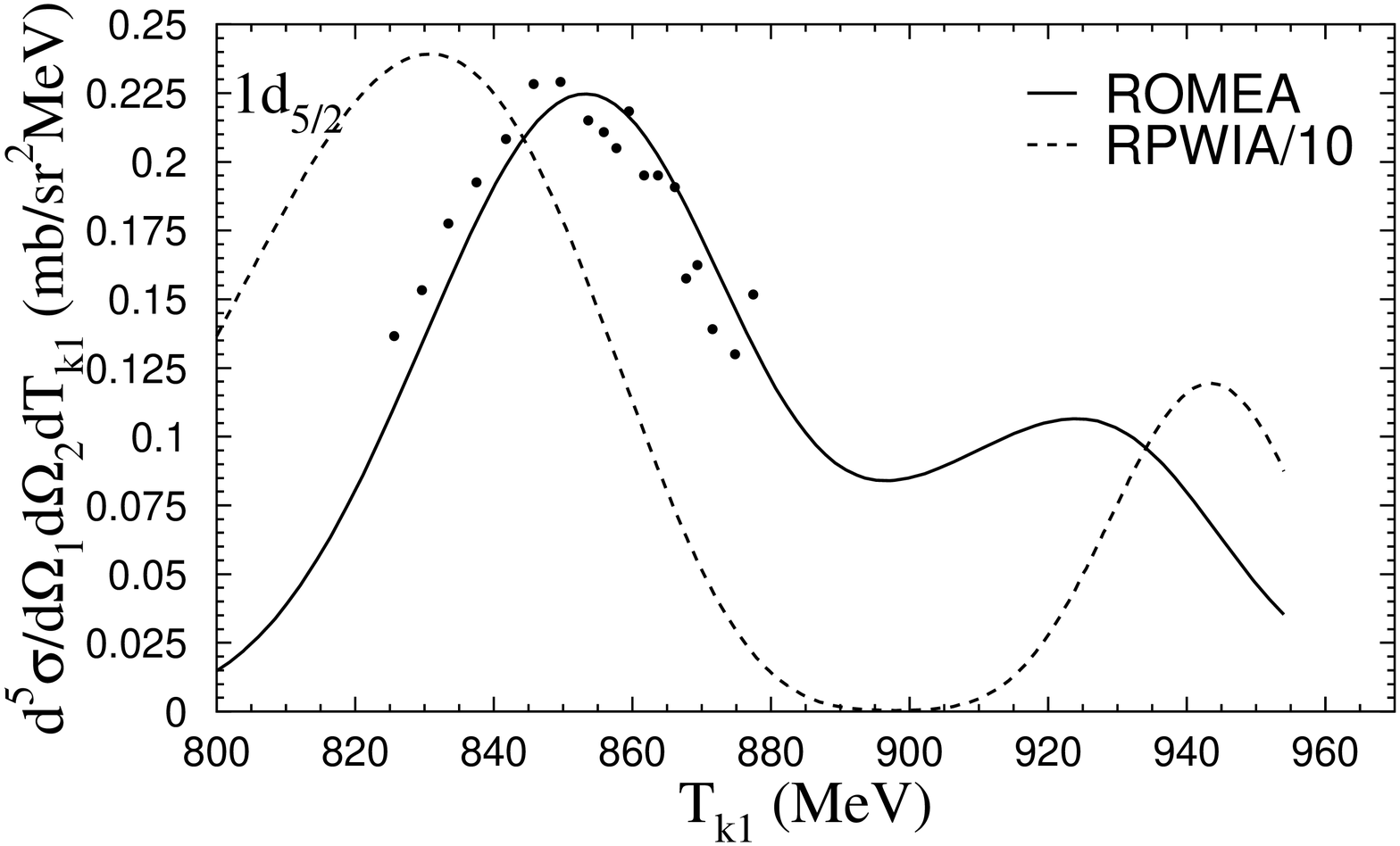}}
\caption{As in Fig.~\ref{fig:diffcross.12c.p.romea.edai} but for the
  $^{40}$Ca$(p,pn)$ reaction.}
\label{fig:diffcross.40ca.n.romea.edai}
\end{center}
\end{figure*}

Figures~\ref{fig:diffcross.12c.p.romea.edai}--\ref
{fig:diffcross.40ca.n.romea.edai} display a selection of differential
cross section results as a function of the kinetic energy of the most
energetic nucleon in the final state.  The EDAI optical potential
\cite{cooper93} was used for the ROMEA calculations.  The other
parametrizations of Ref.~\cite{cooper93} produce similar predictions,
whereas the RMSGA approach fails to give an adequate description of
the data because of the low kinetic energy of the ejected
nucleon.  Since the experiment of Ref.~\cite{belostotsky87} only
measured relative cross sections, the ROMEA results were normalized to
the experimental data.

\begin{figure}
\begin{center}
\includegraphics[width=0.5\textwidth]{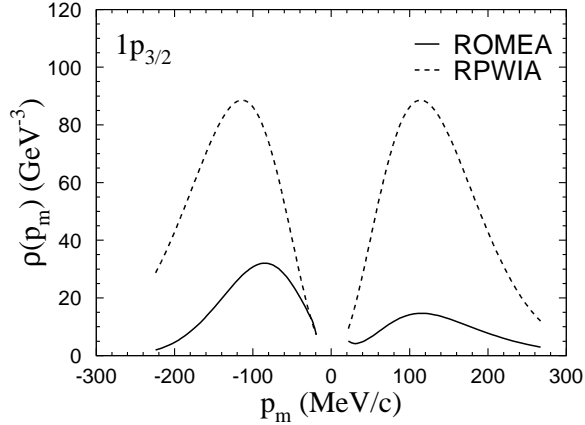}
\caption{The $^{12}$C$(p,2p)$ momentum distribution for the $1p_{3/2}$
  state as a function of the missing momentum.  The solid (dashed)
  curve are ROMEA (RPWIA) calculations.}
\label{fig:momdistr.12c.p.1p32.romea.edai}
\end{center}
\end{figure}

The ROMEA calculations reproduce the shapes of the measured
differential cross sections.  Furthermore, comparison of the RPWIA and
ROMEA calculations shows that the effect of the IFSI is twofold.  First,
IFSI result in a reduction of the RPWIA cross section that is
both level and $A$ dependent.  From the figures it is clear that
ejection of a nucleon from a deeper lying level leads to stronger
initial- and final-state distortions.  This reflects the fact that the
incoming and outgoing nucleons encounter more obstacles when a deeper
lying bound nucleon is probed.  The $A$
dependence also conforms with our expectations, i.e., the IFSI effects
are larger for heavier nuclei.  Besides the attenuation, the IFSI also
make the measured missing momentum different from the initial momentum
of the struck nucleon.  As can be inferred from
Fig.~\ref{fig:momdistr.12c.p.1p32.romea.edai}, this momentum shift
leads to an asymmetry between the positive and negative
missing-momentum side of the momentum distribution.  Note that positive
missing momentum corresponds to $p_{m_x} = k_1 \sin \theta_1 + k_2
\sin \theta_2 \cos \phi_2 > 0$.

\subsection{\label{sec:TRIUMF} The TRIUMF $^{4}$He$(p,2p)$ experiment}

\begin{figure*}
\begin{center}
\resizebox{0.95 \textwidth}{!}
{\includegraphics{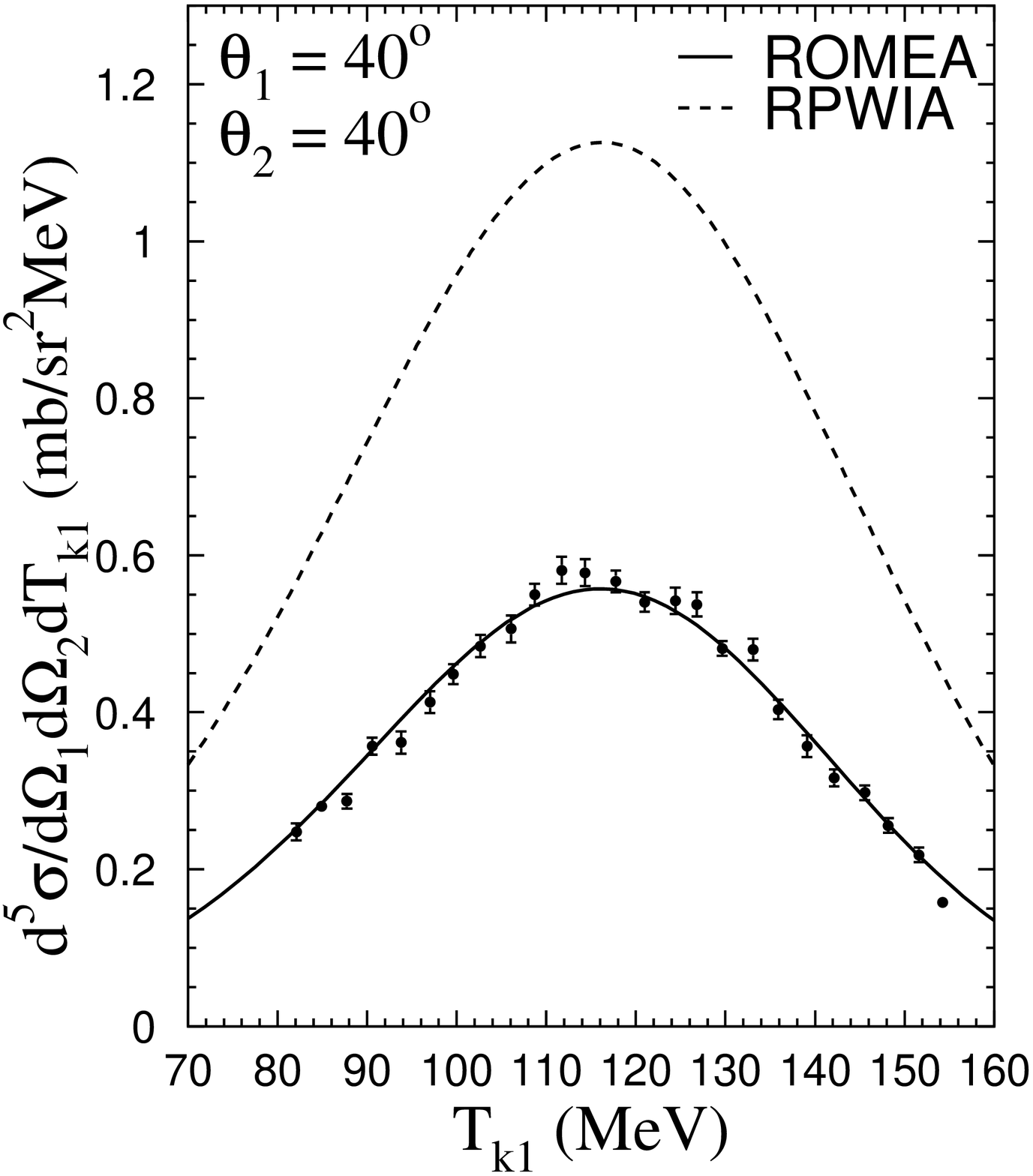}
  \includegraphics{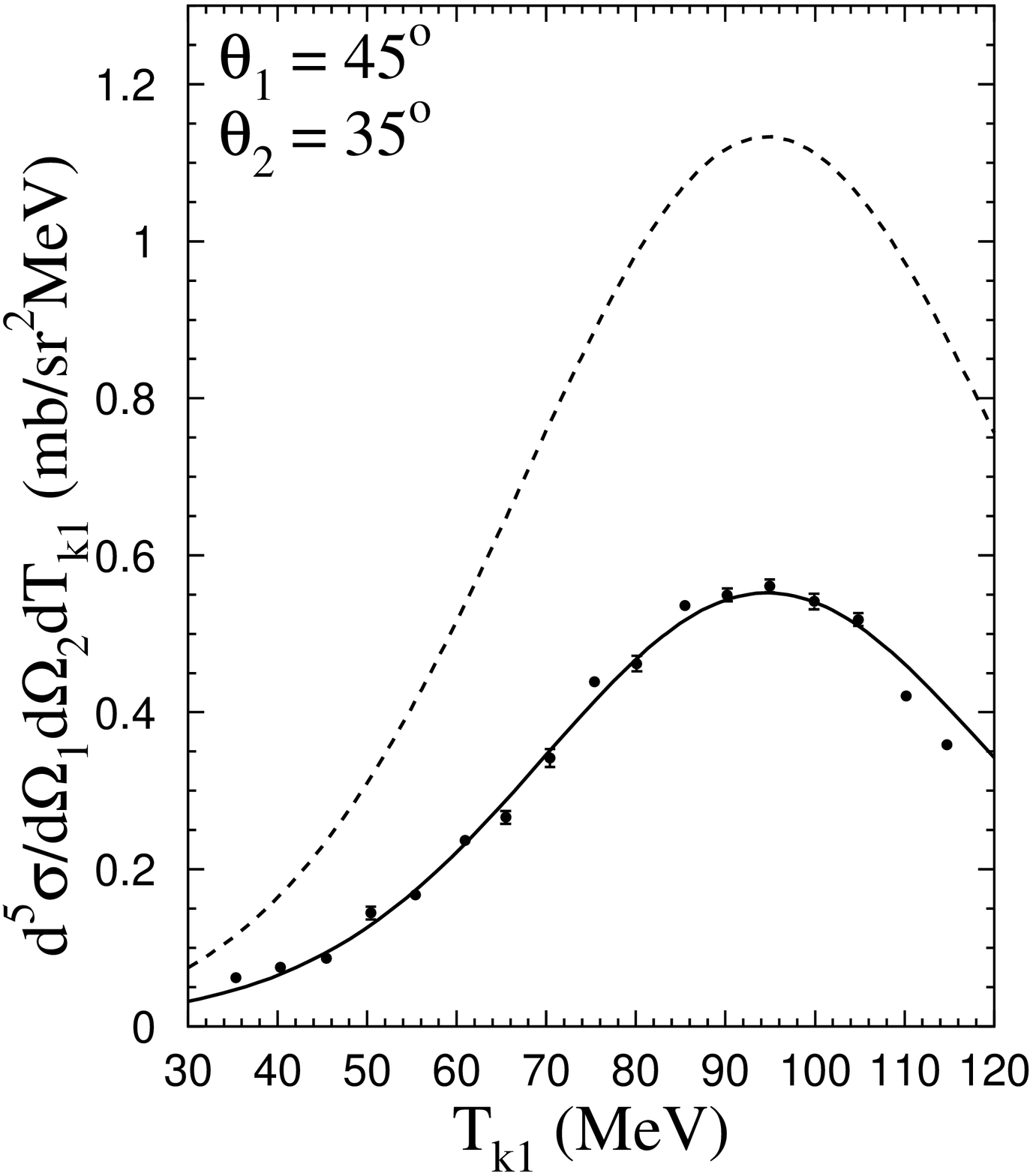}}
\caption{Differential cross section for the $^{4}$He$(p,2p)$ reaction
  at the angle pairs $(40^{\circ}, 40^{\circ})$ and $(45^{\circ},
  35^{\circ})$ at $250$~MeV.  The solid (dashed) curves refer to ROMEA
  (RPWIA) results.  The data are from Ref.~\cite{oers82}.}
\label{fig:diffcross.4he.p.vanOers.250}
\end{center}
\end{figure*}

Finally, we present some results for the
$^{4}$He$(p,2p)$ reaction at an incident proton energy of $250$~MeV.
Figure~\ref{fig:diffcross.4he.p.vanOers.250} compares the data from
the TRIUMF experiment \cite{oers82} with ROMEA calculations using the
optical potential of van Oers \textit{et al.} \cite{oers82}.  The
typical shape for knockout of an $s$-state proton is reproduced by the
ROMEA predictions.  This fair agreement of the ROMEA results with the
data demonstrates that our ROMEA model also works satisfactorily at
lower incident energies.

\section{\label{sec:concl} Conclusions}

A relativistic and cross-section factorized framework to describe the
IFSI in quasielastic $A(p,pN)$ reactions has been outlined.  The
model, which relies on the eikonal approximation, can use either
optical potentials or Glauber multiple-scattering theory to deal with
IFSI.  Thanks to the freedom of choice between these two substantially
different techniques, our model is expected to be applicable at both
intermediate and high incident energies.

The properties of the IFSI factor have been investigated for an
incident proton energy of $1$~GeV.  Not surprisingly, the strongest
attenuation occurs in the nuclear interior and heavier target nuclei
are found to induce larger IFSI effects.  Also, the surface-peaked
character of the $A(p,pN)$ reaction was clearly established to be a
consequence of the IFSI.  Whereas the different types of
optical-potential sets contained in Ref.~\cite{cooper93} yield
comparable IFSI factors, the RMSGA calculations exhibit an unrealistic
behavior for these kinematics.

The ROMEA model has been used to calculate cross sections for the
kinematics of two different experiments~: quasielastic proton
scattering from $^{12}$C, $^{16}$O, and $^{40}$Ca at $1$~GeV, and
$^{4}$He$(p,2p)$ scattering at $250$~MeV.  The predictions are shown
to reproduce the shape of the data reasonably well at both incoming energies,
thereby providing support for the wide applicability range of our
model.

Although the RMSGA approach was deemed unsuitable for the kinematics
discussed here, it should prove useful when trying to describe nuclear
transparency data.  Work in this direction is in progress.

\begin{acknowledgments}
This work was supported by the Fund for Scientific Research, Flanders
(FWO) and the Research Council of Ghent University.
\end{acknowledgments} 

\appendix

\section{\label{app:bound} Relativistic bound-state wave functions}

For spherically symmetric potentials, the solutions $\phi_{\alpha}
\left(\vec{r} \right)$ to a single-particle Dirac equation have the
form \cite{Walecka2001}
\begin{equation}
\phi_{\alpha} \left( \vec{r}, \vec{\sigma} \right) \equiv
\phi_{n \kappa m } \left( \vec{r}, \vec{\sigma} \right) =
\left[ \begin{array}{c}
i \frac {G_{n \kappa}  ( r )} {r} \;
{\cal Y}_{\kappa m} (\Omega, \vec{\sigma}) \\
- \frac {F_{n \kappa} ( r )} {r } \;
{\cal Y}_{- \kappa m} (\Omega, \vec{\sigma}) 
\end{array} \right] \; ,
\label{eq:bound_state}
\end{equation}
where $n$ denotes the principal, $\kappa$ and $m$ the generalized
angular momentum quantum numbers.  The ${\cal Y}_{\pm \kappa m}$ are
the spin spherical harmonics and determine the angular and spin parts
of the wave function,
\begin{eqnarray}
{\cal Y}_{\kappa m} (\Omega, \vec{\sigma}) = \sum_{m_{l} m_{s}}
\left< l m_{l} \frac{1}{2} m_{s} \mid j m \right> Y_{l m_{l}} (\Omega)
\chi_{\frac{1}{2} m_{s}} (\vec{\sigma}) \; , \nonumber \\ 
{\cal Y}_{- \kappa m} (\Omega, \vec{\sigma}) = \sum_{m_{l} m_{s}}
\left< \bar{l} m_{l} \frac{1}{2} m_{s} \mid j m \right>
Y_{\bar{l} m_{l}} (\Omega)
\chi_{\frac{1}{2} m_{s}} (\vec{\sigma}) \; , 
\label{eq:spin_spherical_harm}
\end{eqnarray}
with
\begin{equation}
j = |\kappa| - \frac{1}{2} \; , \hspace{0.8cm} l = \left\{
\begin{array}{ll}
\kappa, & \kappa > 0 \\
-(\kappa+1), & \kappa < 0
\end{array}
\right.
\; ,
\hspace{0.8cm} \bar{l} = 2 j - l = \left\{
\begin{array}{ll}
\kappa - 1, & \kappa > 0 \\
- \kappa, & \kappa < 0
\end{array}
\right.
\; .
\end{equation}

The Fourier transform of the relativistic bound-nucleon wave function
is given by
\begin{equation}
\phi_{\alpha} \left( \vec{p} \right) =
\int d \vec{r}
e^{- i \vec{p} \cdot \vec{r}} \phi_{\alpha} \left( \vec{r} \right)
= \left( - i \right)^l 
(2\pi)^{3/2}
\left[ \begin{array}{c}
g_{n \kappa} ( p ) \; {\cal Y}_{\kappa m} (\Omega_p) \\
- S_{\kappa} \; f_{n \kappa} ( p ) \; {\cal Y}_{- \kappa m} (\Omega_p)
\end{array} \right]
\; ,
\label{eq:Fourier_bound_wave}
\end{equation}
with $S_{\kappa} = \kappa / \left| \kappa \right|$.  The radial
functions $g_{n \kappa}$ and $f_{n \kappa}$ in momentum space are
obtained from their counterparts in coordinate space~:
\begin{subequations}
\begin{equation}
g_{n \kappa} ( p ) =
i \; \sqrt{\frac{2}{\pi}} \int_0^{\infty} r^2 dr \;
\frac{G_{n \kappa} ( r )} {r} j_l ( pr ) \; ,
\end{equation}
\begin{equation}
f_{n \kappa} ( p ) =
i \; \sqrt{\frac{2}{\pi}} \int_0^{\infty} r^2 dr \;
\frac{F_{n \kappa} ( r )} {r} j_{\bar{l}} ( pr ) \; ,
\end{equation}
\end{subequations}
with $j_l ( pr )$ the spherical Bessel function of the first kind. 

\section{\label{app:radial_contrib} Radial contribution to the
  $A(p,pN)$ cross section}

The differential $A(p,pN)$ cross section
(\ref{eq:factorized_ifsi_cross_sect}) is proportional to the distorted
momentum distribution $\rho^{D}$ of Eq.~(\ref{eq:dist_mom_dist}).
When approximating the completeness relation (\ref{eq:completeness})
as
\begin{equation}
\sum_{s}
u ( \vec{p}_{m}, s ) \bar{u} ( \vec{p}_{m}, s ) \approx 1
\; ,
\label{eq:approx_completeness}
\end{equation}
i.e., neglecting the negative-energy term as in Sec.~\ref{sec:RPWIA},
this amounts to
\begin{equation}
\left( \frac{d^{5}\sigma}{dE_{k1} d\Omega_{1} d\Omega_{2}} \right)^{D}
\propto \sum_{m} \overline{\phi_{\alpha_1}^{D}} \: \phi_{\alpha_1}^{D}
\; .
\label{eq:cross_sect_propto_dist_mom_dist}
\end{equation}

Thus, with $D \left( r \right)$ defined as
\begin{equation}
D \left( r \right) \equiv
\int d\Omega \: r^2 \: e^{- i \vec{p}_{m} \cdot \vec{r}} \:
\phi_{\alpha_1} \left( \vec{r} \right) \:
\mathcal{S}_{IFSI} \left( \vec{r} \right)
\; ,
\label{eq:D_r}
\end{equation} 
the function
\begin{eqnarray}
\delta \left( r_{1} \right)
& \equiv & \sum_{m} \frac{1}{\Delta R}
\Biggl[
\int_{0}^{\infty} dr \: \overline{D} \left( r \right)
\int_{0}^{\infty} dr \: D \left( r \right) -
\Biggr.
\nonumber \\ & &
\Biggl. \Bigl(
\int_{0}^{r_{1}} dr \: \overline{D} \left( r \right) +
\int_{r_{1} + \Delta R}^{\infty} dr \: \overline{D} \left( r \right)
\Bigr)
\Bigl(
\int_{0}^{r_{1}} dr \: D \left( r \right) +
\int_{r_{1} + \Delta R}^{\infty} dr \: D \left( r \right)
\Bigr)
\Biggr]
\label{eq:delta_r}
\\ & = &
\sum_{m}
\bigl(
\overline{D} \left( r_{1} \right) \: \phi_{\alpha_1}^{D} +
\overline{\phi_{\alpha_1}^{D}} \: D \left( r_{1} \right)
\bigr)
\nonumber
\end{eqnarray}
represents the contribution of an infinitesimal interval in $r$ around
$r_{1}$ to the $A(p,pN)$ cross section.  This procedure also enables
us to estimate the average density seen through this reaction as
\begin{equation}
\overline \rho \equiv 
\frac{\int_{0}^{\infty} \rho \left( r \right) \: \delta \left( r
  \right) \: dr}{\int_{0}^{\infty} \delta \left( r \right) \: dr}
\; .
\label{eq:average_rho}
\end{equation}



\end{document}